\documentclass{aa}
\usepackage{graphicx}
\usepackage{txfonts}
\usepackage{natbib}
\usepackage{booktabs}
\usepackage[bookmarks=true,bookmarksopen=true,pdfstartview = FitV,pdfpagelayout = SinglePage,colorlinks = true,linkcolor=blue,citecolor=blue,urlcolor = blue]{hyperref} 
\bibpunct{(}{)}{;}{a}{}{,}
\let\oldbibliography\thebibliography
\renewcommand{\thebibliography}[1]{
  \oldbibliography{#1}
  \setlength{\itemsep}{0pt}
} 

\newcommand{\sgra}{Sgr~A$\textrm{*}$}
\newcommand{\radioFreq}{13.37\ \mathrm{GHz}}
\newcommand{\magn}{SGR~J1745-29}
\newcommand{\period}{3.76398106}		
\newcommand{\errGaucheP}{2.1 \times 10^{-7}}	
\newcommand{\errDroiteP}{2.0 \times 10^{-7}}	
\newcommand{\dper}{3.7684 \times 10^{-11}}	
\newcommand{\errGaucheDp}{1.6 \times 10^{-13}}	
\newcommand{\errDroiteDp}{9.9 \times 10^{-14}}	
\newcommand{\RiseFirst}{7700}			
\newcommand{\DecayFirst}{844}			
\newcommand{\DurationFirst}{8418.44}		
\newcommand{\DurationSecond}{964.91}		
\newcommand{\NHFirst}{23.7\,(14.5$--$37.5)}	
\newcommand{\GammaFirst}{3.1\,(2.1$--$4.5)}	
\newcommand{\FluxFirst}{10.1\,(4.9$--$33.5)}	
\newcommand{\LumFirst}{7.7\,(3.7$--$25.6)}	
\newcommand{\NHSecond}{9.8\,(2.0$--$23.5)}	
\newcommand{\GammaSecond}{2.2\,(0.7$--$4.7)}	
\newcommand{\FluxSecond}{6.3\,(3.5$--$25.7)}	
\newcommand{\LumSecond}{4.8\,(2.7$--$19.7)}	
\usepackage{ulem}

\begin{document}

\title{Multiwavelength study of the flaring activity of \sgra\ \\  in 2014 February$-$April}
\author{E.~Mossoux \inst{\ref{inst1}}
\and N.~Grosso \inst{\ref{inst1}}
\and H.~Bushouse \inst{\ref{inst2}}
\and A.~Eckart \inst{\ref{inst4},\ref{inst5}}
\and F.~Yusef-Zadeh \inst{\ref{inst3}}
\and R.~L.~Plambeck \inst{\ref{inst8}}
\and F.~Peissker \inst{\ref{inst4}}
\and M.~Valencia-S. \inst{\ref{inst4}}
\and D.~Porquet \inst{\ref{inst1}}
\and W.~D.~Cotton \inst{\ref{inst7}}
\and D.~A.~Roberts \inst{\ref{inst3}}} 
\institute{Observatoire astronomique de Strasbourg, Universit\'{e} de Strasbourg, CNRS, UMR 7550, 11 rue de l'Universit\'{e}, F-67000 Strasbourg, France. \label{inst1}
\and Space Telescope Science Institute (STScI), 3700 San Martin Drive Baltimore, MD 21218, USA. \label{inst2}
\and Physikalisches Institut der Universit$\mathrm{\ddot{a}}$t zu K$\mathrm{\ddot{o}}$ln, Z$\mathrm{\ddot{u}}$lpicher Str. 77, D-50937 K$\mathrm{\ddot{o}}$ln, Germany. \label{inst4}
\and Max-Planck-Institut f$\mathrm{\ddot{u}}$r Radioastronomie, Auf dem H$\mathrm{\ddot{u}}$gel 69, D-53121 Bonn, Germany. \label{inst5}
\and Department of Physics and Astronomy, CIERA, Northwestern University, Evanston, IL 60208, USA. \label{inst3}
\and Radio Astronomy Laboratory, University of California, Berkeley, CA 94720, USA. \label{inst8}
\and National Radio Astronomy Observatory,  Charlottesville, VA 22903, USA. \label{inst7}}
\date{Received  / Accepted  }

\abstract
{The supermassive black hole named \sgra\ is located at the dynamical center of the Milky Way.
This closest supermassive black hole is known to have a luminosity several orders of magnitude lower than the Eddington luminosity.
Flares coming from the \sgra\ environment can be observed in infrared, X-ray, and submillimeter wavelengths, but their origins are still debated.
Interestingly, the close passage of the Dusty S-cluster Object (DSO)/G2 near \sgra\ may increase the black hole flaring activity and could therefore help us to better constrain the radiation mechanisms from \sgra.} 
{Our aim is to study the X-ray, infrared, and radio flaring activity of \sgra\ close to the time of the DSO/G2 pericenter passage in order to constrain the physical properties and origin of the flares.}
{Simultaneous observations were made with XMM-Newton and WFC3 onboard HST during the period Feb--Apr 2014, in addition to coordinated observations with SINFONI at ESO's VLT, VLA in its A-configuration, and CARMA.}
{We detected two X-ray flares on 2014 Mar. 10 and Apr. 2 with XMM-Newton, three near-infrared (NIR) flares with HST on 2014 Mar. 10 and Apr. 2, and two NIR flares on 2014 Apr. 3 and 4 with VLT.
The X-ray flare on 2014 Mar.\ 10 is characterized by a long rise ($\sim$\RiseFirst{} s) and a rapid decay ($\sim$\DecayFirst{} s). Its total duration is one of the longest detected so far in X-rays.
Its NIR counterpart peaked well before ($4320$~s) the X-ray maximum, implying a dramatic change in the X-ray-to-NIR flux ratio during this event.
This NIR/X-ray flare is interpreted as either a single flare where variation in the X-ray-to-NIR flux ratio is explained by the adiabatic compression of a plasmon, or two distinct flaring components separated by 1.2~h with simultaneous peaks in X-rays and NIR.
We identified an increase in the rising radio flux density at 13.37~GHz on 2014 Mar. 10 with the VLA that could be the delayed radio emission from a NIR/X-ray flare that occurred before the start of our observation.
The X-ray flare on 2014 Apr. 2 occurred for HST during the occultation of \sgra\ by the Earth, therefore we only observed the start of its NIR counterpart.
With NIR synchrotron emission from accelerated electrons and assuming X-rays from synchrotron self-Compton emission, the region of this NIR/X-ray flare has a size of 0.03$-$7 times the Schwarzschild radius and an electron density of $10^{8.5}$--$10^{10.2}$~cm$^{-3}$, assuming a synchrotron spectral index of 0.3$-$1.5.
When \sgra\ reappeared to the HST view, we observed the decay phase of a distinct bright NIR flare with no detectable counterpart in X-rays.
On 2014 Apr. 3, two 3.2-mm flares were observed with CARMA, where the first may be the delayed (4.4~h) emission of a NIR flare observed with VLT.}
{We observed a total of seven NIR flares, with three having a detected X-ray counterpart.
The physical parameters of the flaring region are less constrained for the NIR flare without a detected X-ray counterpart, but none of the possible radiative processes (synchrotron, synchrotron self-Compton, or inverse Compton) can be ruled out for the production of the X-ray flares.
The three X-ray flares were observed during the XMM-Newton total effective exposure of $\sim 256\ \mathrm{ks}$.
This flaring rate is statistically consistent with those observed during the 2012 \textit{Chandra XVP} campaign, implying that no increase in the flaring activity was triggered close to the pericenter passage of the DSO/G2.
Moreover, higher flaring rates had already been observed with Chandra and XMM-Newton without any increase in the quiescent level, showing that there is no direct link between an increase in the flaring rate in X-rays and the change in the accretion rate.}


\titlerunning{The flaring activity of \sgra\ in 2014 Feb.$-$Apr.}
\maketitle

\section{Introduction} 
\sgra, located at the dynamical center of our Galaxy, is currently a dormant supermassive black hole (SMBH) of mass $M$ about $4 \times 10^6\ \mathrm{\textit{M}_{\sun}}$ \citep{schodel02,ghez08,gillessen09}.
Its bolometric luminosity ($L_\mathrm{bol} \sim 10^{36}\ \mathrm{erg}~\mathrm{s^{-1}}$) is lower than the Eddington luminosity ($\mathrm{\textit{L}_{Edd}}=3.3 \times 10^4\,M/M_{\sun}\,L_{\sun}=3\times 10^{44}\ \mathrm{erg}~\mathrm{s^{-1}}$) \citep{yuan03}.
This low luminosity can be explained by radiatively inefficient accretion flow models (RIAF) such as advection-dominated accretion flows (ADAF; \citealt{narayan98}) and jet-disk models.
Because of its proximity ($d=8$~kpc; \citealt{genzel10,falcke13}), \sgra\ is the best target to study the accretion and ejection physics for the case of low accretion rate, which is a regime where SMBH's are supposed to spend most of their lifetime.
Its physical understanding can be applied to a large number of normal galaxies that are supposed to host a SMBH. 

Above \sgra\ quiescent emission, some episodes of increased flux are observed in X-rays, near-infrared (NIR), and sub-millimeter/radio.
These flaring events from \sgra\ were first discovered in X-rays \citep{baganoff01} and were then also observed in NIR \citep{genzel03} and sub-millimeter wavelengths \citep{zhao03}.
NIR flares, which happen several times per day and have various amplitude up to 32~mJy \citep{witzel12}, are interpreted as synchrotron emission from accelerated electrons close to the black hole \citep{eisenhauer05,eckart06}.
In the NIR, the synchrotron emission is optically thick and the spectral index between the $H$ and $L$ band is $\alpha=-0.62 \pm 0.1$ with $S_\nu \propto \nu^\alpha$ \citep{witzel13}.
The X-ray flaring rate is 1.0$-$1.3 flares per day \citep{neilsen13}, but two episodes of higher flaring activity in X-rays have been observed \citep{porquet08,neilsen13}.
Most X-ray flares have moderate amplitude \citep{neilsen13} with 2--45 times the quiescent luminosity of \sgra\ (about $3.6 \times 10^{33}\ \mathrm{erg\ s^{-1}}$ in 2$-$8 keV; \citealt{baganoff03,nowak12}),
but brighter flares with amplitudes up to 160 times the quiescent level have also been observed \citep{porquet03,porquet08,nowak12}.
Several emission mechanism models are proposed in order to explain X-ray flares, such as: synchrotron \citep{dodds-eden09,barriere14},
synchrotron self-Compton \citep{eckart08}, and inverse Compton \citep{yusef-zadeh06,wardle11,yusef-zadeh12} emissions.
During simultaneous NIR/X-ray observations, X-ray flares always have a NIR counterpart and their light curves have similar shapes, with
an apparent delay less than 3~min between the peaks of flare emission \citep{eckart06,yusef-zadeh06c,dodds-eden09}.
The sub-millimeter and radio flare peaks, however, are delayed several tens of minutes and hours,
respectively \citep{marrone08,yusef-zadeh08,yusef-zadeh09}, and are proposed to be due to synchrotron radiation of an expanding relativistic plasma blob with an adiabatic cooling \citep{yusef-zadeh06c}.
Considering the intrinsic size of \sgra\ at a wavelength $\lambda$ of $(0.52 \pm 0.03)\,\mathrm{mas} \times (\lambda/\mathrm{cm})^{1.3\pm0.1}$,
the time lag between the sub-millimeter and radio light curves suggests a collimated outflow \citep{brinkerink15}.
On 2012 May 17, a NIR flare was followed $4.5\pm0.5$~h later by a 7-mm flare that was observed with the Very Long Baseline Array (VLBA) and
localized 1.5~mas southeast of \sgra, providing evidence for an adiabatically expanding jet with a speed of $0.4\pm0.3$~c \citep{rauch15}.
 
\citet{gillessen12} reported the detection of the object named G2 on its way towards \sgra\ in an eccentric keplerian orbit with the 2004 data from the Very Large Telescope (VLT) using the Spectrograph for INtegral Field Observations in the Near-Infrared (SINFONI) and the Nasmyth  Adaptive  Optics  System  (NAOS) and COudé  Near-IR  Camera (CONICA), i.e., NACO.
Their observations of the redshifted emission lines Br$\gamma$, Br$\delta$, and HeI in the NIR between 2004 and 2011 allowed them to determine the pericenter passage of $2013.51 \pm 0.04$.
They developed the first interpretation of the nature of the G2 object based on the observation of these lines: a compact gas blob.
From the $M$-band they showed that G2 has a dust temperature consistent with 450~K.
They predicted that, because G2 moves supersonically through the ambient hot gas, a bow shock should be created close to the pericenter passagei, which should be seen from radio to X-rays.
The observation of such X-ray emission could help to put some constraints on the physical characteristics of the ambient medium around \sgra.
The compact gas blob interpretation was still favored by \citet{gillessen13} who analyzed the Br$\gamma$ line width using data from SINFONI and NACO in March$-$July 2012.
They derived a pericenter passage of $2013.69 \pm 0.04$, adding their observations to those between 2004 and 2011.
A velocity-position diagram of G2 was computed by \citet{gillessen13b} using the emission lines Br$\gamma$, HeI, and Pa$\alpha$ from SINFONI and NACO observations in April 2013.
An elongation of G2 in the direction of its orbit was seen in the velocity-position diagram, which, together with the low dust temperature, favored the interpretation of an ionized gas cloud.

Two other interpretations based on the observations of these emission lines were also developed.
The first one was proposed by \citet{burkert12}: a spherical gas shell, which was supported by a simulation that reproduced the observed elongated structure in the velocity profile. 
They also simulated the effects of tidal shearing produced by Rayleigh-Taylor and Kelvin-Helmholtz instabilities during its approach to \sgra\ \citep{morris12}.
The shearing should produce a fragmentation of the envelope of G2 and provide fresh matter that would accrete onto \sgra.
This should increase the flaring activity of \sgra, depending on the filling factor, or (re-)activate the Active Galactic Nuclei (AGN) phase during the subsequent years.
The other interpretation is a dust-enshrouded stellar source, first developed by \citet{eckart13}, which leads to the second name of G2: a Dusty S-cluster Object (DSO).
This classification is supported by its detection in the $K_\mathrm{s}$- and $K'$-bands in observations from NACO and the NIRC2 camera of the Keck Observatory, respectively.
The $M$-band measurements showed that the integrated luminosity of this object is $5-10\ \mathrm{\textit{L}_{\sun}}$. 
Moreover, the $L$-band emission remained constant and spatially unresolved from 2004 to 2014, which ruled out a coreless model \citep{witzel14}.
The compact nature of the source is also supported by SINFONI observations between February and September 2014 \citep{valencias15}.
They showed that the wide range of Br$\gamma$ line widths ($200-700\,\mathrm{km\ s^{-1}}$) is reproduced well by the emission from a pre-main sequence star, because the magnetospheric accretion of circumstellar matter on the photosphere of these young stars emits the Br$\gamma$ line.
The tidal stretching of the accretion disk around the star as DSO/G2 approaches pericenter may explain the increase of the Br$\gamma$ line width.
A star with a mass of $1-2\ \mathrm{\textit{M}_{\sun}}$ and a luminosity less than $10\,\mathrm{\textit{L}_{\sun}}$ agrees with the dust temperature of 450~K found by \citet{gillessen12}.
As \citet{valencias15} observed the blueshifted Br$\gamma$ line after 2014 May, they were able to improve the estimation of the time of the pericenter passage to $2014.39 \pm 0.14$ and a distance of $\sim$163~au (4075 gravitational radius) from \sgra.
For comparison, the B0 spectral-type star S2 with a 15.2-year orbit around \sgra\ has a 1.3 times smaller pericenter distance \citep{schodel02}.
The absence of a redshifted counterpart after the pericenter passage favored the interpretation of the nature of DSO/G2 as a compact object and still ruled out the coreless model.

The multiwavelength campaign presented here was designed in 2012 to study the impact of the passage of the DSO/G2 object close to the SMBH (based on the pericenter date predicted by \citealt{gillessen12}) from the NIR/X-ray flaring activity of \sgra.
We report the results of joint observations of \sgra\ between February and April 2014 with the X-ray Multi-Mirror mission (XMM-Newton) and the
Hubble Space Telescope (HST) (XMM-Newton AO-12; PI: N. Grosso), close to the pericenter passage of DSO/G2.
We also obtained coordinated observations with the VLT, the Combined Array for Research in Millimeter-wave Astronomy (CARMA), and the Karl Jansky Very Large Array (VLA) to investigate NIR flaring emission and delayed millimeter/radio flaring emission.
In Sect.~\ref{observation} we present the observations and data reduction.
In Sect.~\ref{analyze} we report the analysis of these observations.
In Sect.~\ref{x_to_nir} we determine the X-ray emission related to each NIR flare observed during this campaign.
In Sect.~\ref{phys_param} we constrain the physical parameters of the flaring region associated with the NIR flares and their X-ray counterparts.
In Sect.~\ref{discussion} we discuss the X-ray flaring rate observed during this campaign.
Finally, in Sect.~\ref{conclusion} we summarize our main results and discuss their possible implications.

\section{Observations and data reduction}
\label{observation}
\begin{figure}
\centering
\includegraphics[angle=90,width=9cm]{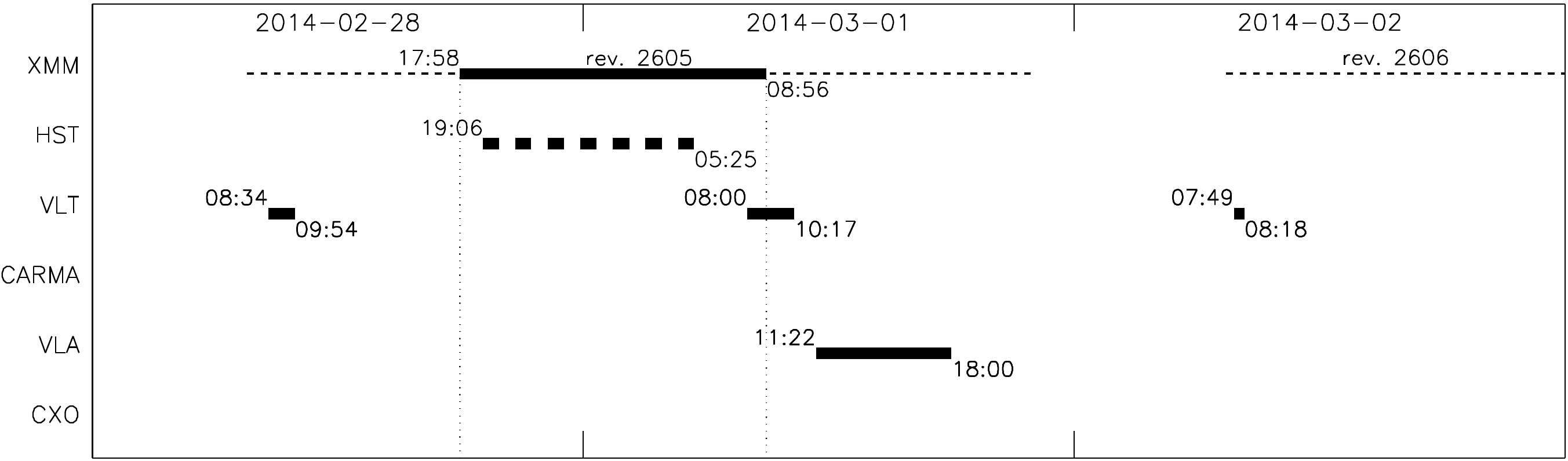}\\
\includegraphics[angle=90,width=9cm]{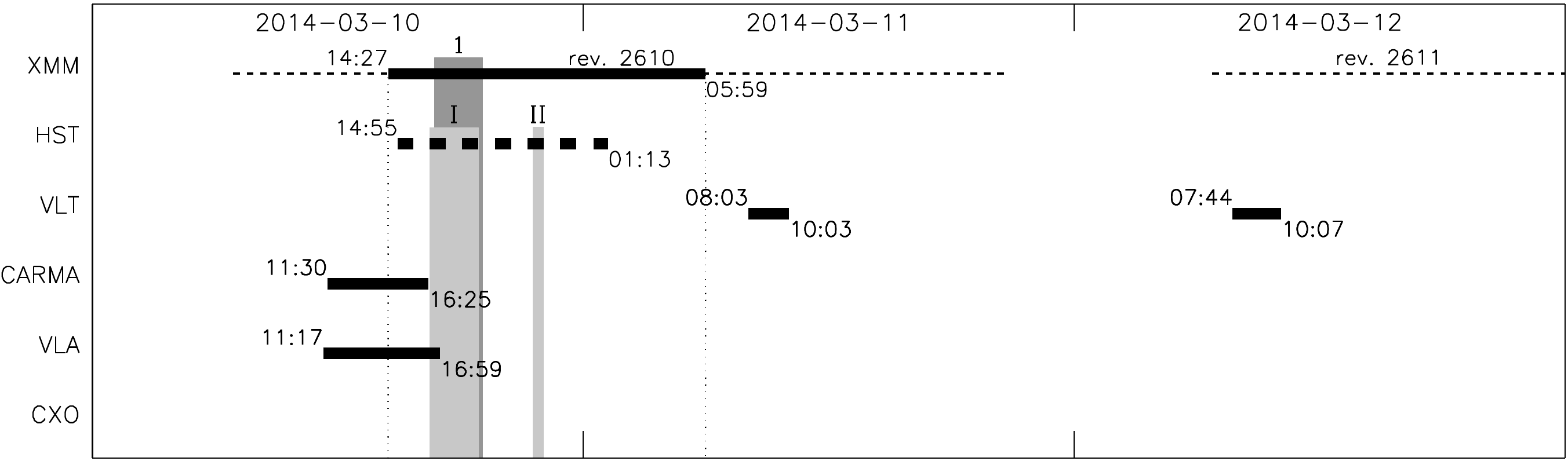}\\
\includegraphics[angle=90,width=9cm]{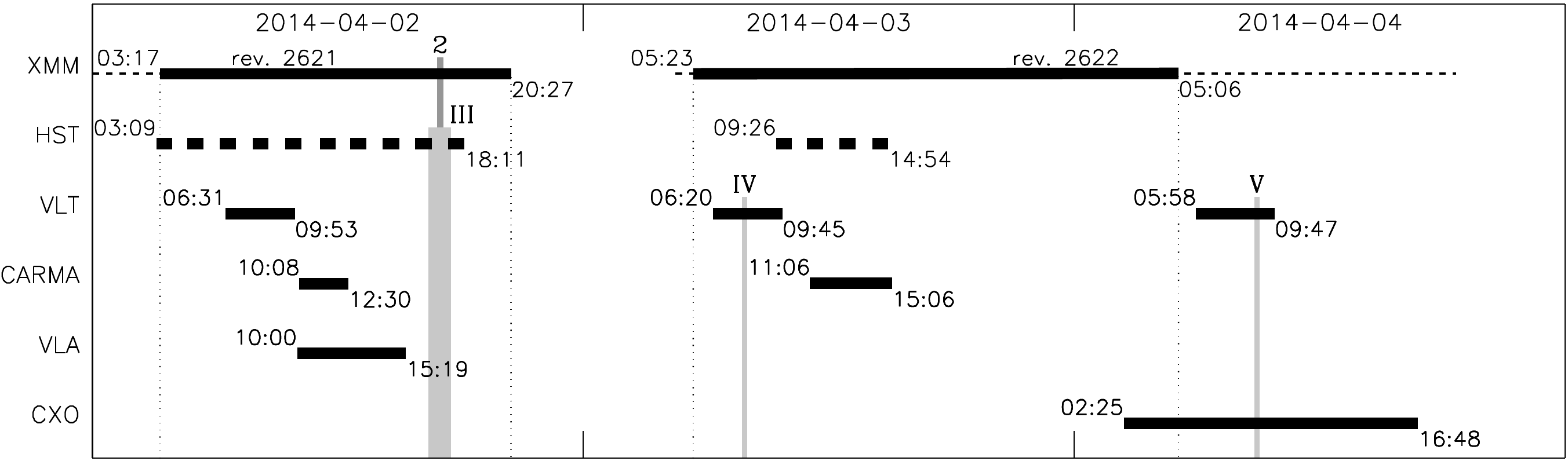}
\caption{Time diagram of the 2014 Feb.$-$Apr. campaign.
The horizontal dashed lines are the XMM-Newton orbital visibility times of \sgra\ labeled with revolution numbers.
The thick solid lines are the time slot of the observations for each instrument with start and stop hours.
The vertical dotted lines are the limits of the XMM-Newton observations.
The vertical gray blocks are the X-ray (Arabic numerals) and near-IR (Roman numerals) flares reported in this work.}
\label{fig:obs_log}
\end{figure}

Here we present the schedule of the coordinated observations of the 2014 Feb.$-$Apr. campaign (Fig.~\ref{fig:obs_log}) 
followed by a description of the data reduction for each facility used during this campaign. 

\subsection{XMM-Newton observations}
\label{xmm}
Table~\ref{table:1} reports the log of the XMM-Newton campaign for 2014 Feb.$-$Apr (AO-12; PI: N. Grosso).
The last X-ray observation is an anticipated Target of Opportunity (ToO) that was triggered to observe the new flaring magnetar \magn{} (AO-12; PI: G.L. Isra$\ddot{\mathrm{e}}$l).
We only use the data from the EPIC camera since the optical extinction towards the Galactic center is too high to get optical or soft X-ray photons from \sgra\ with the Optical-UV Monitor or the Reflection Grating Spectrometers.

\begin{table}
\caption{XMM-Newton observation log for the 2014 Feb.$-$Apr. campaign.}
\tiny
\label{table:1}
\centering
\begin{tabular}{@{}ccccc@{}}
\hline
\hline
 ObsID & Orbit & Start Time & End Time & Duration \\
 & & (UT) & (UT) & (ks) \\
\hline
 0723410301 & 2605 & Feb. 28, 17:59:00 & Mar. 01, 08:53:14 & 53.654 \\
 0723410401 & 2610 & Mar. 10, 14:28:16 & Mar. 11, 05:55:49 & 55.653 \\
 0723410501 & 2621 & Apr. 02, 03:18:22 & Apr. 02, 20:18:01 & 61.178 \\
 0690441801 & 2622 & Apr. 03, 05:23:33 & Apr. 04, 05:02:52 & 85.159 \\
\hline
\end{tabular}
\normalsize
\end{table}
During the first three XMM-Newton observations, the two EPIC/MOS cameras \citep{turner01} and the EPIC/pn camera \citep{strueder01} observed in frame window mode.
During the last observation, the two MOS cameras were in small window mode and the pn camera observed in frame window mode.
All observations were made with the medium filter.
The effective observation start and end times are reported in Table \ref{table:1} in Universal Time (UT).
During these observations, the conversion from the Terrestrial Time (TT) registered aboard XMM-Newton to UT is $\mathrm{UT}=\mathrm{TT}-67.108\mathrm{s}$ (NASA's HEASARC Tool: xTime\footnote{The website of xTime is: \href{http://heasarc.gsfc.nasa.gov/cgi-bin/Tools/xTime/xTime.pl}{http://heasarc.gsfc.nasa.gov/cgi-bin/Tools/xTime/xTime.pl}}).
The total effective exposure for the four XMM-Newton observations during this campaign is $\approx 256\ \mathrm{ks}$.

The XMM-Newton data reduction is the same as presented in \citet{mossoux14}.
We used the Science Analysis Software (SAS) package (version 13.5) with the 2014 Apr.\ 4 release of the Current Calibration files (CCF) to reduce and analyze the data.
The tasks \texttt{emchain} and \texttt{epchain} were used to create the event lists for the MOS and pn camera, respectively.
The soft proton flare count rate in the full detector light curve in the 2$-$10 keV energy range was high (up to $0.02\ \mathrm{count\ s^{-1}\ arcmin^{-2}}$ in EPIC/pn) only during the last two hours of the third observation.

As we looked for variability of the X-ray emission from \sgra, we extracted events of the source+background region from a disk of $10\arcsec$-radius
centered on the VLBI radio position of \sgra: RA(J2000)=$17^{\mathrm{h}}45^{\mathrm{m}}40\fs{}0409$, Dec(J2000)=$-29^{\circ} 00\arcmin 28\farcs{}118$ \citep{reid99}.
The contribution of the background events was estimated by extracting a $\approx 3\arcmin \times 3\arcmin$ region at $\approx 4\arcmin$ -north of \sgra\ on the same CCD where the X-ray emission is low.
For the last observation, the background extraction region was a $\approx 3\arcmin \times 3\arcmin$ area at $\approx 7\arcmin$ -east of \sgra\ on the adjacent CCD because of the small window mode.

The light curves of the source+background and background regions were created from events with \texttt{PATTERN$\leq 12$} and \texttt{\#XMMEA\_SM} and \texttt{PATTERN$\leq 4$} and \texttt{FLAG==0} for the MOS and pn cameras, respectively.
These light curves are computed in the 2$-$10 keV energy range using a time bin of 300~s.
The task \texttt{epiclccorr} applies relative corrections to those light curves.
We then summed the background-subtracted light curves of the three cameras to produce the total EPIC light curves. 
Missing values were inferred using a scaling factor between the pn camera and the sum of the MOS1 and MOS2 cameras.
This factor was computed during the full time period where all detectors are observing and leads to a number of pn counts that is equal,
on average, to 1.46$\pm$0.03 times the sum of the number of MOS counts.

To perform the timing analysis of the light curves we adapted the Bayesian-blocks method developed by \citet{scargle98} and refined by \citet{scargle13} to the XMM-Newton event lists, using a two-step algorithm to correct for any detector flaring background \citep{mossoux14,mossoux15}\footnote{\citet{worpel15} tested different photon-weighting in order to subtract the background during observations of bursting and eclipsing objects in X-rays.
The equation for the Voronoi time-interval weighting used in \citet{mossoux14,mossoux15} follows the recipe of \citet{scargle13b}, which is identical to the ``alternative'' photon-weighting described in Sect.~4.6 of \citet{worpel15} since the photon-weighting is equal to the inverse of the Voronoi time-interval weighting.
We can see in Fig.~12 of \citet{worpel15} that the method of \citet{mossoux14,mossoux15} (labeled $h$ in this figure) locates the eclipses, as well as their weighted-photon method (labeled $f$ in this figure).
As noticed by \citet{worpel15}, their weighted-photon method may produce both negative and implausibly high count rates.
Indeed, the method of \citet{mossoux14,mossoux15} produces much fewer Bayesian blocks with negative count rates and no implausibly high count rates in comparison to their weighted-photon method (see for comparison panel $h$ and $f$ of Fig.~12 of \citealt{worpel15}).
This last point is crucial for flare and burst detection.}.
We used the false detection probability $p_\mathrm{1} = \exp(-3.5)$ \citep{neilsen13,mossoux14} and geometric priors of 7, 6.9, and 6.9 for pn, MOS1, and MOS2, respectively.
We created smoothed light curves by applying a density estimator \citep{silverman86,Feigelson12} and using the same method as in \citet{mossoux14} to correct the exposure time and the background contribution to the source+background event list.
The amplitude and time of the flare maximum were computed on the smoothed light curve with a window width of 100s and 500s and a time grid interval of 10s.

\subsection{HST observations}
\label{hst}
The NIR observations of \sgra\ were obtained with the Wide Field Camera 3 (WFC3) on HST, under joint XMM-Newton/HST programs 13403 (AO-12, PI: N. Grosso) and 13316 (Cycle 21, PI: H. Bushouse) in order to measure the delay between X-ray flares and their NIR counterparts. 
\sgra\ was observed in four visits with $7-10$ consecutive HST orbits, whose observation start and end times are reported in UT in Table~\ref{table:hst}.
The total effective exposure for these four HST visits during the 2014 Feb.$-$Apr. campaign is about $69\ \mathrm{ks}$.
Exposures were taken constantly during each part of these windows in which \sgra\ was visible to HST, usually resulting in an uninterrupted cadence of exposures lasting for 40--50 minutes at a time, and then interrupted for the remaining 40--50 minutes of each HST orbit in which \sgra\ is occulted by the Earth.
The four visits were planned to have the maximum number of consecutive orbits before HST entered the South Atlantic Anomaly (SAA), in order to maximize the simultaneous observing time in NIR and X-ray.

\begin{table}
\caption{Observation log of WFC3 on board HST for the 2014 Feb.$-$Apr. campaign.}
\tiny
\label{table:hst}
\centering
\begin{tabular}{@{}cccc@{}}
\hline
\hline
Visit & Start Time & End Time & Number of orbits \\
 & (UT) & (UT) & \\
\hline
1 & Feb. 28, 19:06 & Mar. 01, 05:25 & $\ \ $7 \\
2 & Mar. 10, 14:55 & Mar. 11, 01:13 & $\ \ $7 \\
3 & Apr. 02, 03:09 & Apr. 02, 18:11 & 10 \\
4 & Apr. 03, 09:26 & Apr. 03, 14:54 & $\ \ $4 \\
\hline
\end{tabular}
\normalsize
\end{table}
Each WFC3 exposure was taken with the IR channel of the camera, which has a $1024 \times 1024$~pixel HgCdTe array, with a pixel scale of $\sim0\farcs{}13$. 
We used the F153M filter, which is a medium-bandwidth filter ($\Delta \lambda=0.683\ \mu$m) with an effective wavelength $\lambda_\mathrm{eff}=1.53157\ \mu$m (from the Spanish Virtual Observatory\footnote{The website of the Spanish Virtual Observatory is: \href{http://svo.cab.inta-csic.es/main/index.php}{http://svo.cab.inta-csic.es/main/index.php}}). 
Each exposure used the predefined readout sequence ``SPARS25'' with NSAMP=12 or 13, which produces non-destructive readouts of the detector every 25 secs throughout the exposure, and a total of 12 or 13 readouts, resulting in a total exposure time of $275-300$~s after discarding the first short ($2.932$~s) readout.
The exposures were obtained in a 4-point dither pattern centered on \sgra, with a spacing  of $\sim0.6$~arcsec ($\sim 4$ pixels) per step to improve the sampling of the Point Spread Function (PSF) of FWHM$=0\farcs{}145$ (1.136 detector pixels) at 1.50~$\mu$m \citep{wfc3_uhb}\footnote{For comparison, the FWHM of the NICMOS Camera 1 is $0\farcs{}16$ (3.75 detector pixels) at 1.60~$\mu$m \citep{yusef-zadeh06}, i.e., better sampled than the FWHM of the WFC3 camera.}.
All of the WFC3 exposures were calibrated using the standard STScI calibration pipeline task \texttt{calwf3}.
Once the pointing information was set for each WFC3 exposure, we could safely use the known relative position of \sgra\ for positioning a photometry
aperture (\sgra\ itself cannot be easily identified in the WFC3 images because it is in the PSF wings of the star S2 located at $0\farcs{}15$ during our observational epoch according to the orbital elements of \citealt{gillessen09}).

The absolute coordinates of HST exposures are limited by uncertainties in the positions of the guide stars that are used to acquire and track the target. 
We therefore used the radio position of IRS-16C (also known as S96; \citealt{yusef-zadeh14}), a star near \sgra, as an astrometric reference to accurately register the pointing of each WFC3 exposure. 
The radio position of IRS-16C came from VLA observations in February 2014, which is nearly co-eval with the HST observations. 

The accumulating, non-destructive readouts of each calibrated exposure were ``unraveled'' by taking the difference of adjacent readouts, which results in a series of independent samples taken at 25 sec intervals, thereby increasing the time resolution for the subsequent photometric analysis. 
Photometry of \sgra\ was performed with the IRAF routine \texttt{phot}, using a 3-pixel ($\sim0.4$~arcsec) diameter circular aperture centered on the known radio coordinates of \sgra\ \citep{petrov11,yusef-zadeh14}.

Initial analysis of the photometry results for \sgra\ and other stars in the field revealed an overall tendency for the fluxes of individual sources to gradually decrease on the order of $\sim$3\%\ during the course of each individual exposure (i.e., across the span of multiple readouts).
We believe this effect is due to persistence within an individual exposure, as the total signal level reaches fairly high levels by the end of each 
$\sim$5 min exposure.
We measured this trend for stars near \sgra\ and applied the results to the \sgra\ photometry to remove the effect. When applied to other stars in the field, the corrected photometry was constant, on average, throughout each exposure.
The error on the photometry obtained in each of the four visits, within an individual 25 s readout interval, is 0.0044, 0.0046, 0.0022, and 0.0042~mJy, respectively, which has been estimated from the standard deviation of the flux density of a reference star.
For comparison, similar observations obtained in the past using NICMOS camera 1 have an uncertainty within a bin of 32 s of 0.002~mJy at 1.60~$\mu$m \citep{yusef-zadeh06}.

Aperture and extinction corrections were also applied to the \sgra\ photometry.
The aperture correction was determined by measuring the curves of growth of several isolated stars in the field, using a series of apertures of increasing size. 
The correction factor for an aperture diameter of 3~pixels is 1.414.
The extinction correction was derived from $A(H)=4.35 \pm 0.12$~mag and $A(K_\mathrm{s})=2.46 \pm 0.03$~mag \citep{schodel10} with $\lambda_\mathrm{eff}(\mathrm{NACO\ }H)=1.63725\ \mu$m and $\lambda_\mathrm{eff}(\mathrm{NACO}\ K_\mathrm{s})=2.12406\ \mu$m (from the Spanish Virtual Observatory), respectively, assuming a power law leading to $A(\lambda) \propto \lambda^{-2.19 \pm 0.06}$.
Thus, the computed extinction for the effective wavelength of the WFC3 F153M filter ($\lambda_\mathrm{eff}=1.53157\ \mu$m) used is $5.03 \pm 0.20$~mag, which corresponds to a multiplicative factor of $103.2 \pm 19.0$ to correct the observed flux density for extinction.

\subsection{VLT observations}
\label{vlt_obs}
\begin{table}
\caption{Coordinated observation log with SINFONI at ESO's VLT for the 2014 Feb.--Apr. campaign.}
\tiny
\label{table:vlt}
\centering
\scalebox{.9}{
\begin{tabular}{@{}ccccc@{}}
\hline
\hline
 \multicolumn{1}{c}{Date} & \multicolumn{1}{c}{Start Time} & \multicolumn{1}{c}{End Time} & \multicolumn{1}{c}{Number of Exposures$\ $\tablefootmark{f}}  & \multicolumn{1}{c}{Total Exposure} \\
 & \multicolumn{1}{c}{(UT)} & \multicolumn{1}{c}{(UT)} & \multicolumn{1}{c}{(Used/Total)} & \multicolumn{1}{c}{(s)} \\
\hline
 Feb. 27$\ $\tablefootmark{a$\ \ \ $} & 08:20:42 & 09:48:55 & $\ \ \ $4/4 & 1600 \\
 Feb. 28$\ $\tablefootmark{b$\ \ \ $} & 08:34:58 & 09:54:37 & $\ \ \ $0/7 & $\ \ \ \ \ \ $0 \\
 Mar. 01$\ $\tablefootmark{b,d} & 08:00:14 & 10:17:59 & $\ $0/12 & $\ \ \ \ \ \ $0 \\
 Mar. 02$\ $\tablefootmark{b$\ \ \ $} & 07:49:06 & 08:18:54 & $\ \ \ $0/3 &  $\ \ \ \ \ \ $0 \\
 Mar. 11$\ $\tablefootmark{a$\ \ \ $} & 08:03:55 & 10:03:28 & 11/11 & $4400$ \\
 Mar. 12$\ $\tablefootmark{a$\ \ \ $} & 07:44:35 & 10:07:45 & 13/13 & $5200$ \\
 Apr. 02$\ $\tablefootmark{c,e} & 06:31:39 & 09:53:52 & 16/18 & $6400$ \\
 Apr. 03$\ $\tablefootmark{c,e} & 06:20:46 & 09:45:02 & 18/18 & $7200$ \\
 Apr. 04$\ $\tablefootmark{c$\ \ \ $} & 05:58:19 & 09:47:58 & 21/21 & $8400$ \\
\hline
\end{tabular}
}
\tablefoot{
\tablefoottext{a}
{ESO program 092.B-0920(A) (PI: N. Grosso);}
\tablefoottext{b}
{ESO program 091.B-0183(H) (PI: A. Eckart);}
\tablefoottext{c}
{ESO program 093.B-0932(A) (PI: N. Grosso);}
\tablefoottext{d}
{Partially-simultaneous observation with XMM-Newton;}
\tablefoottext{e}
{Simultaneous observation with XMM-Newton;}
\tablefoottext{f}
{Each exposure has a duration of 400~s.}
}
\normalsize
\end{table}
Near-infrared integral-field observations of the Galactic Center were performed using SINFONI at the VLT in Chile \citep{eisenhauer03,bonnet04}. 
\sgra\ was monitored nine times in 2014 Feb.$-$Apr..
Table~\ref{table:vlt} summarizes the observing log, including the amount of exposures that were selected for the analysis. 
The selection criteria is described below. 
These observations were planned to be coordinated with those carried out with XMM-Newton.
Two of these observations were simultaneous with XMM-Newton observations and one was partially simultaneous.
They are part of the ESO programs 092.B-0183(H) (PI: A. Eckart), 093.B-0932(A) (PI: N. Grosso), and 092.B-0920(A) (PI: N. Grosso) presented in \citet{valencias15} for the DSO/G2 study.

The SINFONI instrument is an integral-field unit fed by an adaptive optics (AO) module.
The AO module was locked on a bright star 8$\farcs$85 east and 15$\farcs$54 north of \sgra.
The $H+K$ grating used in these observations covers the $1.45 \mu\rm{m} - 2.45 \mu\rm{m}$ range and exhibits a spectral resolution of $R \sim 1500$ (which corresponds to approximately $200\,\rm{km\,s^{-1}}$ at $2.16\,\mu\rm{m}$).
The smallest SINFONI field of view ($0\farcs8 \times 0\farcs8$) was jittered around the position of S2.
Observations of different B- and G-type stars were performed for further telluric corrections.

Exposure times of 400~s were used to observe the Galactic center region,followed or preceded by observations on a dark cloud located about 12$\arcmin$45$\arcsec$ west and 5$\arcmin$36$\arcsec$ north of the \sgra\ sky position.
These integration times were chosen to fully sample the variations of \sgra\  flux density over typical flare lengths, while optimizing the quality of the data.

The data processing and calibration was performed as described in \citet{valencias15} and it is outlined here for completeness.
First, bad lines were corrected using the procedure suggested in the SINFONI user manual. 
Then, a rough cosmic-ray correction in the sky and target exposures was performed using the algorithm of \citet{pych04}.
Some science and calibration files showed random patterns that were detected and removed following the algorithms proposed by \citet{smajic14}.
Afterwards, the SINFONI pipeline was used for the standard reduction steps (e.g., flat fielding and bad pixel corrections) and for the wavelength calibration.
A deep correction of cosmic rays and the atmospheric refraction effects were done using our own DPUSER routines \citep[Thomas Ott, MPE Garching; see also][]{eckart91}.

The quality of individual exposures was judged based on the point-spread function (PSF) at the moment of the observation.
The PSF was estimated by fitting a 2D Gaussian to the bright star S2.
Data cubes where the full width at half maximum of the Gaussian was higher than 96~mas (or $7.65$ detector pixels) were discarded in the analysis.
The 2014 Feb. 28, Mar. 1, and Mar. 2 observations are thus not used because of their poor quality.
On 2014 Apr. 2 two data cubes of larger field-of-view were used for pointing.
They were not included in the light curves since they map regions just beside the central S-cluster.
Flux calibration on individual data cubes was performed using aperture photometry on the deconvolved K-band image.
The deconvolution was done using the Lucy–Richardson algorithm in DPUSER. 
For calibration we used the stars S2 ($K_{\rm s}=14.13$), S4 ($K_{\rm s}=14.61$), S10 ($K_{\rm s}=14.12$), and S12 ($K_{\rm s}=15.49$), and adopted the $K_{\rm s}$-band extinction correction  $A(K_{\rm s})=2.46\pm0.03$~mag \citep{schodel10}.
Additional information on the flux estimation is given by \citet{witzel12}.
The final flux densities were extracted by fitting a 2D Gaussian to the calibrated continuum images for all time steps.

\subsection{VLA observations}
\label{vla_obs}
\begin{table}
\caption{VLA observation log for the 2014 Feb.$-$Apr. campaign.}
\tiny
\label{table:vla}
\centering
\begin{tabular}{@{}cccc@{}}
\hline
\hline
 Date & Start Time & End Time & Band\tablefootmark{a} \\
 & (UT) & (UT) & \\
\hline
 2014 Mar. 01 & 11:22:08 & 18:01:07 & $\ \ \ \ X$ \\
 2014 Mar. 10 & 11:17:00 & 17:25:24 & $\ \ Ku$ \\
 2014 Apr. 02 & 10:00:15 & 15:52:48 & $C$, $L$\\
\hline
\end{tabular}
\tablefoot{\tablefoottext{a}
{We report in this paper the $X-$, $Ku-$, $C-$, and $L-$band observations obtained only at 8.56, 13.37, 5.19 and 1.68~GHz, respectively.}
}
\normalsize
\end{table}
Radio continuum observations were carried out with the Karl G. Jansky Very Large Array (VLA) on 2014 March 1, March 10 and April 2 (observing program 14A-231). 
The VLA was in its A-configuration during these three days of observations, with start and stop times reported in Table~\ref{table:vla}. 
In all observations, we used 3C286 to calibrate the flux density scale, both 3C286 and NRAO530 to calibrate the bandpass, and J1744-3116 to calibrate the complex gains. 

On 2014 Mar. 1 we observed \sgra\  at 8$-$10~GHz ($X$-Band) using the 8-bit sampler system with 2~GHz total bandwidth, each consisting of 64 channels each 2~MHz wide. 
On 2014 Mar. 10 we used the same correlator setup as 2014 Mar. 1, except using the $Ku$-Band between 13 and 15~GHz. 
On 2014 Apr. 2 we used the two bands 5$-$7~GHz ($C$-band) and 1$-$2~GHz ($L$-band), and alternated between these bands every 7 minutes.  
The $C$-band correlator was set-up similarly to that of $X$-band. 
The $L$-band correlator, however, used 1~GHz of bandwidth, which consisted of 16 IFs with channel widths of 1~MHz each. 
After primary calibration using OBIT \citep{cotton08}, a self-calibration procedure was applied using AIPS in phase only, to remove atmospheric phase errors.

\subsection{CARMA observations}
\label{carma_obs}
\begin{table}
\caption{CARMA 95~GHz observation log for the 2014 Feb.$-$Apr. campaign.}
\tiny
\label{table:carma}
\centering
\begin{tabular}{@{}ccc@{}}
\hline
\hline
 Date & Start Time & End Time \\
 & (UT) & (UT) \\
\hline
 2014 Mar. 10 & 11:14:46 & 16:29:42 \\
 2014 Apr. 02 & 09:54:18 & 15:14:31 \\
 2014 Apr. 03 & 10:52:01 & 15:10:17 \\
\hline
\end{tabular}
\normalsize
\end{table}
Observations of \sgra\  at 95~GHz (corresponding to 3.2 mm) were obtained with CARMA on 2014 Mar. 10, Apr. 2, and Apr. 3 (see Table~\ref{table:carma}).
The array was in the C-configuration, with antenna separations ranging from 30--350 meters.
The correlator processed frequencies range was 88.76$-$93.24~GHz in the lower sideband of the receivers and 96.76$-$101.24~GHz in the upper sideband.
The spectral resolution was 25~MHz after Hanning smoothing.
Channels corresponding to strong absorption lines of HCO$^+$ (89.19~GHz), HNC (90.65~GHz), and CS (97.98~GHz) were dropped from \sgra\  data before averaging to get the continuum flux density.
Only visibility data corresponding to telescope separations larger than $20$~k$\lambda$ were used for the flux measurements, to reduce contamination from extended emission near \sgra.

Observations of 3C279 were used to calibrate the instrumental passband.  
The flux density scale was established from observations of Neptune, assuming it is a 2$\farcs$2 diameter disk with brightness temperature 123~K (consistent with the Butler-JPL-Horizons 2012 model shown in ALMA memo 594). 
Observations of a secondary flux calibrator (the blazar 1733-130, a.k.a. NRAO 530) were interleaved with the \sgra\  observations every 15 minutes to monitor the antenna gains. 
The flux density of 1733-130 was measured to be 2.7$\pm$0.3~Jy on 2014 Mar. 10, and 2.5$\pm$0.3~Jy on Apr. 2 and Apr. 3, relative to Neptune.

The data on Mar. 10 were obtained in turbulent weather and are of poor quality, therefore we do not use it in this work.
On 2014 Apr. 2 we only use the data before the beginning of the snow at about 12:30~UT.

\section{Data analysis}
\label{analyze}
\subsection{XMM-Newton data}
\begin{figure*}
\centering
\includegraphics*[width=6.5cm, angle=90]{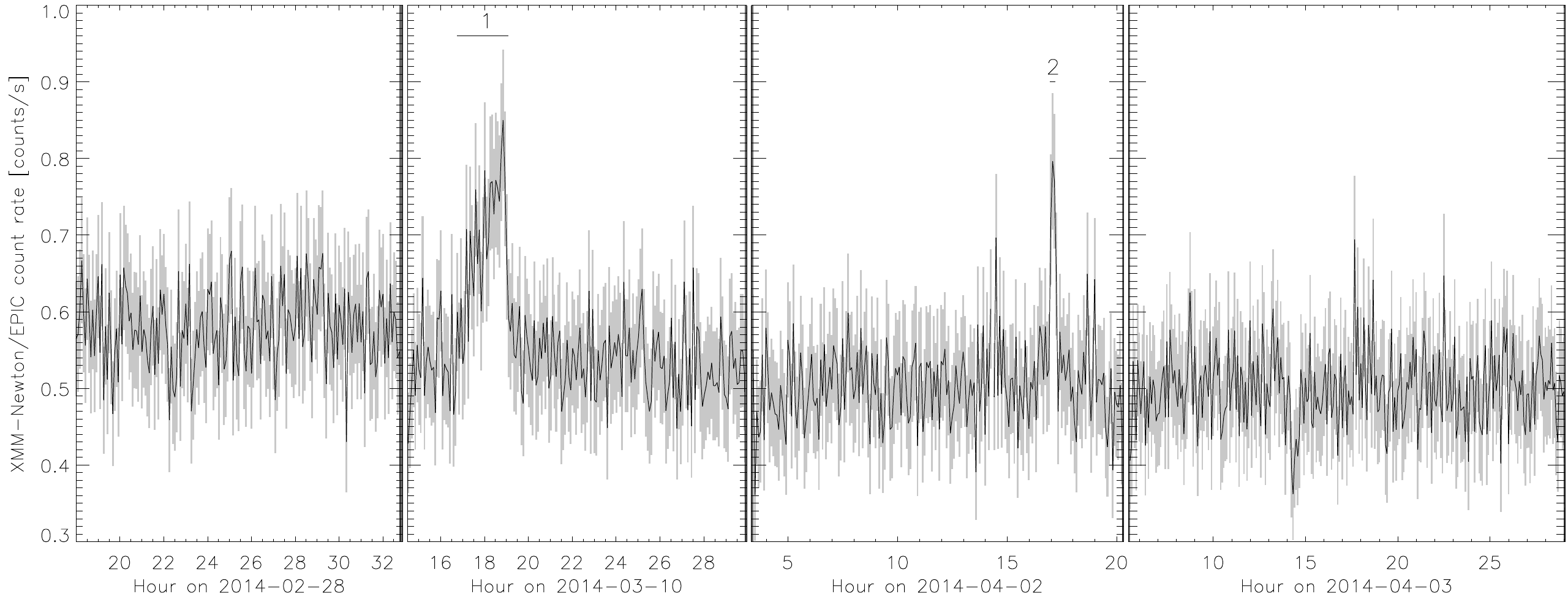}
\caption{XMM-Newton/EPIC (pn+MOS1+MOS2) light curves of \sgra\  in the 2-10 keV energy range obtained 2014 Feb.$-$Apr.
The time interval used to bin the light curve is 300~s.
The X-ray flares are labeled with Arabic numerals.
The horizontal lines below these labels indicate the flare durations.}
\label{lc_total}
\end{figure*}
Figure~\ref{lc_total} shows the XMM-Newton/EPIC (pn+MOS1+MOS2) background-subtracted light curves of \sgra\  binned to 300~s in the 2$-$10 keV energy range.
The non-flaring level (i.e., the longest interval of the Bayesian blocks) during 2014 Feb.$-$Apr. is about 3 times the typical value of 0.18~$\mathrm{count}\ \mathrm{s^{-1}}$ (e.g., \citealt{porquet08,mossoux14}).
This is due to the flaring magnetar \magn{} located only 2$\farcs$4 from \sgra\  \citep{rea13}.
Because the radius enclosing 50\% of the energy for EPIC/pn at 1.5~keV on-axis is about 10$\arcsec$ \citep{ghizzardi02}, we extract the events from a 10$\arcsec$-radius circle centered on \sgra\  as done in previous studies.
This extraction region therefore includes events from \magn{}, which artificially increases the non-flaring level of \sgra\  (Fig.~\ref{lc_total}).

\subsubsection{Impact of the magnetar on the flare detection}
\label{magnetar}
\citet{degenaar13b} reported a large flare towards \sgra\  detected by Swift on 2013 Apr. 24.
The detection of a hard X-ray burst by BAT near \sgra\  on 2013 Apr. 25 led \citet{kennea13} to attribute this flux increase to a new Soft Gamma Repeater unresolved from \sgra: \magn{}.
The X-ray spectrum of this magnetar is an absorbed blackbody with $N_\mathrm{H}=13.7^{+1.3}_{-1.2} \times 10^{22}\ \mathrm{cm^{-2}}$ and $kT_\mathrm{BB}=1.06 \pm 0.06$ keV \citep{kennea13}.
But the Chandra X-ray Observatory (CXO) results between 1 and 10~keV from \citet{coti15} show that the temperature of the blackbody emitting region decreases with time: $kT_\mathrm{BB}/\mathrm{keV}=(0.85 \pm 0.01)-(1.77 \pm 0.04)\times 10^{-4} (t-t_0)$ with $t_\mathrm{0}$ the time of the peak outburst (i.e., 2013 Apr. 24 or 56406 in MJD).
They show that before 100~d from outburst, the magnetar luminosity between 1 and 10~keV is characterized by a linear model plus an exponential decay whose e-folding time is $37 \pm 2$~d.
After 100~d from the burst activation, the magnetar flux is well fitted by an exponential with an e-folding time of $253 \pm 5$~d.
This flux decay is one of the slower decays observed for a magnetar.
Thanks to 8 months of observations with the Green Bank Telescope and 18 months of observations with the Swift's X-Ray Telescope, the evolution of the X-ray flux and spin period of the magnetar {bf have been} well constrained by \citet{lynch14}.
The X-ray flux between 2 and 10 keV in a $20\arcsec$-radius extraction region centered on the magnetar decreases as the sum of two exponentials: $F(t)=(1.00 \pm 0.06) \,e^{-(t-t_\mathrm{0})/(55 \pm 7\ \mathrm{d})}+(0.98 \pm 0.07) \,e^{-(t-t_\mathrm{0})/(500 \pm 41\ \mathrm{d})}$ in unit of $10^{-11}\ \mathrm{erg\ s^{-1}\ cm^{-2}}$ with $t_\mathrm{0}$ the same as in \citet{coti15}.

We determined the exponential decay of the magnetar flux between 2 and 10~keV by applying a chi-squared fitting of the non-flaring level of each observation computed using the Bayesian-blocks algorithm: on Feb. 28, Mar. 10, and Apr. 2 and 3 the non-flaring level is $0.562 \pm 0.003$, $0.528 \pm 0.004$, $0.489 \pm 0.003$ and $0.499 \pm 0.002\ \mathrm{EPIC\ count}\ \mathrm{s^{-1}}$, respectively.
The magnetar flux variation can be described as $N(t)=N_\mathrm{0}\,e^{-(t-t_\mathrm{0})/\tau}$ with $t$ the time corresponding to the middle of each observation, $t_0$ and $N_0$ the time and count rate of the non-flaring level of the first observation, and $\tau$ the decay time scale.
Our best fit parameters with corresponding 1-$\sigma$ uncertainties are: $N_\mathrm{0}=0.558 \pm 0.003 \ \mathrm{count\ s^{-1}}$ and $\tau=281 \pm 15$ days.
The decay time scale is about 2 times shorter than those computed from the formula of \citet{lynch14} for this date.
However, as we can see in Fig.~2 of \citet{lynch14}, the magnetar flux is not a perfect exponential decay and has some local increase of the flux, in particular during our observing period.
This is seen in the last XMM-Newton/EPIC pn observation on 2014 Apr. 3, which is characterized by two blocks whose change point is at 16:27:48 (UTC).
The corresponding count rates for the first and second blocks are $0.254 \pm 0.03$ and $0.299 \pm 0.03\ \mathrm{pn\ count}\ \mathrm{s^{-1}}$.
By folding light curves for each block on this date with the magnetar spin period of 3.76398106~s computed in Appendix~\ref{appendix_d}, we see that the pulse shape has not changed, but the flux increased by a factor of about $1.2$, as determined by the Bayesian-blocks algorithm.
Moreover, the Chandra monitoring of DSO/G2 shows that there is no significant increase of \sgra\  flux on 2014 Apr. 4 \citep{atel6242}.

This contamination of the non-flaring level implies a decrease of the detection level of the faintest and shortest flares, as explained in details in Appendix~\ref{appendix_b}.
Comparing the detection probability of an XMM-Newton observation with the distribution of flares during the 2012 \textit{Chandra XVP} campaign \citep{neilsen13}, we estimate that we lost no more than one flare during our four XMM-Newton observations due to the magnetar contribution.

\subsubsection{X-ray flare detection}
\label{flare_xmm}
\begin{figure}
\centering
\includegraphics[trim = 10cm 0cm 5cm 0cm, clip,width=4.5cm,angle=90]{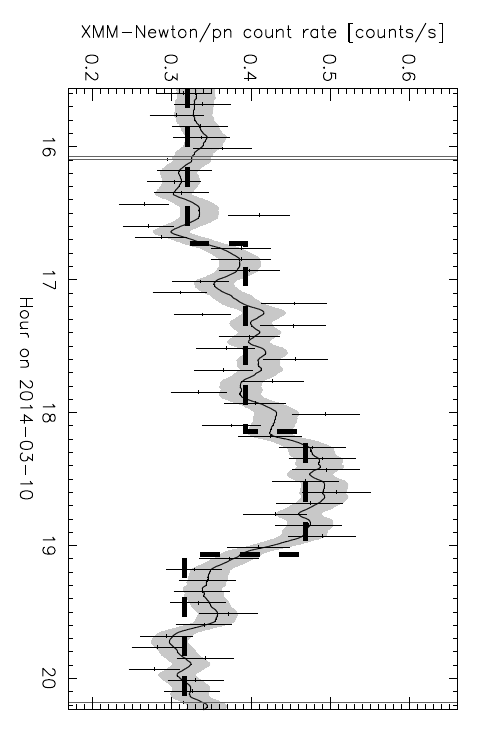}\\
\includegraphics[trim = 15cm 0cm 0cm 0cm, clip,width=4.5cm,angle=90]{27554_fig3.pdf}
\caption{XMM-Newton light curve binned on 500s of the 2014 Mar. 10 flare from \sgra\  in the $2-10$~keV energy range. 
\textit{Top panel:} The crosses are the data points of the EPIC/pn light curve.
The dashed lines represent the Bayesian blocks.
The solid line and the gray curve are the smoothed light curve and the associated errors ($h=500s$).
\textit{Bottom panel:} The total (pn+MOS1+MOS2) light curve.
The horizontal dashed line and the solid line are the sum of the non-flaring level and the smoothed light curve for each instrument.
The vertical gray stripe is the time during which the camera did not observe.}
\label{flare1}
\end{figure}

\begin{figure}
\centering
\includegraphics[trim = 10cm 0cm 5cm 0cm, clip,width=4.5cm,angle=90]{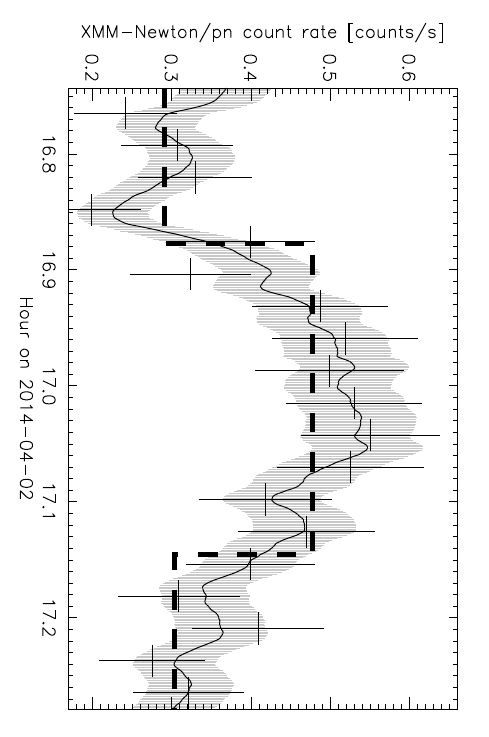} \\
\includegraphics[trim = 15cm 0cm 0cm 0cm, clip,width=4.5cm,angle=90]{27554_fig4.pdf}
\caption{XMM-Newton light curve binned on 100s of the 2014 Apr. 2 flare from \sgra\  in the $2-10$~keV energy range. 
The window width of the smoothed light curve is 100~s.
See caption of Fig.~\ref{flare1} for panel description.}
\label{flare2}
\end{figure}

By applying the Bayesian-blocks analysis on the EPIC event lists, we are able to detect two flares: one on 2014 Mar. 10 and one on 2014 Apr. 2.
These flares are labeled 1 to 2 in Fig. \ref{lc_total}.
Figures~\ref{flare1} and \ref{flare2} focus on the EPIC (pn+MOS1+MOS2) and EPIC/pn flare light curves with a bin time interval of 500 and 100~s, respectively.
The comparison of the flare light curves observed by each EPIC cameras can be found in appendix~\ref{appendix_a}. 
The second flare is detected by the Bayesian-blocks algorithm in pn, but not in MOS1 or MOS2.
This is explained by the lower sensitivity of the MOS cameras, resulting in a lower detection level of the algorithm (see Fig. \ref{detection_level}).
Table \ref{table:2} gives the temporal characteristics of these X-ray flares. 

\begin{table}
\caption{Characteristics of the X-ray flares observed by XMM-Newton in 2014 after removing the magnetar contribution.}
\centering
\scalebox{0.685}{
\label{table:2}
\begin{tabular}{@{}ccccccccc@{}}
\hline
\hline
Flare & Date & Start Time\tablefootmark{a} & End Time\tablefootmark{a} & Duration & Total\tablefootmark{b} & Peak\tablefootmark{c} \\
(\#) & (yy-mm-dd) & (hh:mm:ss) & (hh:mm:ss)  & (s) & (cts) & ($\mathrm{count}\ \mathrm{s^{-1}}$) \\
\hline  
 1 & 2014-03-10 & 16:44:48 & 19:05:07 & \DurationFirst{} & $900 \pm 60$ & $0.159 \pm 0.032$\\
 2 & 2014-04-02 & 16:52:38 & 17:08:42 & $\ \ $\DurationSecond{} & $180 \pm 12$ & $0.252 \pm 0.058$ \\
\hline
\end{tabular}
}
\tablefoot{
\tablefoottext{a}
{Start and end times (UT) of the flare time interval defined by the Bayesian-blocks algorithm \citep{scargle13b} on the EPIC/pn data;}
\tablefoottext{b}
{Total EPIC/pn counts in the 2$-$10 keV energy band obtained in the smoothed light curve during the flare interval (determined by Bayesian blocks) after subtraction of the non-flaring level obtained with the Bayesian-blocks algorithm;}
\tablefoottext{c}
{EPIC/pn count rate in the 2$-$10 keV energy band at the flare peak (smoothed light curves) after subtraction of the non-flaring level.}
}
\normalsize
\end{table}

\begin{figure}[!ht]
\centering
\begin{tabular}{@{}c@{}c@{}}
\includegraphics[width=0.5\columnwidth, trim = 0 3.6cm 0 0, clip=true, keepaspectratio]{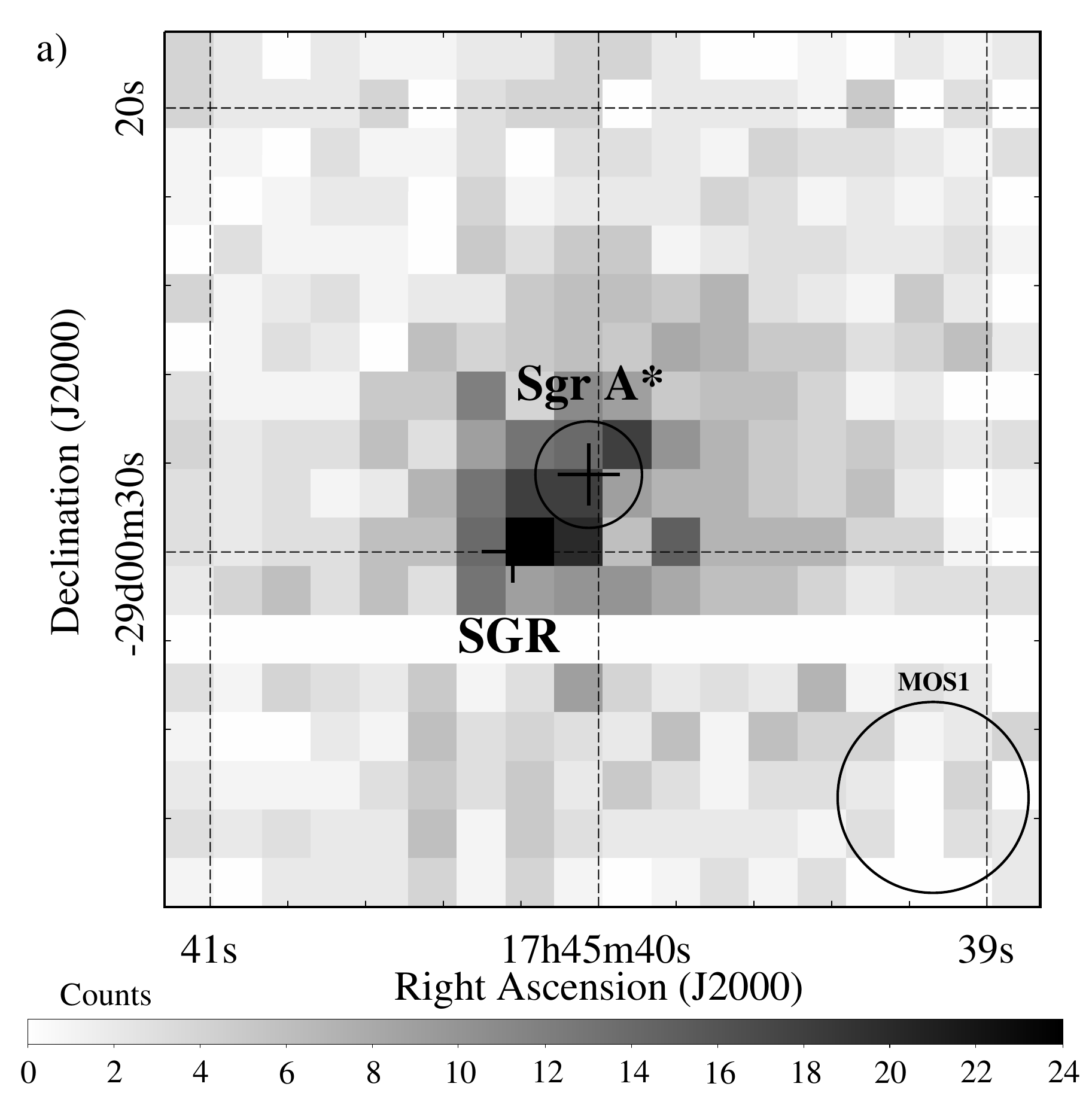}&
\includegraphics[width=0.5\columnwidth, trim = 0 3.6cm 0 0, clip=true, keepaspectratio]{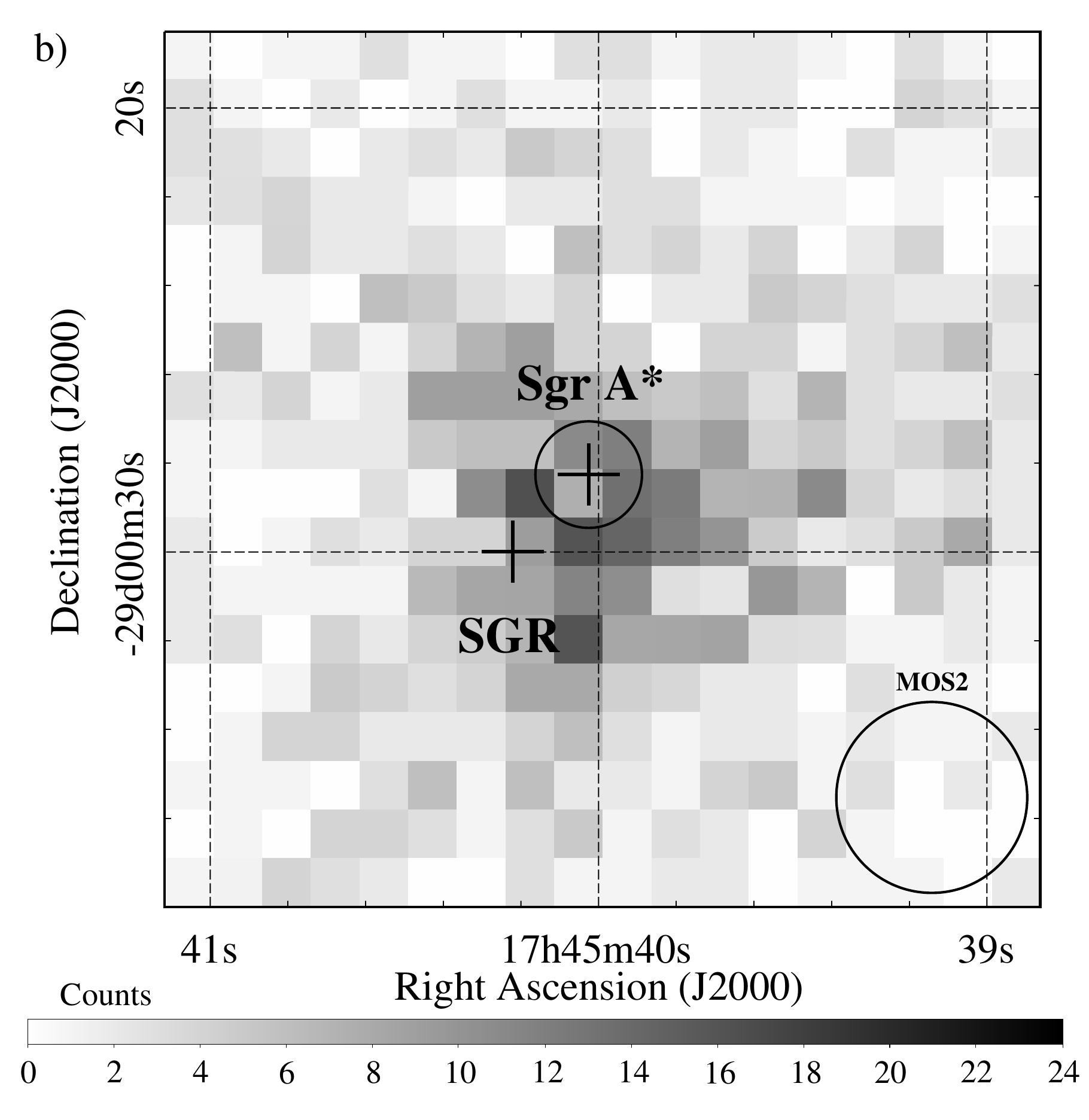}\\
\includegraphics[width=0.5\columnwidth, trim = 0 3.6cm 0 0, clip=true, keepaspectratio]{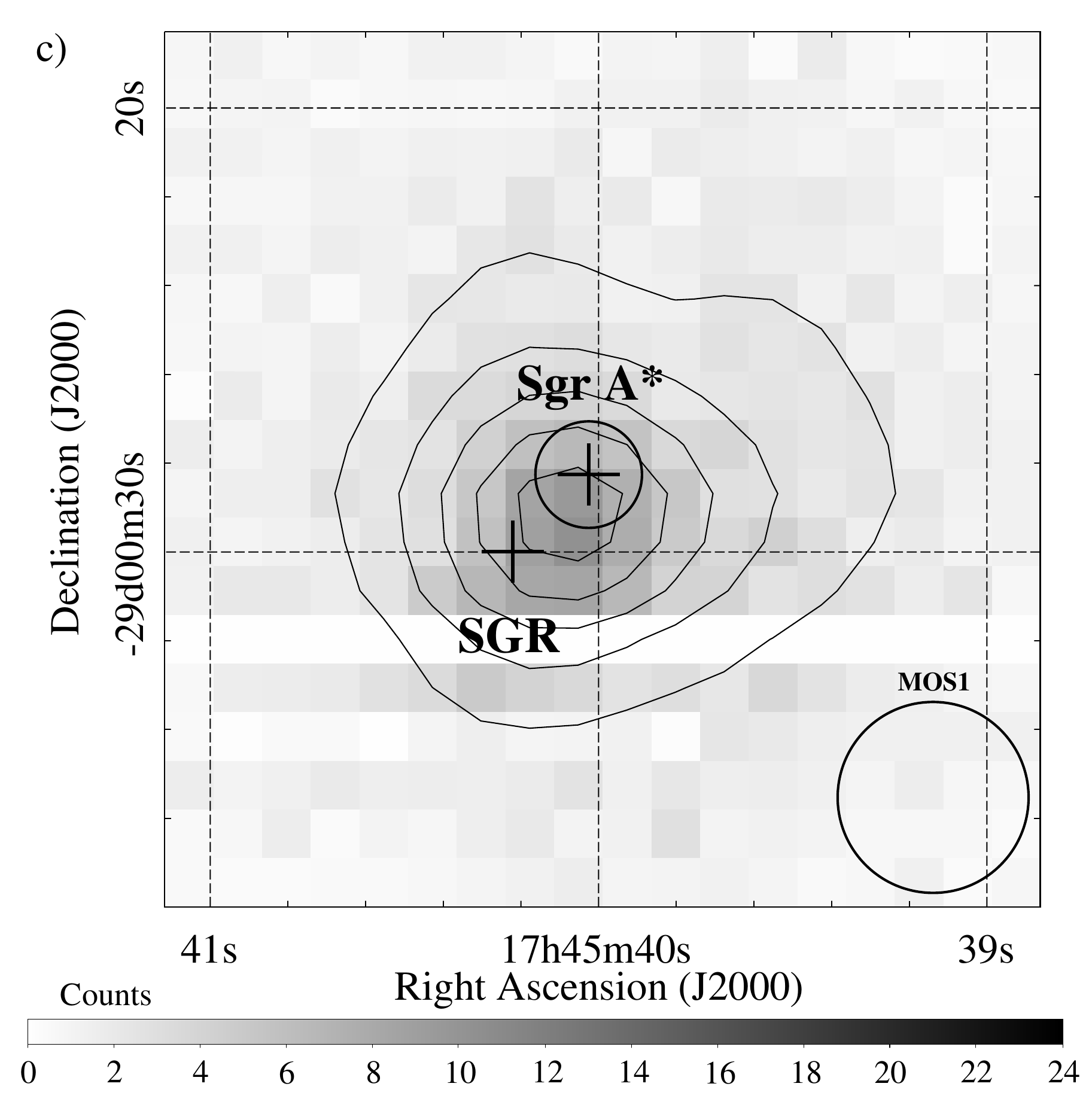}&
\includegraphics[width=0.5\columnwidth, trim = 0 3.6cm 0 0, clip=true, keepaspectratio]{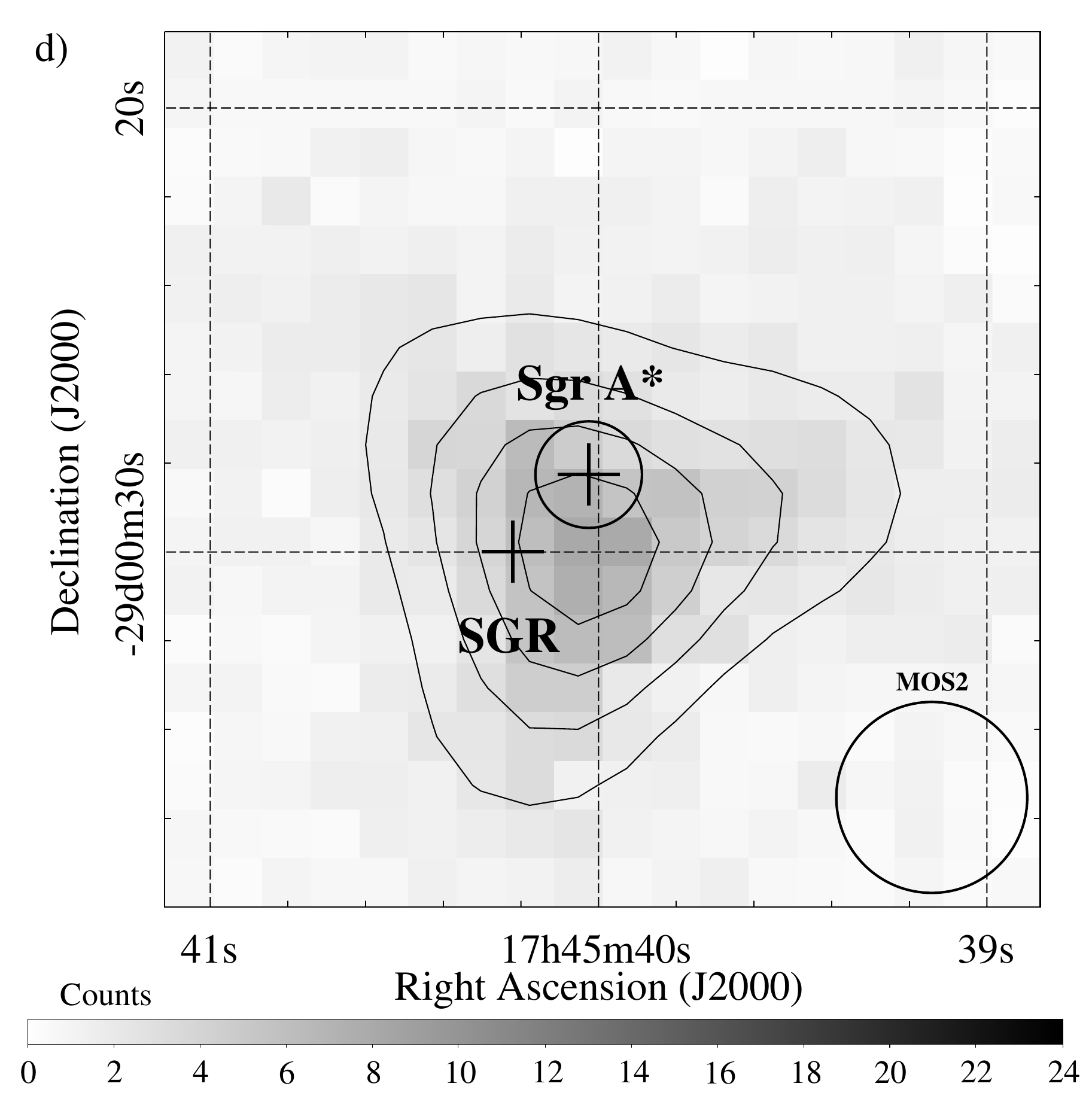}\\
\includegraphics[width=0.5\columnwidth, trim = 0 0cm 0 0, clip=true, keepaspectratio]{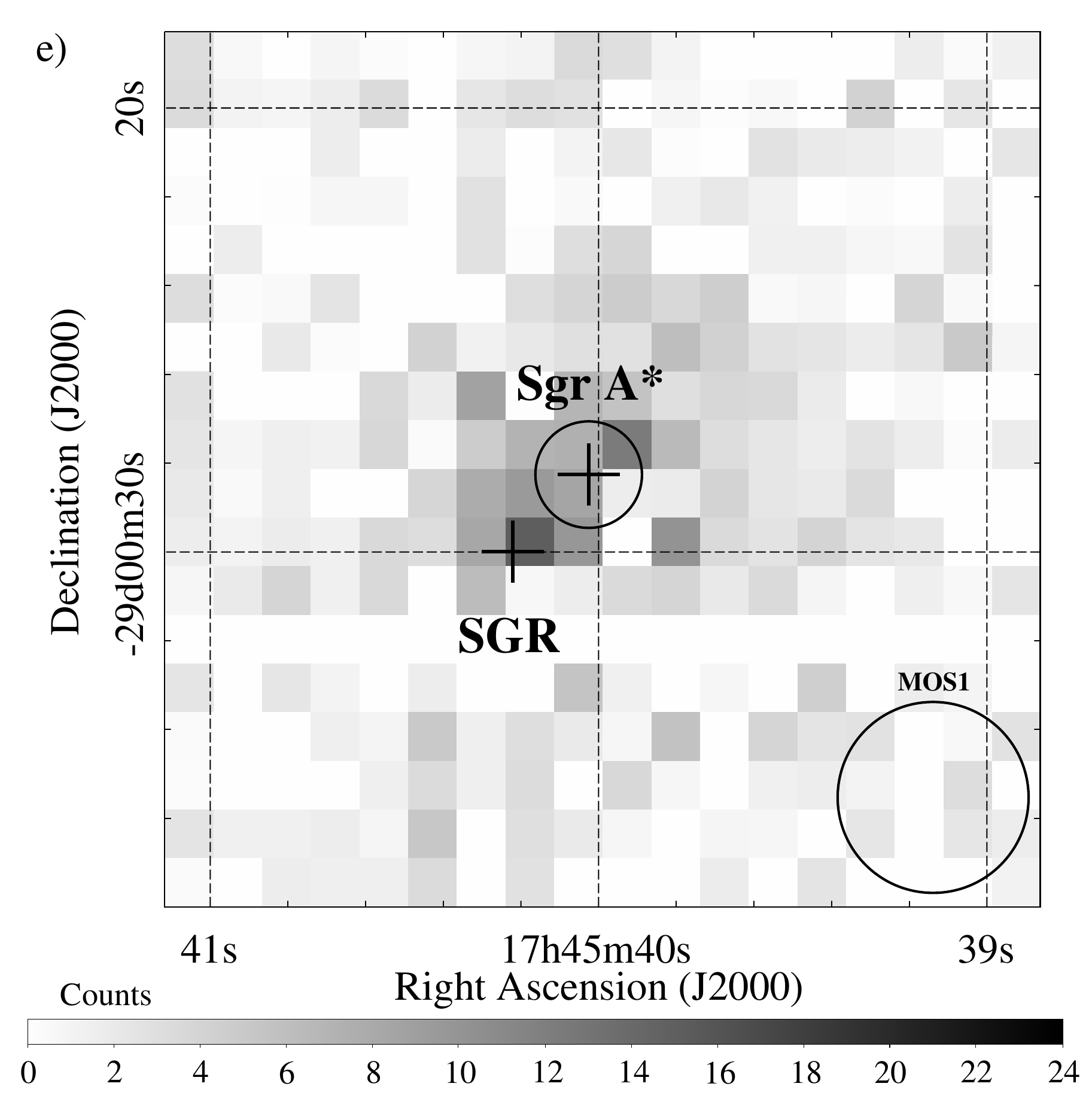}&
\includegraphics[width=0.5\columnwidth, trim = 0 0cm 0 0, clip=true, keepaspectratio]{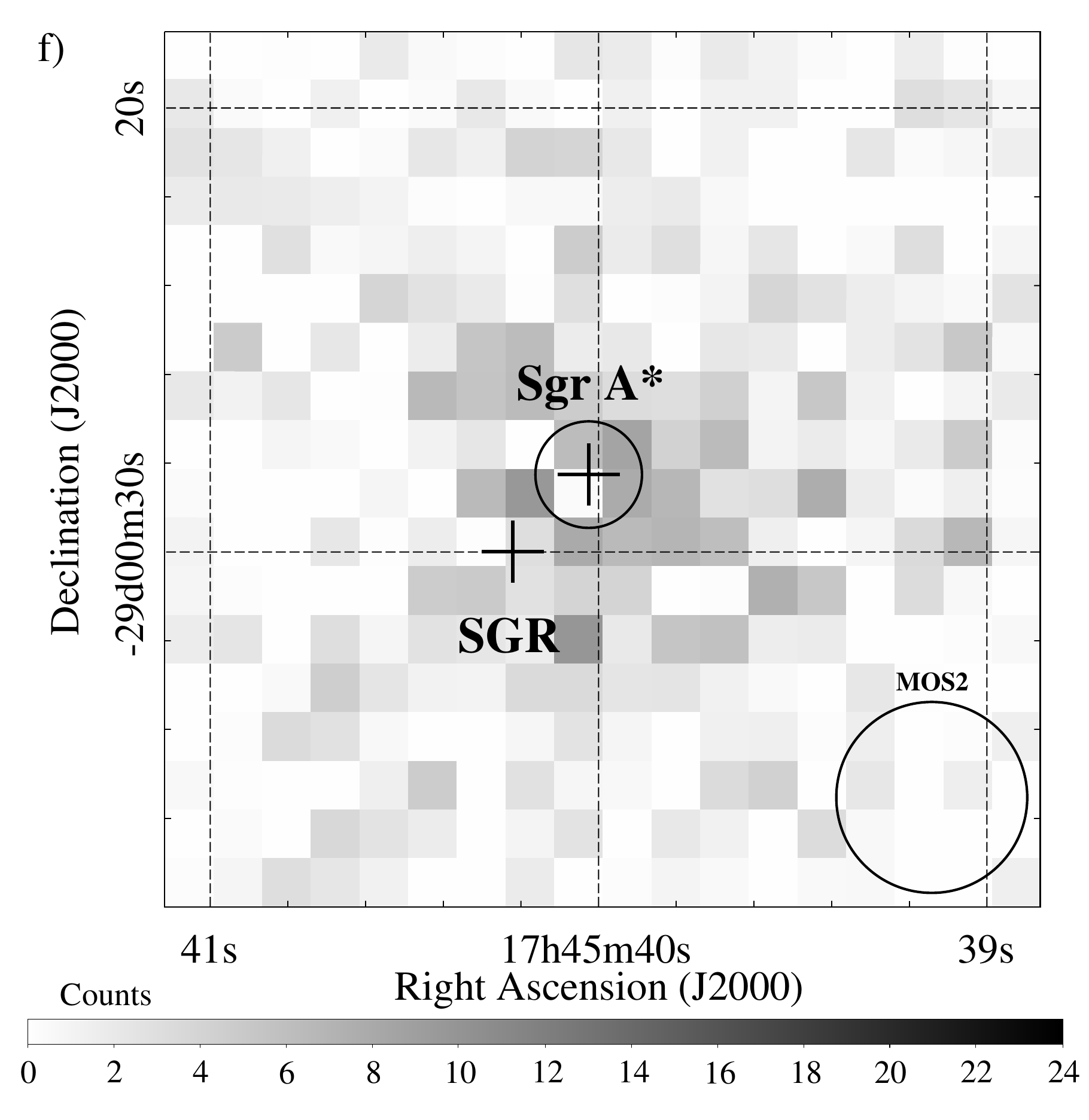}\\
\includegraphics[width=0.5\columnwidth, trim = 0 0cm 0 0, clip=true, keepaspectratio]{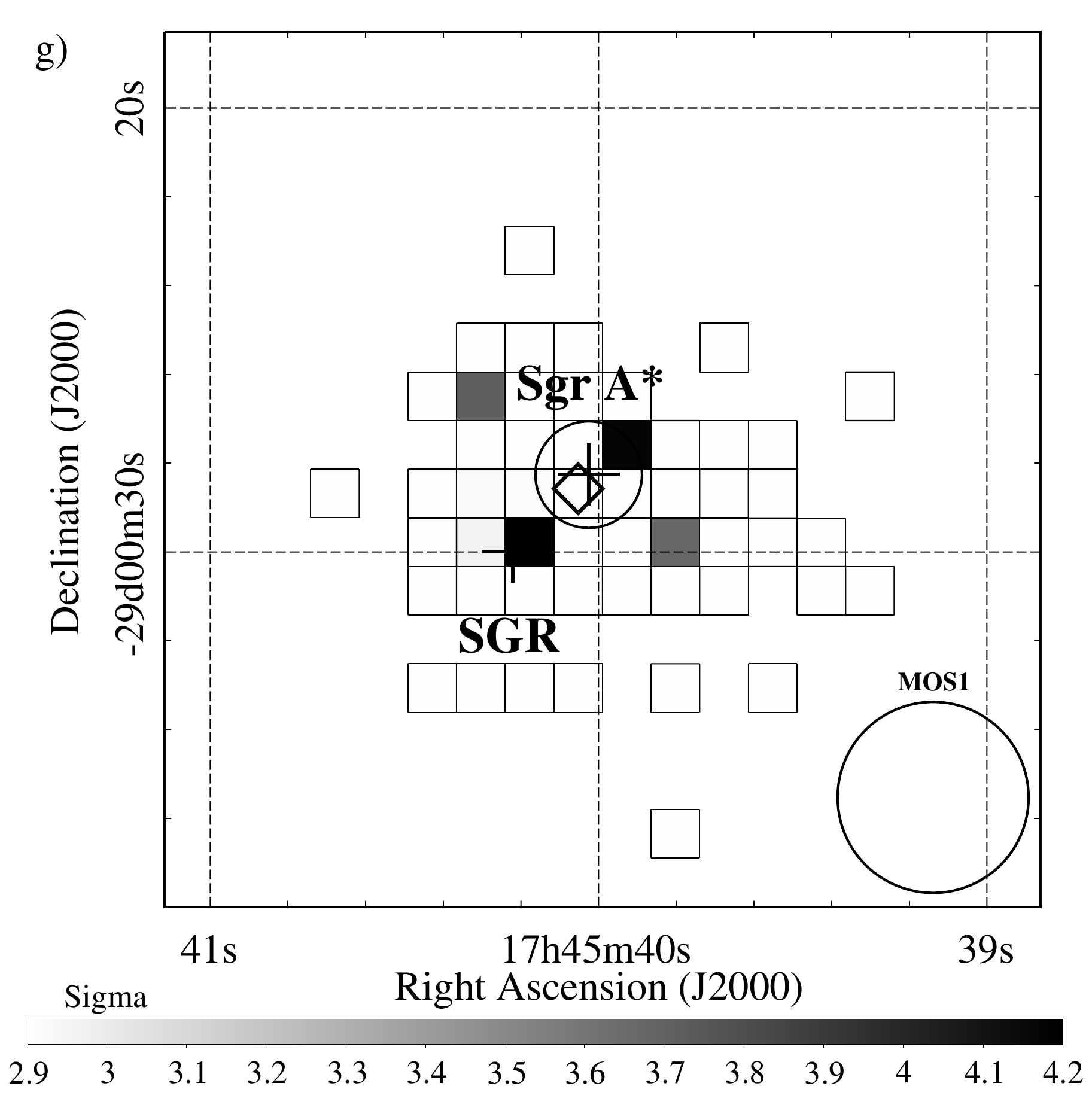}&
\includegraphics[width=0.5\columnwidth, trim = 0 0cm 0 0, clip=true, keepaspectratio]{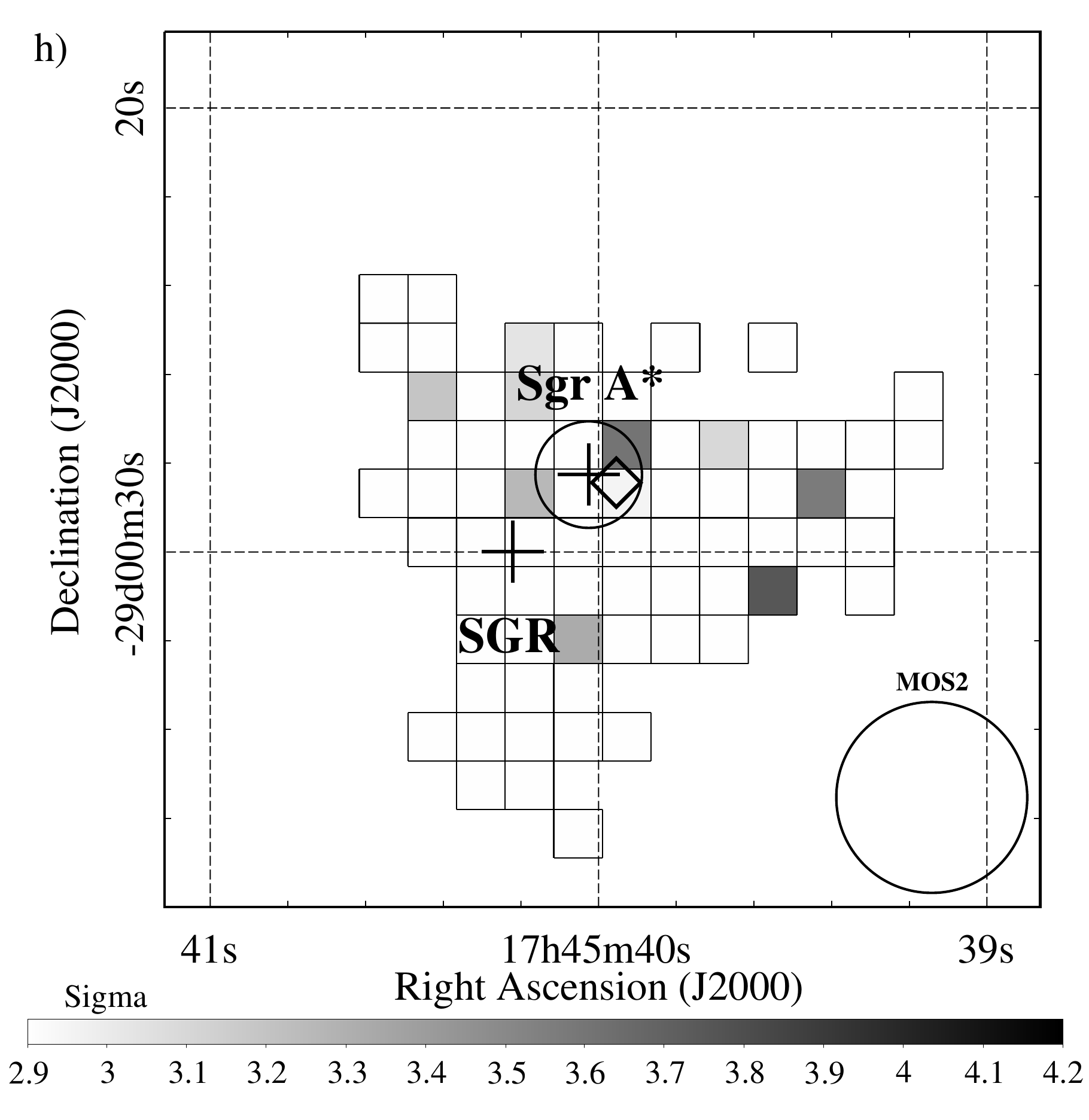}\\
\end{tabular}
\caption{XMM-Newton/MOS1 (left column) and MOS2 (right column) images of \sgra\  on 2014 Mar.\ 10.
The energy range is 2--10~keV.
The field of view is $20\arcsec\times20\arcsec$, the pixel size is $1\farcs1\times1\farcs1$.
The same linear color-scale is used for Fig.~a--f and Fig.~g--h.
In all panels, the black circle in the right-bottom corner is the instrument 
angular-resolution (FWHM); the crosses are the positions of SGR~J1745-2900 \citep{bower15} 
and \sgra\  \citep{petrov11}, surrounded by a circle giving the absolute-astrometry 
uncertainty of EPIC ($1\sigma=1\farcs2$; \citealt{EPIC_calibration_status_document}).
{\sl Panels a and b:} count numbers observed during the flaring period.
{\sl Panels c and d:} count numbers observed during the non-flaring period scaled-down 
to the flaring-period exposure. 
The contour map shows count numbers smoothed on four pixels with a Gaussian, 
starting from 2~counts with step of 1~count.
{\sl Panels e and f:} count excesses during the flaring period.
{\sl Panels g and h:} statistically significant count excesses
($\ge3\sigma$; computed on the boxed-pixel area with the Bayesian method of 
\citealt{kraft91}), the diamond is the corresponding 
count-weighted barycenter of these detections.
}
\label{figure:flaring_images}
\end{figure} 

We removed the magnetar contribution from the \sgra\  EPIC/pn event list in order to increase the detection level of the flares.
This was done by computing the period and period derivative of \magn{} and filtering out time intervals where the magnetar flux is less than
50\% of its total flux (see Appendix \ref{appendix_d} for details).
We only work with EPIC/pn, because it has a better temporal resolution (73.4~ms) than the EPIC/MOS cameras (2.6~s; \citealt{XMM_UHB}).
By applying the Bayesian-blocks analysis on the filtered pn event lists, we find no additional flares, and the start and end times of the already detected flares do not change significantly.

The flare detected on 2014 Mar. 10 is characterized by a long rise ($\sim \RiseFirst{}$ s) and a rapid decay ($\sim \DecayFirst{}$ s).
This is one of the longest flares ever observed in X-ray, with a duration of about 8.5~ks.
For comparison, the largest flare observed during the \textit{Chandra XVP} 2012 campaign has a duration of $7.9$~ks and the first flare detected from \sgra\  observed by \citet{baganoff01} had a duration of $\sim 10$~ks.
In EPIC/pn, the Bayesian-blocks algorithm divides the flare into two blocks, but in EPIC/MOS1 and MOS2 this flare is described with only one Bayesian block.

To localize the origin of this flaring emission we focus on the MOS observations, which provide a good sampling of the X-ray PSF (FWHM~$\sim 4\farcs3$) thanks to their $1\farcs1\times1\farcs1$~pixels.
We first compute sky images that match the detector sampling for the flaring and non-flaring periods, and then we look for any significant excess counts during the flaring period compared to the non-flaring one, using the Bayesian method of \citet{kraft91}.

We have suppressed the randomization of the event position inside the detector pixel during the production of the event list, therefore the event is assigned to the center of the detector pixel and its sky coordinates are reconstructed from the spacecraft attitude with an angular resolution
of $0\farcs05$. 
We filter the X-ray events using the (softer) \texttt{\#XMMEA\_EM} flag (e.g., bad rows are filtered out, keeping adjacent rows) and we select only events with the best positioning (single-pixel events, corresponding to \texttt{pattern=0}) and 2--10~keV energy.
We first assess the mean sky position of the detector pixel that was the closest to \sgra\  by comparing the event sky positions with the pattern of the spacecraft offsets from the mean pointing that we derived from the attitude history file (\texttt{*SC*ATS.FIT}). 
We then compute images and exposure maps centered on this sky position with $1\farcs1\times1\farcs1$~sky-pixels for the flaring and non-flaring periods (see Appendix~\ref{appendix_a} for the definition of the Bayesian blocks). 
There is no moir\'e effect in these images, because the mean position-angle of the detector (90\fdg78) is very close to 90\degr. 
Panels~a and b of Fig.~\ref{figure:flaring_images} show the MOS1 and MOS2 count numbers during the flaring period. 
Following \citet{kraft91} we denote this image $N$. The horizontal row with no counts in the MOS1 image is due to a bad row. 
Panels~c and d of Fig.~\ref{figure:flaring_images} show the MOS1 and MOS2 count numbers during the non-flaring period, scaled-down to the flaring-period exposure using the exposure map ratios. 
This image is our estimate of the mean count numbers during the non-flaring period. 
Following \citet{kraft91} we denote this image $B$, as background.
Panels~e and f of Fig.~\ref{figure:flaring_images} show the difference between the previous panels, shown only for potential count excesses ($N-B>0$). 
Following \citet{kraft91} we denote this image $S$, as source.
Poisson statistics are required due to the low number of counts, hence we have to carefully determine the confidence limits of the observed count excesses to select only pixels that exclude null values at the confidence level $CL$. 

Since the Bayesian method of \citet{kraft91} requires that the background estimate is close to the true value (see also \citealt{helene83}), we limit our statistical analysis to the pixels where the count number during the non-flaring period is larger or equal to 20, in order to reduce the Poisson noise (see the boxed pixel areas in panels g and h of Fig.~\ref{figure:flaring_images}). 
We compute the confidence level for each count excess using Eq.~(9) of \citet{kraft91}\footnote{Following \citet{kraft91}, we first determine the confidence interval [$s_\mathrm{min}$,~$s_\mathrm{max}$] of $S \equiv N-B$ at the confidence level $CL$ where, for a count excess, $s_\mathrm{max}$ is defined as $f_{N,B}(s_\mathrm{max}) \equiv f_{N,B}(s_\mathrm{min})$ and $s_\mathrm{min}=0$, with $f_{N,B}(S)\equiv exp(-(S+B))\,(S+B)^N/(N!\,\Sigma_{n=0}^N exp(-B)B^n/n!)$ is the posterior probability distribution function.
We then compute $CL=\int_{s_\mathrm{min}}^{s_\mathrm{max}}\!f_{N,B}(s)\,ds$ and its Gaussian equivalent in units of $\sigma$ given by $\phi^{-1}((1-CL)/2)$, with $\phi^{-1}$ being the reciprocal of the cumulative distribution function of the normal distribution.} 
and convert it to a Gaussian equivalent in units of $\sigma$. 
Panels~g and h of Fig.~\ref{figure:flaring_images} show pixels with confidence levels that are larger or equal to $3\sigma$. The barycenters of these pixels weighted by their count excesses (diamonds in panels~g and h of Fig.~\ref{figure:flaring_images}) are consistent with the position of \sgra\  when considering the absolute astrometry uncertainty of EPIC, which confirms that the flaring emission detected on 2014 Mar.\ 10 came from \sgra.

\subsubsection{Spectral analysis of the X-ray flares}
\label{spectrum_xmm}
\begin{table}
\caption{Spectral properties of the X-ray flares observed by XMM-Newton. }
\centering
\scalebox{0.65}{
\label{table:spectre}
\begin{tabular}{@{}cccccc@{}}
\hline
\hline
Flare day & $N_\mathrm{H}$\tablefootmark{a} & $\Gamma$\tablefootmark{b} & $F\mathrm{^{unabs}_{2-10keV}}$~\tablefootmark{c} & $L\mathrm{^{unabs}_{2-10keV}}$~\tablefootmark{d} & $\chi^2_\mathrm{red}\ $\tablefootmark{h}\\
 (yy-mm-dd) & ($10^{22}\ \mathrm{cm^{-2}}$) &  & ($10^{-12}\ \mathrm{erg\ s^{-1}\ cm^{-2}}$)  & ($10^{34}\ \mathrm{erg\ s^{-1}}$)\\
\hline  
 2014-03-10\tablefootmark{e} & $\NHFirst{}$ & $\GammaFirst{}$ & $\FluxFirst{}$ & $\LumFirst{}$ & 1.65\\
 2014-04-02\tablefootmark{e} & \enskip$\NHSecond{}$ & $\GammaSecond{}$ & $\FluxSecond{}$ & $\LumSecond{}$ & 1.72\\
 2002-10-03\tablefootmark{f} & $16.1\,(13.9$--$18)$ & $2.3\,(2.0$--$2.6)$ & $26.0\,(22.5$--$30.6)$ & $19.8\,(17.1$--$23.3)$ & \\
 2007-04-04\tablefootmark{g} & $16.3\,(13.7$--$19.3)$ & $2.4\,(2.1$--$2.8)$ & $16.8\,(13.8$--$21.4)$ & $12.8\,(10.5$--$16.3)$ & \\
\hline
\end{tabular}
}
\tablefoot{
\tablefoottext{a}
{Hydrogen column density;}
\tablefoottext{b}
{Photon index of the power law;}
\tablefoottext{c}
{Unabsorbed average flux between 2 and 10 keV;}
\tablefoottext{d}
{Unabsorbed average luminosity between 2 and 10 keV assuming a distance of 8~kpc;}
\tablefoottext{e}
{Spectral properties of the EPIC/pn spectrum computed using the MCMC method. The range given between parenthesis represents the 90\% confidence interval;}
\tablefoottext{f}
{Spectral properties of the EPIC (pn+MOS1+MOS2) spectrum. See \citet{porquet03} and \citet{nowak12};}
\tablefoottext{g}
{Spectral properties of the EPIC (pn+MOS1+MOS2) spectrum. See \citet{porquet08} and \citet{nowak12};}
\tablefoottext{h}
{Reduced $\chi^2$ for 3 degrees of freedom.}
}
\normalsize
\end{table}
\begin{figure}
\centering
\includegraphics[trim= 0cm 0cm 0cm 2cm, clip,width=9cm]{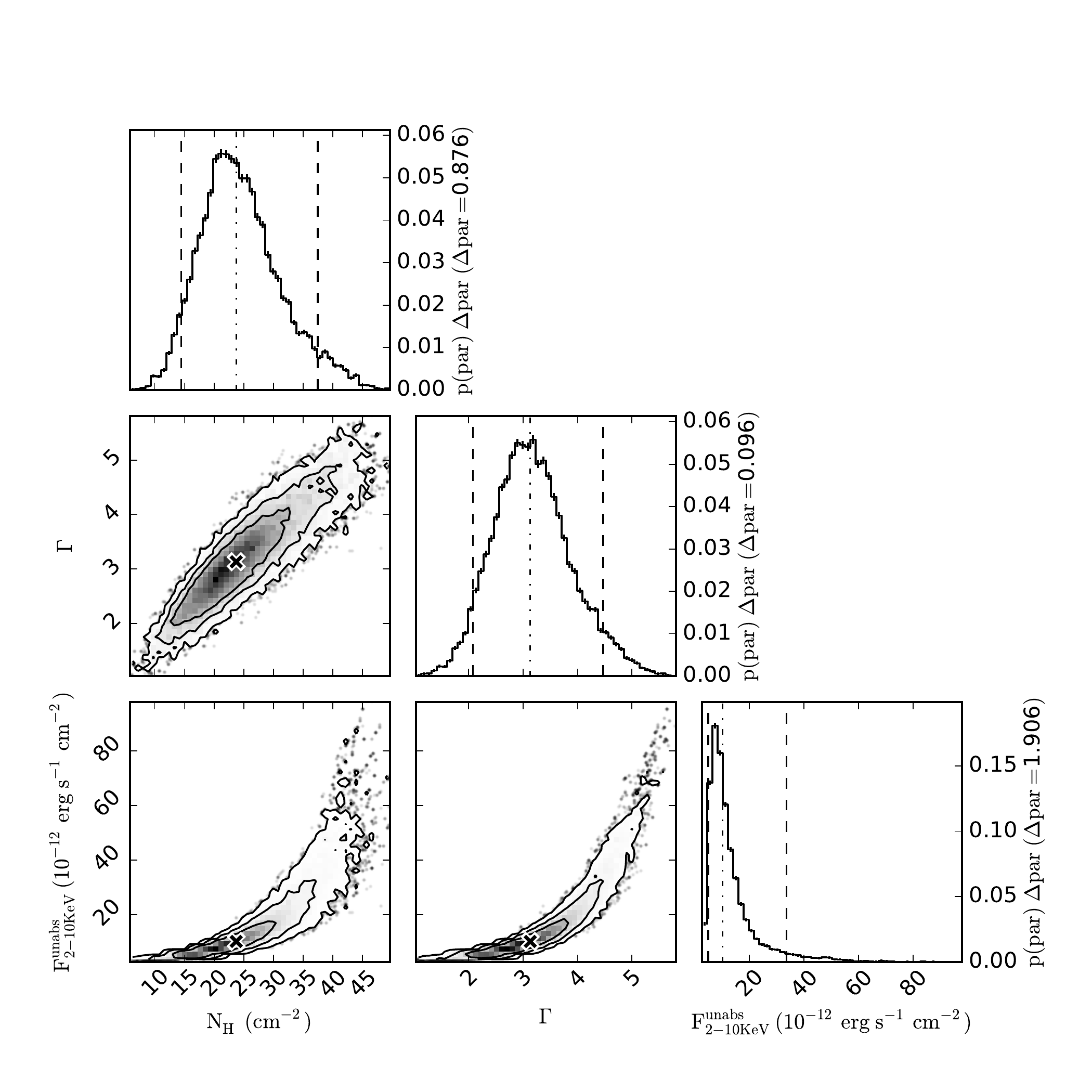}\\
\vspace*{-0.5cm}
\includegraphics[trim= 0cm 0cm 0cm 2cm, clip,width=9cm]{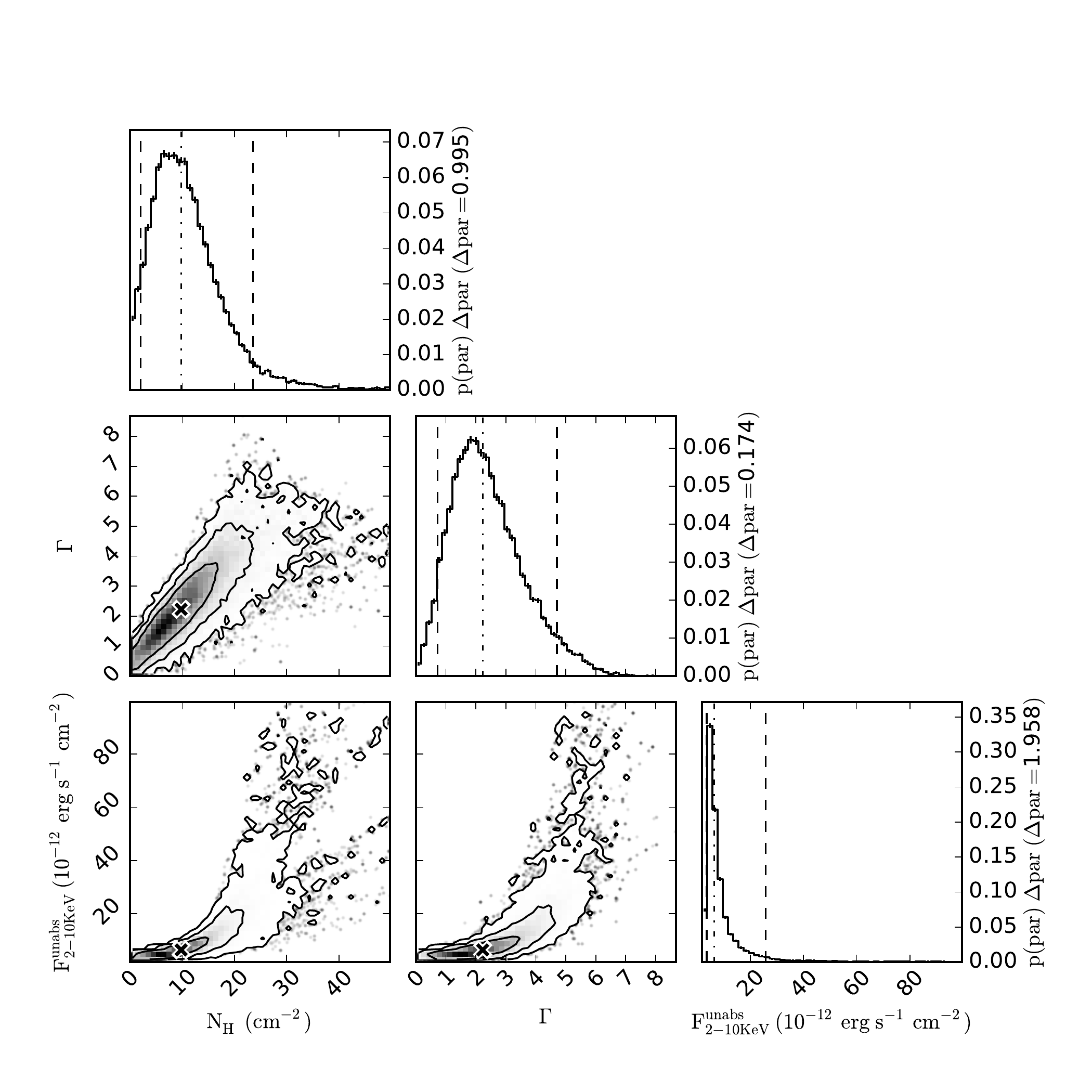}
\caption{Best-fit parameters of the 2014 Mar. 10 (\textit{Top}) and 2014 Apr. 2 (\textit{Bottom}) flares.
The diagonal plots are the marginal density distribution of each parameter.
The median values of each parameter are represented by the vertical dotted lines in diagonal plots and by a cross in other panels; the vertical dashed lines define the 90\% confidence interval (see Table~\ref{table:spectre} for the exact values).
The contours are 68\%, 90\% and 99\% of confidence levels.}
\label{fig:triangle}
\end{figure}
\begin{figure}
\centering
\includegraphics[trim= 0.5cm 0.5cm 0.5cm 0.7cm, clip,width=6.1cm]{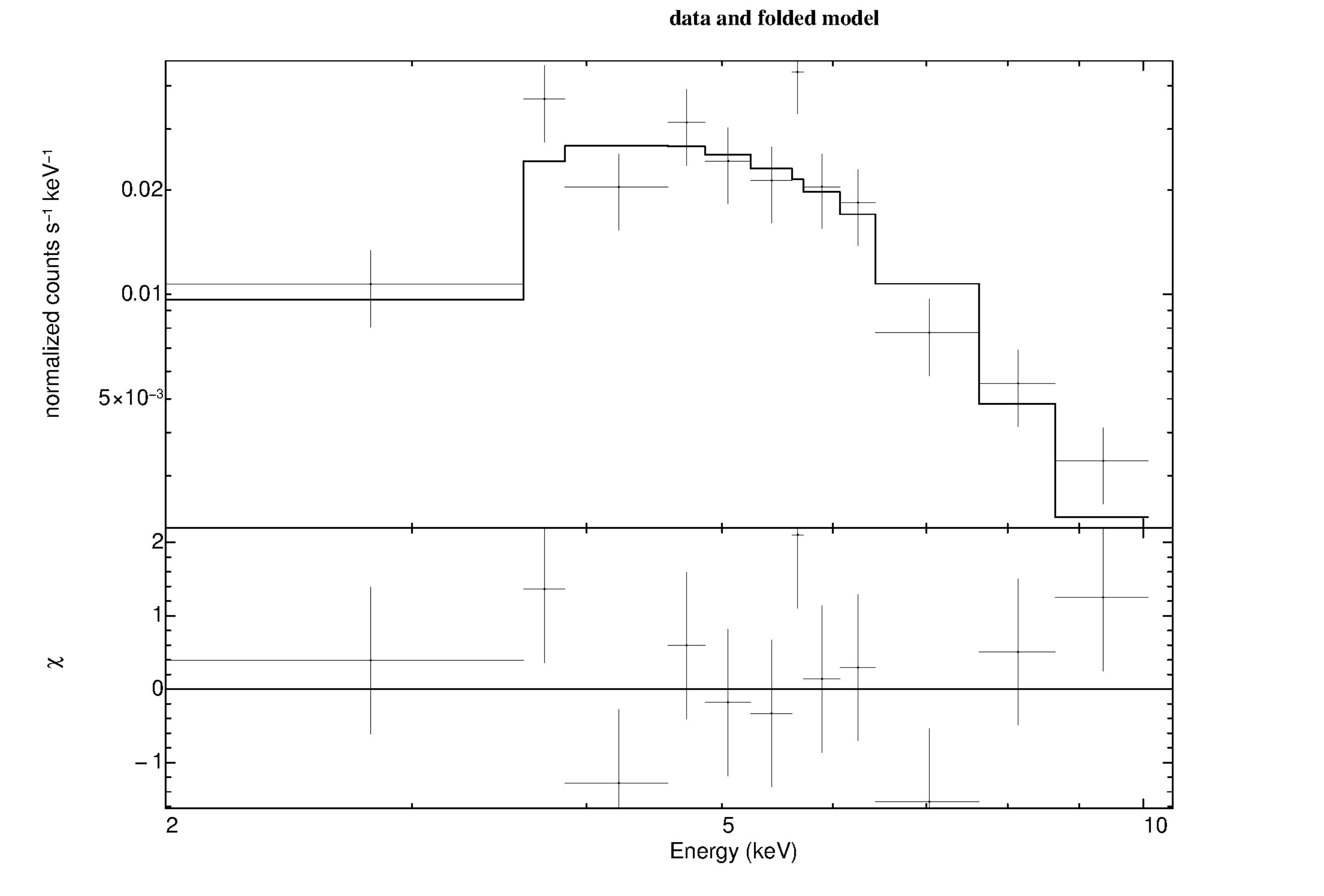}\\
\includegraphics[trim= 0.5cm 0.5cm 0.5cm 0.7cm, clip,width=6.1cm]{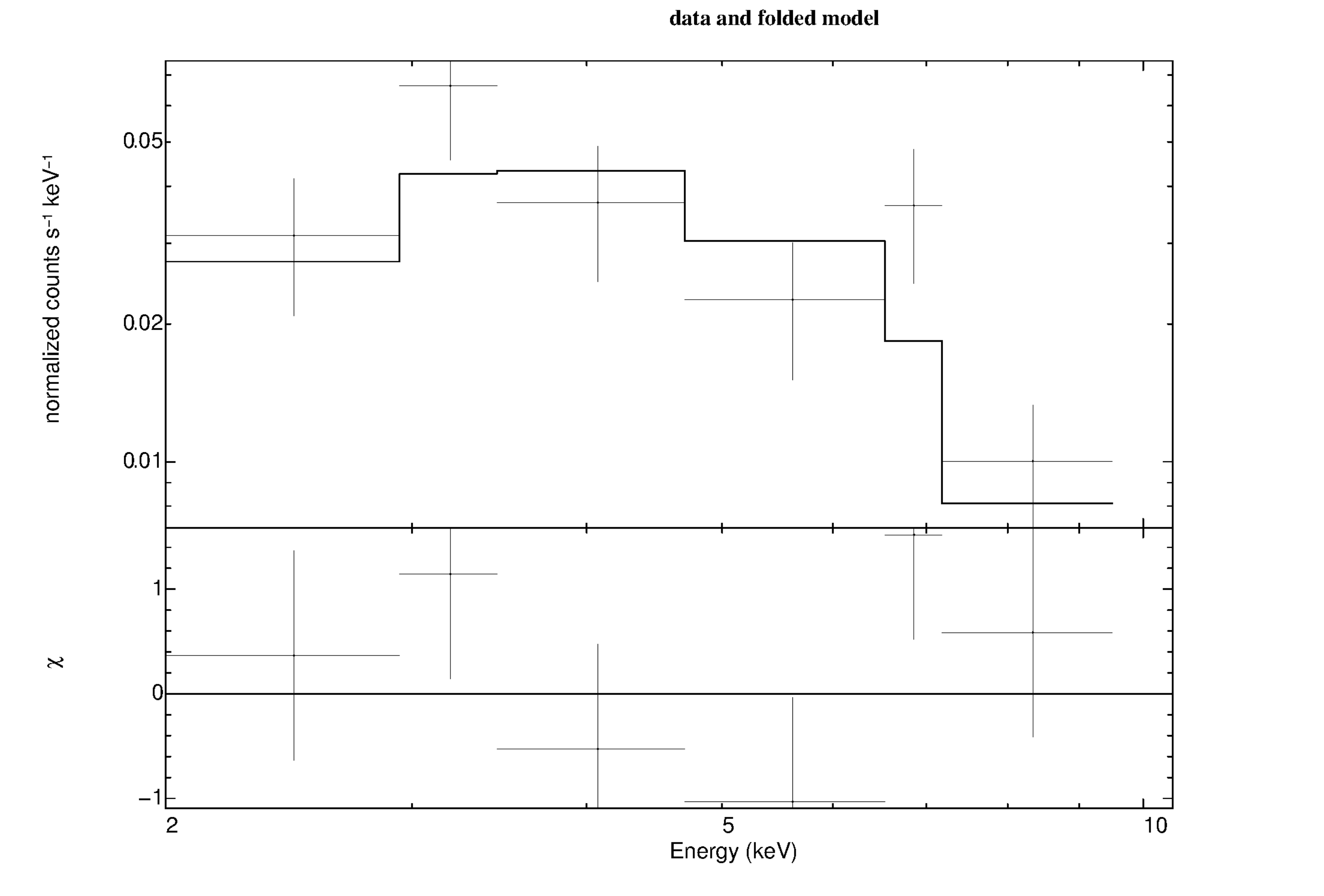}
\caption{XMM-Newton/EPIC pn spectrum of the 2014 Mar. 10 (\textit{Top}) and 2014 Apr 2 (\textit{Bottom}) flares.
The model is the best spectrum obtained with MCMC (see text for details).
The lower panel in the two graphs is the residual.
The horizontal and vertical lines are the spectral bins and the error on the data, respectively.}
\label{fig:spectre}
\end{figure}
To analyze the spectrum of the two flares seen by XMM-Newton on 2014 Mar. 10 and 2014 Apr. 2, we extracted events from a circle of 10$\arcsec$ radius centered on the \sgra\  radio position, as we did for the temporal analysis.
The X-ray photons were selected with \texttt{PATTERN$\leq 4$} and \texttt{FLAG==0} for the pn camera.
We did not work with photons from MOS1 and MOS2, because the number of events is too small to constrain the spectral properties. 
The source+background time interval is the range between the beginning and the end of the flare computed by the Bayesian-blocks algorithm (see Table \ref{table:2}).
The background time interval is the whole observation minus the time range during the flare.
We also rejected 300s on either side of the flare to avoid any bias.
This extraction is the same as used in \citet{mossoux14}.
We computed the spectrum, ancillary files, and response matrices with the SAS task \texttt{especget}.

The model used to fit the spectrum with XSPEC (version 12.8.1o) is the same as that in \citet{mossoux14}: an absorbed power law created using \textit{TBnew} \citep{wilms00} and \textit{pegpwrlw} with a dust scattering model from \textit{dustscat} \citep{predehl95}.
\textit{TBnew} uses the cross-sections from \citet{verner96}.
Interstellar medium abundances of \citet{wilms00} imply a decrease of the column density by a factor of 1.5 \citep{nowak12}. 
The extracted spectrum was grouped using the SAS task \texttt{specgroup}.
The spectral binning begins at 2~keV with a minimum signal-to-noise ratio$\,$\footnote{The equation computing the signal-to-noise ratio is the same as in \texttt{specgroup} and in ISIS \citep{isis}. We therefore use the same grouping as in \citet{mossoux14}.} of 4 and 3 for the first and second flares, respectively.
The number of net counts during the first flare is 900 (see Table~\ref{table:2}) and the number of spectral bins is 12.
This gives an average of about 75 counts in each spectral bin.
If we perform the same computation for the second flare, which has 180 net counts for 6 spectral bins, we have 31 counts per spectral bin.

We used the Markov Chain Monte Carlo (MCMC) algorithm to constrain the three parameters of the model: the hydrogen column density ($N_H$), the photon index of the power law ($\Gamma$), and the unabsorbed flux between 2 and 10 keV ($F^\mathrm{unabs}_\mathrm{2-10\ keV}$).
The MCMC makes a random walk of $nstep$ steps in parameter space for several walkers ($nwalkers$), which evolve simultaneously.
The position of each walker at a step in the parameter space is determined by the positions of the walker at the previous step.
Convergence was achieved using the probability function of the parameters.
The resulting MCMC chain reports all these steps.
This method give us a complete view of the spectral parameters distribution and correlation.

We use Jeremy Sanders' {\tt XSPEC\_emcee}\footnote{\href{https://github.com/jeremysanders/xspec\_emcee}{https://github.com/jeremysanders/xspec\_emcee}} program that allows MCMC analyses of X-ray spectra in XSPEC using {\tt emcee}\footnote{\href{http://dan.iel.fm/emcee/current/user/line/}{http://dan.iel.fm/emcee/current/user/line/}} \citep{foreman-mackey13}, an extensible, pure Python implementation of \citet{goodman10}'s affine invariant MCMC ensemble sampler.
We follow the operating mode explained in the \texttt{XSPEC\_emcee} homepage to find the optimal value for the MCMC sampler parameters.
Two criteria must be fulfilled to have a good sampling in the chain: the chain length must be greater than the autocorrelation time and the mean acceptance fraction must be between 0.2 and 0.5 \citep{foreman-mackey13}.
We created a chain containing 30 walkers.
The Python function \texttt{acor} computes the auto-correlation time ($\tau_{acor}$) needed to have an independent sampling of the target density.
The burn-in period ($nburn$) and chain length ($nstep$) are defined as $20 \times \tau_{acor}$ \citep{sokal96} and $30 \times nburn$ \citep{foreman-mackey13}, respectively.
For the spectral model used here, $\tau_{acor}=$5.1 and 5.3 for the 2014 Mar. 10 and 2014 Apr. 2 flares, respectively.
Thus we used $nburn=102$, $nwalkers=30$, and $nstep=3060$ for the March 10 flare and $nburn=106$, $nwalkers=30$, and $nstep=3180$ for the April 2 flare.
The mean acceptance fraction is around 0.6 for the two flares, which is a reliable value.

The diagonal plots in Fig.~\ref{fig:triangle} are the marginal distribution of each parameter (i.e., the probability to have a certain value of one parameter independently from others).
The other panels in Fig.~\ref{fig:triangle} represent the joint probability for each pair of parameters.
The contours indicate the parameter region where there are 68\%, 90\% and 99\% of the points (i.e., $nwalkers \times nstep$). 
The best-fit parameter values are the median (i.e., 50th percentile) of each parameter obtained from the marginal distribution.
We also define a 90\% confidence range for each parameter as the 5th and 95th percentile of the marginal distribution.
These numbers are reported in Table~\ref{table:spectre}.
The corresponding best spectrum is over-plotted on the data in Fig.~\ref{fig:spectre}.

We can compare the spectral parameters of this flare with those of the two brightest flares detected with XMM-Newton, which have the better constrained spectral parameters thanks to the high throughput and no pileup \citep{porquet03,porquet08}.
Their spectral properties are reported in Table~\ref{table:spectre}.
The magnetar has a soft spectrum, which implies that the soft part ($0.5-3$ keV) of the background is very high.
Thus we have only one spectral bin in this energy band (see Fig.~\ref{fig:spectre}), implying that the hydrogen column density is not well constrained.
The hydrogen column density and the photon index of the two brightest flares are well within the 90\% confidence range of the 2014 Mar. 10 and 2014 Apr. 2 flares even if the parameters of the latter are less constrained than the former.

\begin{figure}
\centering
\includegraphics[width=8.cm,angle=90]{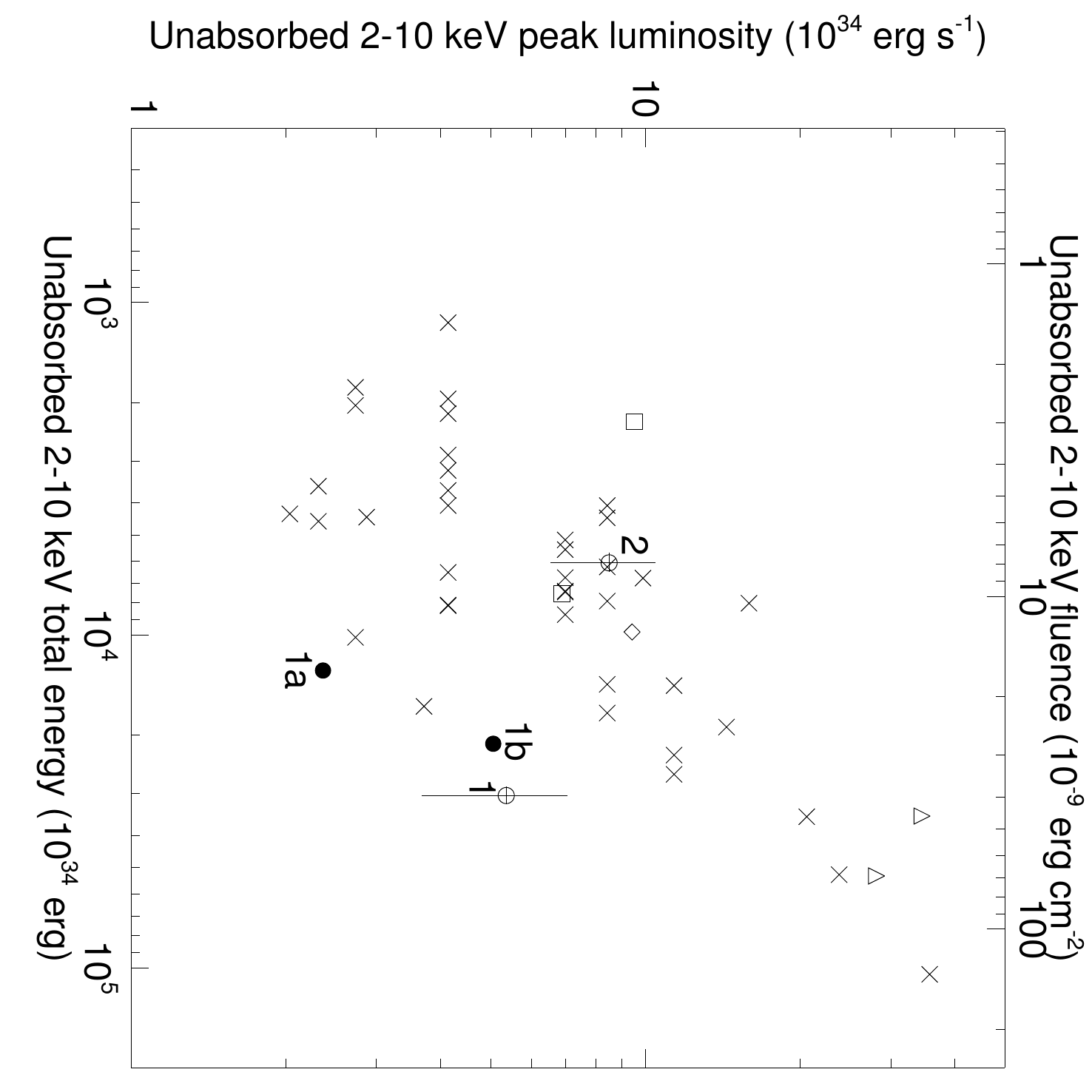}
\caption{Unabsorbed total energy vs. unabsorbed peak luminosity of the X-ray flares (adapted from \citealt{mossoux14}).
The top x-axis is the unabsorbed fluence.
The crosses represent the X-ray flares from the \textit{Chandra XVP} campaign \citep{neilsen13}, 
the triangles are the two brightest flares seen with XMM-Newton \citep{porquet03,porquet08}, 
the diamond and the two squares are the 2011 March 30 flare and its subflares, respectively \citep{mossoux14}.
The empty circles are X-ray flares 1 and 2 of this work with their $1\sigma$ error bars.
The filled circles are the components 1a and 1b of flare 1 (see Section~\ref{two_flares}).}
\label{fig:neilsen13}
\end{figure}

\begin{figure*}
\centering
\includegraphics[width=6.5cm,angle=90]{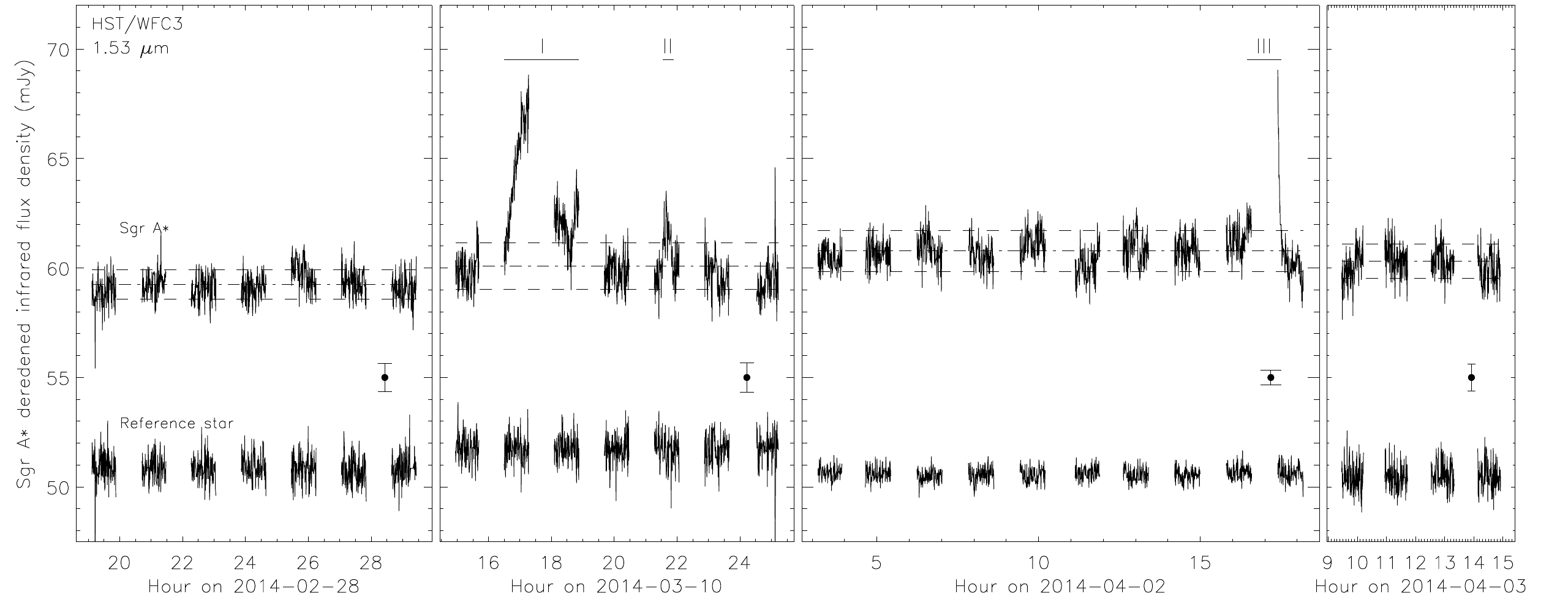}
\caption{Light curves of \sgra\  obtained with WFC3 on board HST during 2014 Feb.$-$Apr.
The NIR flares are labeled with Roman numerals.
The horizontal lines below these labels indicate the flare durations.
The error bar in each panel is standard deviation of the photometry. }
\label{ir_lc}
\end{figure*}

\begin{figure}
\centering
\includegraphics[trim=  0cm 0.cm 0.cm 0cm, clip,width=8.cm,angle=90]{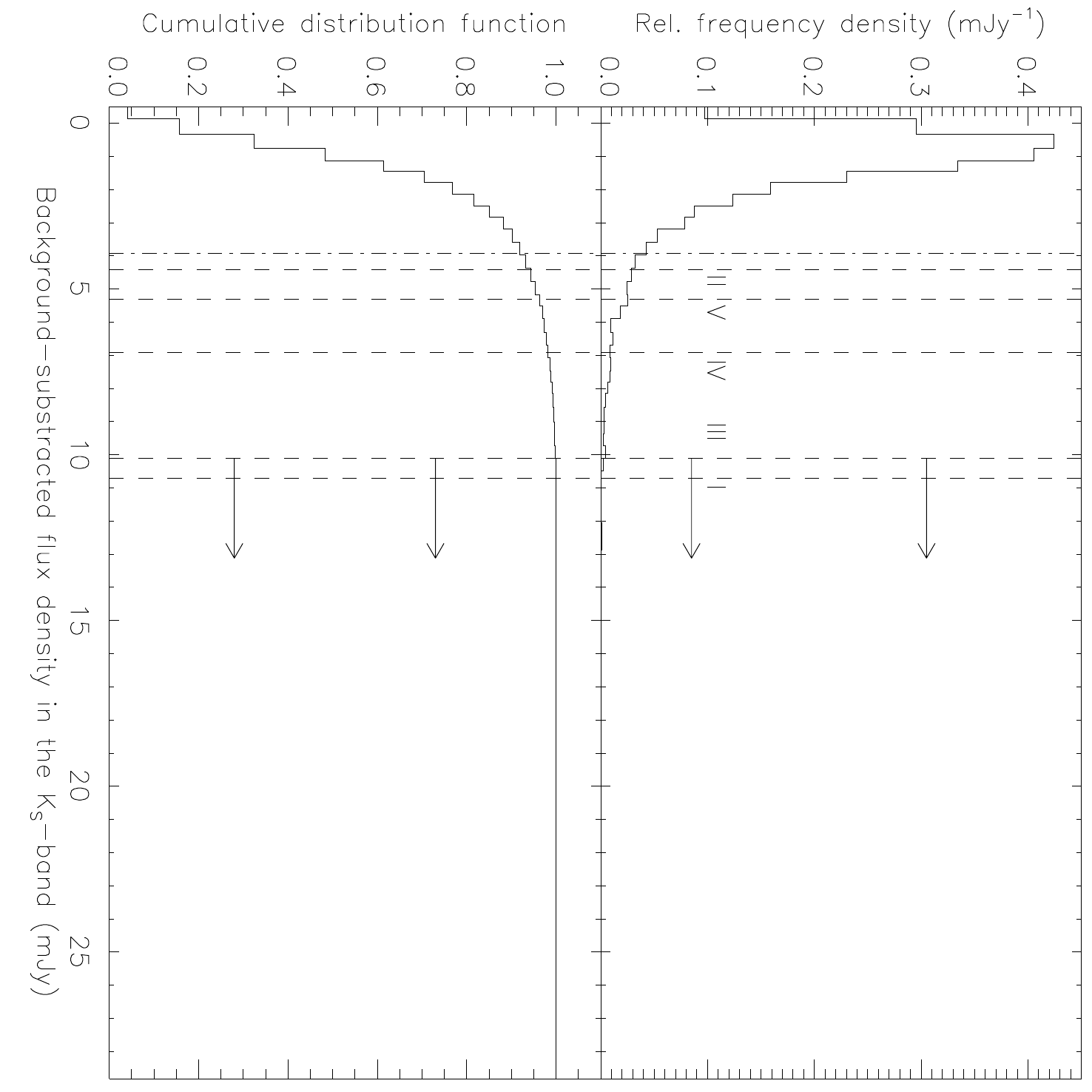}
\caption{Histogram of the NIR flux densities from \sgra\  observed in the $K_\mathrm{s}$-band with NACO at ESO's VLT (adapted from Fig.~3 of \citealt{witzel12}).
{\sl Top panel:} The solid line is the normalized distribution of the NIR flux densities corrected from the background emission.
The dashed lines are the amplitude of the HST flares \textrm{I}, \textrm{II} and the lower limit of the amplitude of the flare \textrm{III} extrapolated to the $K_\mathrm{s}$-band.
We also represented the amplitude above the 3$\sigma$ limit of the VLT flares \textrm{IV} and \textrm{V} extrapolated to the $K_\mathrm{s}$-band.
The dot-dashed line is the detection limit corresponding to 3 times the standard deviation of the quiescent flux density of HST on 2014 Mar. 10. 
{\sl Bottom panel:} The cumulative distribution function of the NIR flux densities from \sgra\  corrected from the background emission.}
\label{witzel}
\end{figure}

Assuming the typical spectral parameters of the X-ray bright flares, i.e., $\Gamma=2$ and $N_\mathrm{H}=14.3 \times 10^{22}\ \mathrm{cm^{-2}}$ \citep{porquet03,porquet08,nowak12}, we determined with XSPEC and the pn response files in the 2--10~keV energy range an unabsorbed-flux-to-count-rate ratio of $4.41 \times 10^{-11}  \,\mathrm{erg\,s^{-1}\,cm^{-2}}/\mathrm{pn\ count\,s^{-1}}$ (corresponding to an absorbed-flux-to-count-rate ratio of $2.01 \times 10^{-11}  \,\mathrm{erg\,s^{-1}\,cm^{-2}}/\mathrm{pn\ count\,s^{-1}}$).
From the 8~kpc distance and the total number of counts (Table~\ref{table:2}), we determine a total energy of $30.4 \pm 1.9\times 10^{37}$ and $6.0 \pm 0.4\times 10^{37}\ \mathrm{ergs}$ ($1\sigma$ error) for the 2014 Mar. 10 and Apr. 2 flares, respectively.
These values can be compared to flares previously observed with Chandra and XMM-Newton.
Figure~\ref{fig:neilsen13} shows the total energy of these flares versus the unabsorbed peak luminosity.
Flare 1 is one of the most energetic flares, due to its very long duration.
The peak amplitude and total energy of flare 2 is close to the median values observed for this flare sample.

\subsection{HST data}
The HST light curves of \sgra\  and a reference star for the four visits are shown in Fig.~\ref{ir_lc}.
The error bar in each panel represents the typical uncertainty on the photometry derived for the reference star (standard deviation of the photometry).
The deredened non-flaring flux density of \sgra\  and the corresponding error, computed using a $1\sigma$-clipping method, are $59.3 \pm 0.7$, $60.1 \pm 0.9$, $60.8 \pm 1.1$ and $60.3 \pm 0.8$~mJy on 2014 Feb. 28, Mar. 10, Apr. 2, and Apr. 3, respectively (horizontal dot-dashed line of Fig.~\ref{ir_lc}).
The beginning and end of each flare is set by the $1\sigma$ limit on the flux density whose maximum amplitude is larger than $3\sigma$.
We only considered flux-density increases that lasted longer than 25~s, in order to discard any calibration glitchs.
All observed NIR flares are labeled with Roman numerals.

The $\sim$10~hour visit on 2014 Mar. 10 detected two NIR flares.
The first one (labeled \textrm{I}) peaks at 8.2$\sigma$ and has an X-ray counterpart.
It lasts from 16:29:51 to 18:52:36 (1$\sigma$ limit).
We can see in Table~\ref{table:2} that it begins and ends $\sim14$~min before the X-ray flare.
As for the X-ray flare, its shape is not a Gaussian, as it has a dip during the third HST orbit.
Two interpretations can be made to explain this shape. First, this flare could be a single flare and the variation from Gaussian shape during the third orbit can be seen as substructures, as is the case for some NIR flares \citep{dodds-eden09}.
The second interpretation is that this NIR flare is in fact two distinct flares with a return below the $1\sigma$ limit between $\sim$18:30 and $\sim$18:39.
The time delay between the two maxima in this scenario would be about 90~min.
From 21:32:33 to 22:02:58 on 2014 Mar. 10, we can see that there is a second NIR flare (labeled \textrm{II}), which has no X-ray counterpart.
Its maximum is about 3.4$\sigma$.

On 2014 Apr. 2 we caught the end of a NIR flare (labeled \textrm{III}), lasting until 17:31:15.
Its amplitude is larger than 8.8$\sigma$, since its maximum occured during the Earth occultation of \sgra.
Its beginning could correspond with the small increase in flux density seen just before the start of the Earth occultation of \sgra, which would lead to an upper limit on its duration of 3360~s.
The duration of this NIR flare~\textrm{III} and its possible relation with X-ray flare 2 will be discussed in Sect.~\ref{x_to_nir}.

The amplitudes of these flares can be compared to the sample of flux densities from \sgra\  observed in the $K_\mathrm{s}$-band with NACO at ESO's VLT and reported by \citet{witzel12}.
They constructed a histogram of all flux densities from the light curves, without distinction between the quiescent and flaring periods.
This observed distribution of the flux density has a relative maximum at 3.57~mJy.
Below this amplitude the distribution decreases, because of the detection limit of NACO.
Above 3.57~mJy, the distribution is highly asymmetric, with a rapid decay of the frequency density followed by a long tail to 32~mJy.
Figure~\ref{witzel} compares the amplitude of the flares detected with HST during this campaign with the relative frequency density given in \citet[][Fig.~3]{witzel12}.
The normalized distribution of the NIR flux densities observed with NACO (top panel of Fig.~\ref{witzel}) is corrected for the background emission of 0.6~mJy \citep{witzel12}.
The amplitude of the flares detected with HST are extrapolated to the $K_\mathrm{s}$-band using the $H-L$ spectral index of \sgra\  computed in \citet{witzel13}, which is $\alpha=-0.62$.

The detection threshold of HST, which we define as the 3$\sigma$ limit (dot-dashed line in Fig.~\ref{witzel}), corresponds to 8\% of the amplitude sample observed with NACO (bottom panel of Fig.~\ref{witzel}).
The amplitude of NIR flare~\textrm{II} is about 7 times smaller than the amplitude of the brightest flare observed with NACO, whereas the amplitude of flare~\textrm{I} is only 3 times smaller than the amplitude of this event.
We can only measure a lower limit on the amplitude of NIR flare~\textrm{III}, since its maximum occured during the Earth occultation.
This lower limit is nearly as large as those of flare~\textrm{I}.

\begin{figure}
\centering
\includegraphics[width=4.5cm, angle=90]{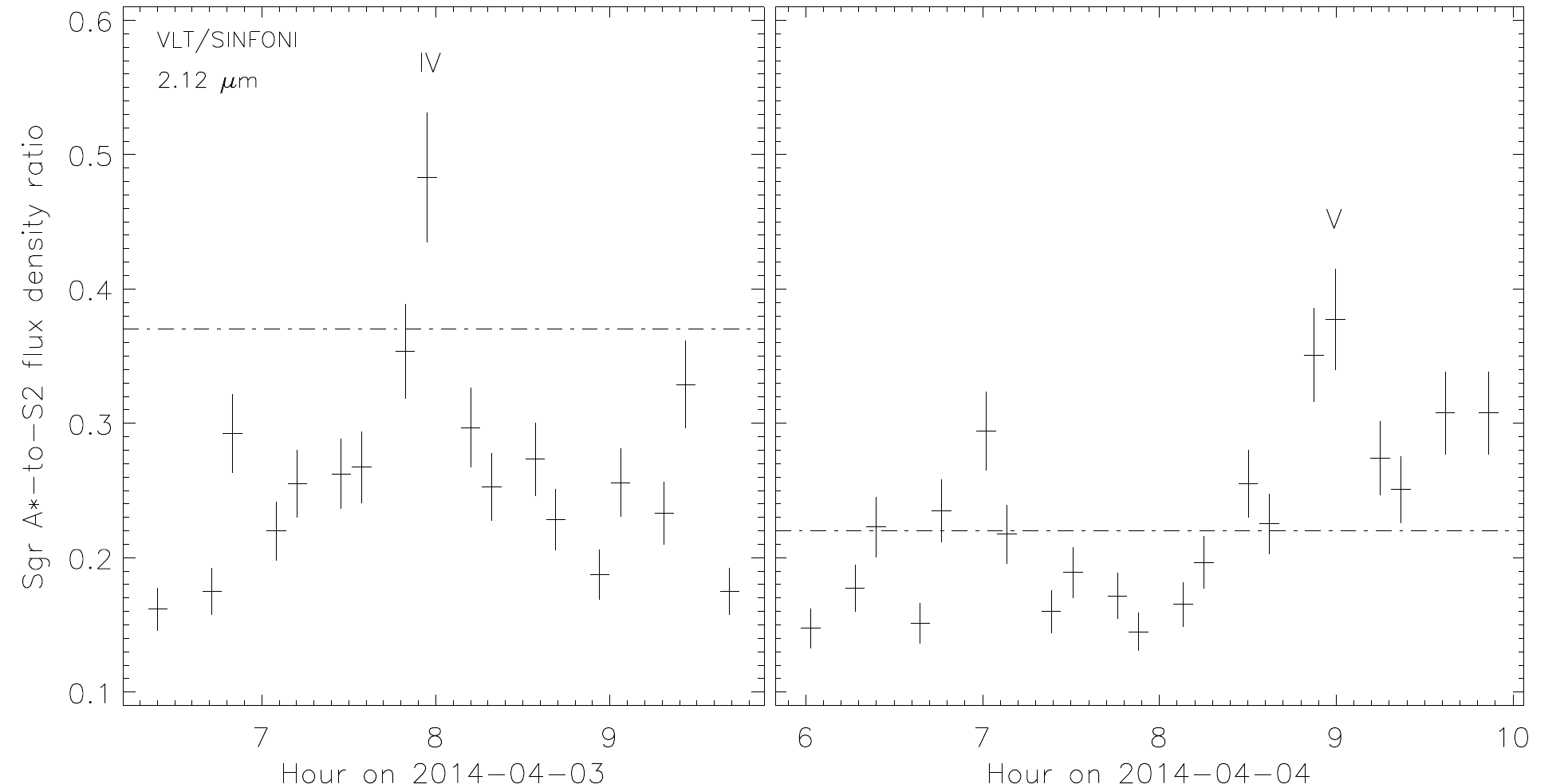}
\caption{Light curves of \sgra\  obtained with SINFONI at ESO's VLT during 2014 Apr. 3 and 4. 
The dash-dotted lines represent the $3\sigma$ detection level of \sgra.
The horizontal segments indicate the exposure length of 400~s.
The NIR flares are labeled with Roman numerals.}
\label{lc_vlt}
\end{figure}

\subsection{VLT data}
\label{vlt}
Fig.~\ref{lc_vlt} shows the ratio between \sgra\  and S2 flux densities for the observations where a NIR flare was detected.
Making a very conservative estimation, the $3\sigma$ detection levels of \sgra\  in the 2014 Apr. 3 and 4 data yield flux density ratios of $F(\mathrm{Sgr\ A^*})/F(\mathrm{S2}) \approx 0.37$ and $0.22$, respectively (dash-dotted lines of Fig.~\ref{lc_vlt}).
A flare (labeled \textrm{IV}) is observed on 2014 Apr. 3 with a peak amplitude of $\sim3.9\sigma$.
We clearly see its rise and decay phase below the $3\sigma$ detection level.
On 2014 Apr. 4, a smaller flare (labeled \textrm{V}) is seen around 9:00 UT with a peak amplitude of $\sim 5.1\sigma$.

Using Eq.~2 of \citet{witzel12}, with $K_\mathrm{s}(\mathrm{S2})=14.13\pm0.01$ and $A(K_\mathrm{s})=2.46\pm0.03$ \citep{schodel10}, we have $F(\mathrm{S2})=14.32\pm0.26$~mJy.
The amplitude of the two NIR flares detected with SINFONI are thus $6.92\pm0.13$ and $5.30\pm0.09$~mJy for 2014 Apr. 3 and 4, respectively.
We consider that all the SINFONI light curve variations above our $3\sigma$ detection limit can be attributed to \sgra\  activity.
We can therefore compare these flux densities with the sample of flux densities observed with NACO after the background subtraction of 0.6~mJy (Fig.~\ref{witzel}).
The 2014 Apr. 3 and 4 flares are within 4\% of the largest amplitude, and are 5 and 6 times smaller than the brightest amplitudes observed with NACO, respectively.
The $3\sigma$ detection level corresponds to $3.15\pm0.06$~mJy, which is comparable to the 11\% of the largest flux density observed with NACO (Fig.~\ref{witzel}).

\subsection{CARMA data}
\label{carma}
\begin{figure}
\centering
\includegraphics*[width=4.5cm, angle=90]{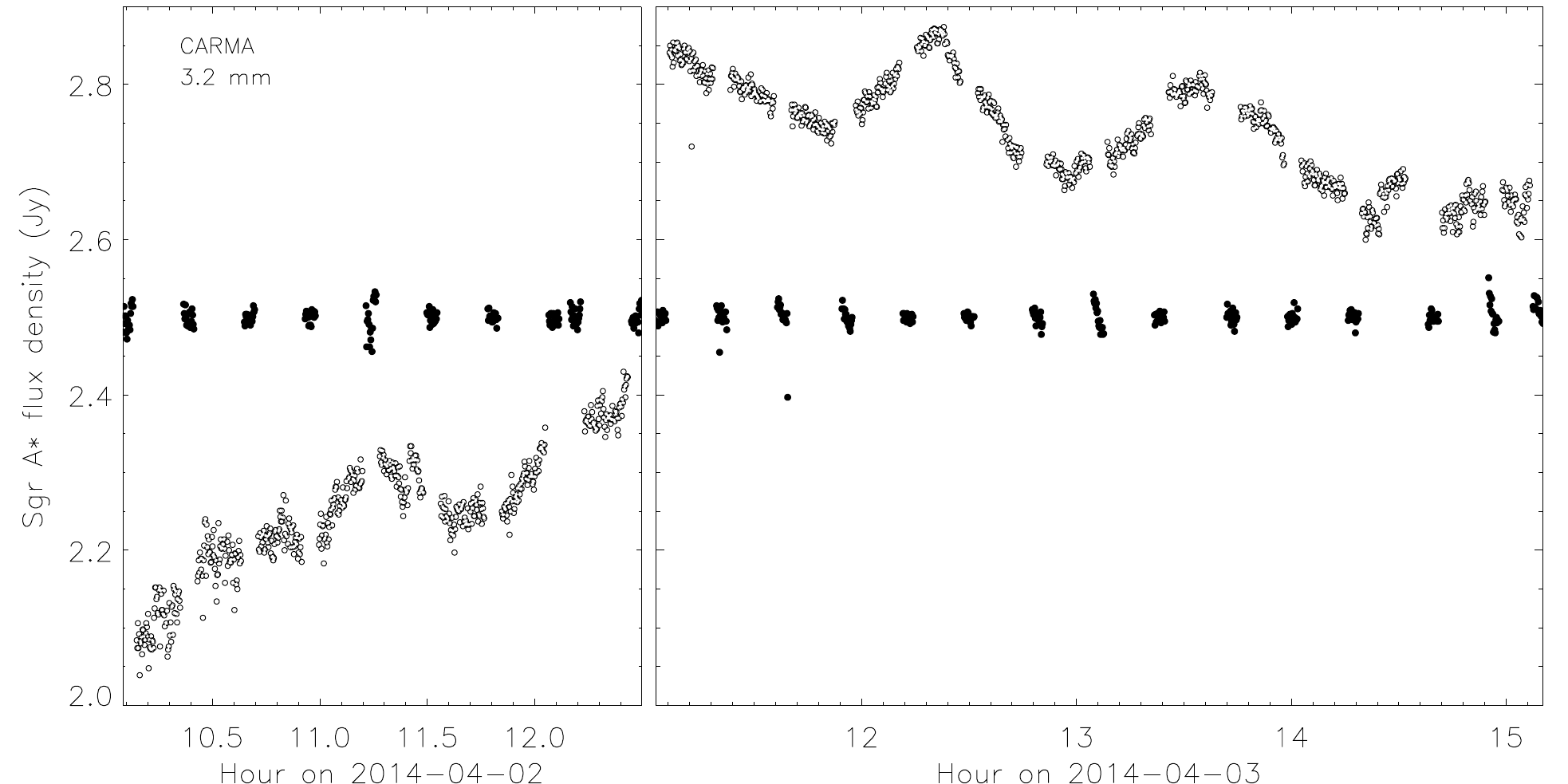}
\caption{CARMA light curves at 3.2~mm (95~GHz) of \sgra\  (white circle) and 1733-130 (black circle) in April 2014. 
The dash-dotted line represents the mean flux density.}
\label{lc_carma}
\end{figure}

\begin{figure*}
\centering
\includegraphics[width=5.15cm,angle=90]{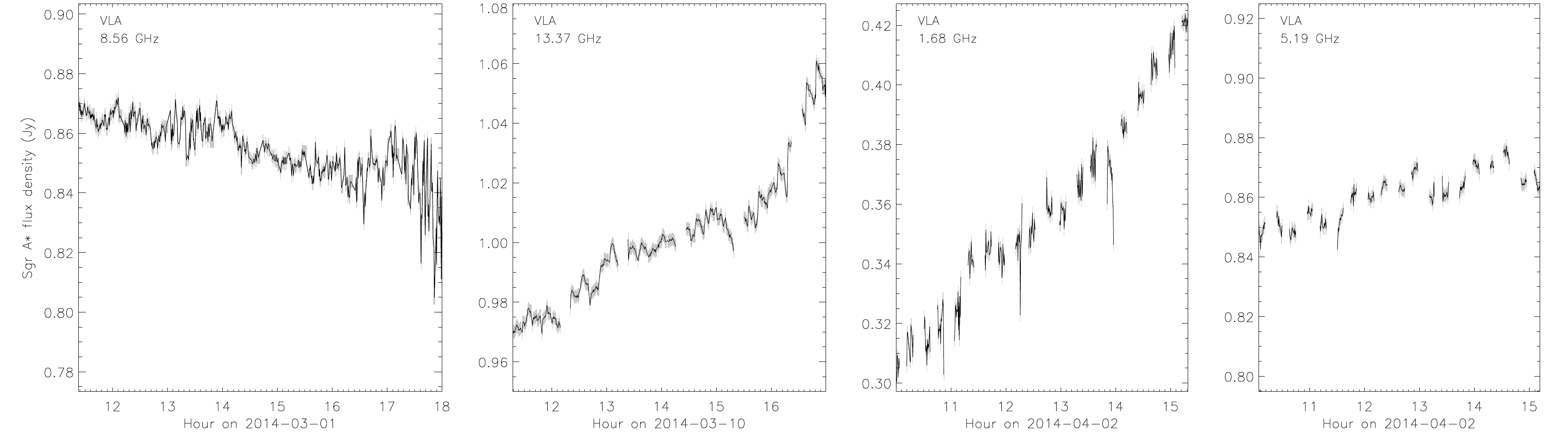}
\caption{VLA light curves of \sgra\  obtained on 2014 Mar. 1 (8.56~GHz), Mar. 10 ($\radioFreq{}$) and Apr. 2 (1.68 and 5.19~GHz).
The y-axis covers the same range of flux density for all observation and is centered on the mean of the minimum and maximum flux density in each panel. }
\label{radio_lc}
\end{figure*}
The flux densities at 95~GHz (3.2~mm) of \sgra\ and 1733-130 shown in Fig.~\ref{lc_carma} are computed for each 10~s integration on 2014 Apr. 2 and 3.
On 2014 Apr. 2 the flux density of \sgra\  increases slowly.
A bump is seen at 11.3~h, but it could not be associated with the observed NIR or X-ray flares, since the CARMA observation occurred before the flares observed with HST and XMM-Newton.

On 2014 Apr. 3 the flux density decreases slowly, with two bumps occuring at 12.4 and 13.6~h.
The maximum of the NIR flare~\textrm{IV} observed with VLT occurred at 7.9~h on the same date.
One of these episodes of radio flux density variation could be the delayed emission from this NIR flare, which would indicate a time delay of 4.4 or 5.6~h for the first and second bumps, respectively.
The delays previously measured between the X-rays and the 850~$\mu$m light curves range between 1.3 and 2.7~h \citep[e.g.,][]{yusef-zadeh06,yusef-zadeh08,marrone08}.
Assuming the expanding plasmon model, the delay between the NIR and and the longer wavelength (3.2mm) emission must be larger than these values, leading to a time delay consistent with those measured for these two bumps.
One time-delay measurement was made between the X-rays and the 7~mm light curve, leading to a delay of about 5.3~h \citep{yusef-zadeh09}.
This measure seems to reject the second bump as being the delayed sub-mm emission from the VLT flare, since the delay is too long.
The first bump, therefore, could be the delayed millimeter emission of the NIR flare~\textrm{IV}.
The second bump could then be the delayed millimeter emission of a NIR flare whose peak is lower than the $3\sigma$ detection level of VLT or which occurred after the end of the VLT observation and during Earth occultation for HST.

The flux density of \sgra\  during these observations increases with frequency as $S_{\nu} \propto \nu^{0.2}$. 
For comparison, previous observations of \sgra\  between 43.3, 95.0, and 151~GHz (corresponding to 7.0, 3.2, and 2.0~mm; Table 2 of \citealt{falcke98}) give a similar spectra index of $0.58\pm0.23$.

\subsection{VLA data}
\label{vla}
We obtained light curves of \sgra\  from all three days of VLA observations, selecting (for the purpose of simplification) only one intermediate frequency channel with 30~s of averaging (analysis of the full radio dataset will be given elsewhere). 
In all observations we selected visibilities greater than 100~k$\lambda$ in order to minimize contamination from extended thermal emission from Sgr~A West. 
The radio light curves for the frequencies obtained with the VLA in configuration~A on 2014 Mar.$-$Apr. are shown in Fig.~\ref{radio_lc}.

We interleaved the CARMA and VLA $L$- and $C$-band observations from 2014 Apr. 2 in order to search for a time delay between the 1.68~GHz and 5.14~GHz, and the 1.68~GHz and 95~GHz light curves, using the z-transformed discrete correlation function (ZDCF; \citealt{alexander97}).
The cross-correlation graphs show no significant maximum of the likelihood function, implying that we can not derive any time delay between these light curves.

\begin{figure}
\centering
\includegraphics[height=7.5cm,angle=90]{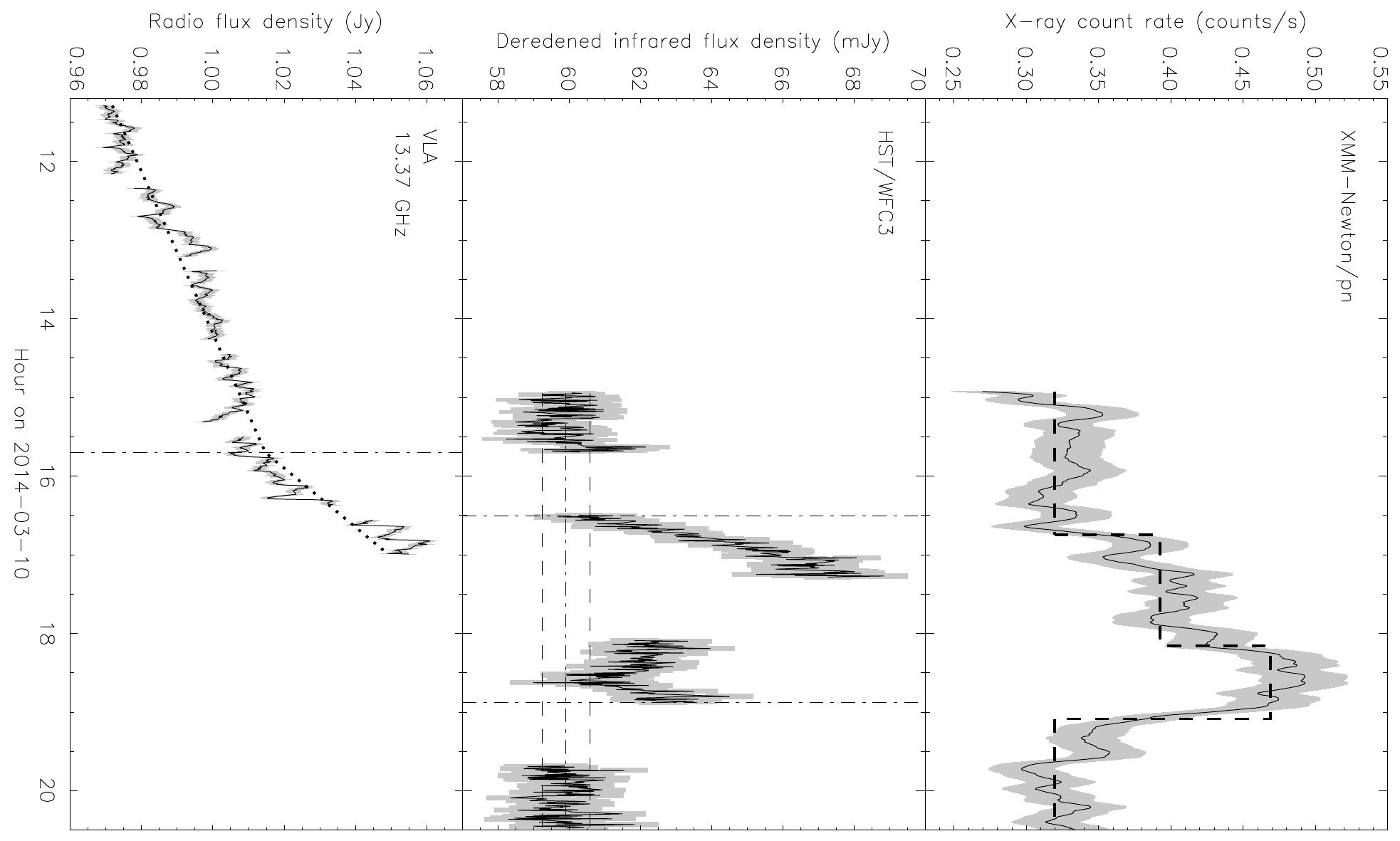}
\caption{Simultaneous X-ray, NIR and radio observations of flare \textrm{I}/1 from \sgra\  on 2014 Mar. 10.
\textit{Top panel:} The EPIC/pn smoothed light curve computed with a window width of 500~s and its error in gray.
The dashed lines are the Bayesian blocks.
\textit{Middle panel:} The deredened HST light curve and its error in gray.
The vertical dot-dashed lines are the beginning and the end of the flares.
\textit{Bottom panel:} The VLA light curve at 13.37~GHz.
The vertical dot-dashed line is the time of the change of slope.
The dashed broken line is the fit.}
\label{comp_03_10}
\end{figure}

The light curves on 2014 Mar. 1 and Apr. 2 display a steady decrease and increase of flux density.
The light curve on 2014 Mar. 10 shows an obvious break in its rising flux density around 16~h, with a clear increase of the rising slope.
To better constrain the time of this slope change, we fit the VLA light curve with a broken line.
The break is located at $15.7\pm0.2$~h with a slope increasing from $9.7\pm0.1$ to $27\pm1\ \mathrm{mJy\ h^{-1}}$ ($\chi^2_{red}=2828$ with 508~d.o.f.), which is significant. 
We therefore tentatively attribute it to the onset of a radio flare, since we have only partial temporal coverage of this radio event.
For comparison purposes, the light curves of the 2014 Mar. 10 flare observed with VLA, WFC3, and XMM-Newton are shown in Fig~\ref{comp_03_10}. 

The radio flare observed at 13.37~GHz (2.2~cm) could be the delayed emission from a NIR/X-ray flare that occurred either at the beginning of the observation with an amplitude lower than the detection limits of WFC3 and XMM-Newton, or before the start of our HST and XMM-Newton observations.
The latter would imply a delay larger than 2.2h.
As explained previously in Sect.~\ref{carma}, the largest time delay that has been measured between X-ray and sub-mm flares is 5.3~h \citep{yusef-zadeh09}.
Considering the expanding plasmon model \citep{yusef-zadeh06c}, the delay between the X-ray and centimeter light curves must be larger than 5.3~h, and therefore the possibility of a non-detected NIR/X-ray flare is likely excluded.

\section{Determination of the X-ray emission related to the NIR flares}
\label{x_to_nir}
In the following subsections we determine the X-ray emission related to each NIR flare observed with HST or VLT, with which we associate either one of the X-ray flares detected with XMM-Newton or an upper limit on the amplitude of a non-detected X-ray flare.

\subsection{The NIR flare \textrm{I} on 2014 Mar. 10}
\label{march_10}
To compare the NIR and X-ray light curves of the 2014 Mar. 10 flare, we express the NIR and X-ray flux in the same units.
To convert the X-ray count rate to flux, we use the unabsorbed-flux-to-count-rate ratio derived in Section~\ref{spectrum_xmm}.

The NIR flux of \sgra\  is obtained from the flux density $S_{\nu}$ by integrating over the F153M filter, using the filter 
profile $T$ (Spanish Virtual Observatory).
To be consistent with the HST photometric calibration (Vega system), we assume a Rayleigh-Jeans regime ($S_{\nu} \propto \nu^{2}$):
\begin{equation}
\frac{F_\mathrm{IR}}{\mathrm{erg\,s^{-1}\,cm^{-2}}}=\int{T\,S_{\nu}\,\left(\frac{\nu}{\nu_\mathrm{eff}}\right)^2 \, d\nu}\, ,
\end{equation} 
with $\nu_{\mathrm{eff}}$ the effective frequency given by the Spanish Virtual Observatory. 

\begin{figure}
\centering
\includegraphics[width=8.5cm,angle=90]{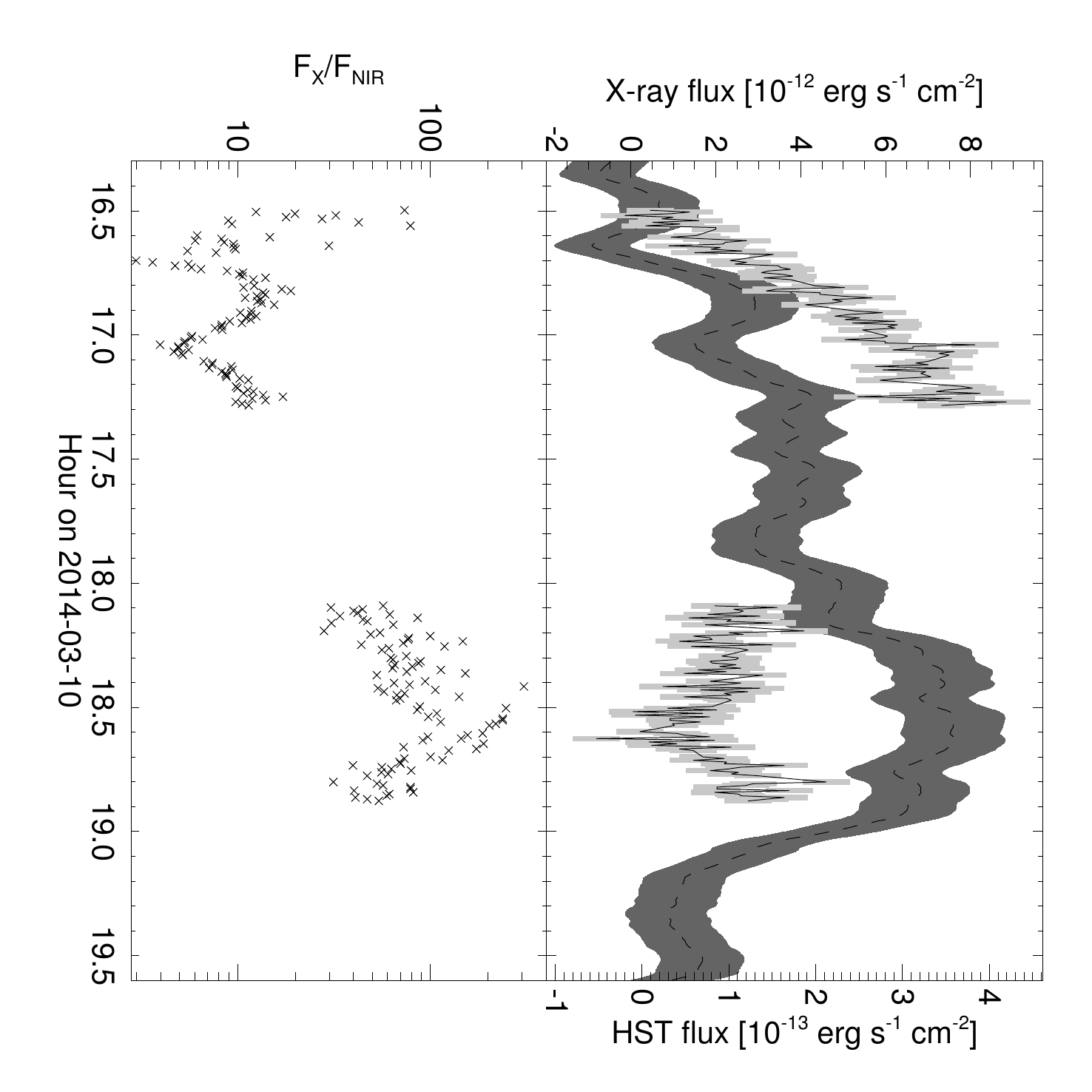}
\caption{Evolution of the ratio between NIR and X-rays during flare \textrm{I}/1 on 2014 Mar. 10.
\textit{Top panel:} The dashed line surrounded by the dark gray error bars corresponds to the smoothed light curve of the X-ray flare and its flux can be seen on the left y-axis.
The solid line and the light gray error bars is the NIR light curve whose flux is read on the right y-axis.
\textit{Bottom panel:} The flux of the X-ray light curve divided by the NIR one.}
\label{ir_x_ratio}
\end{figure}

The ratio between the X-ray and NIR flux during the flare is shown in Fig.~\ref{ir_x_ratio} (the error bars are on the order of the symbol size).
The NIR flux is always lower than the X-ray flux, but during the third orbit of the HST visit the X-ray contribution increased by a factor
of 10 compared to the NIR.
We can test two interpretations: a single flare with non-simultaneous X-ray and NIR peaks, or two distinct flares with simultaneous NIR and X-ray peaks. 

\subsubsection{A single flare with non-simultaneous peaks in NIR and X-rays}
Considering that the NIR flare is produced by synchrotron emission, there are three radiative processes that can explain the X-ray flare
production: synchrotron (SYN; \citealt{dodds-eden09,barriere14}), inverse Compton (IC; \citealt{yusef-zadeh06,wardle11,yusef-zadeh12}), and
synchrotron self-Compton (SSC; \citealt{eckart08}) emission.
In this section, we discuss whether each process can explain the entire observed NIR/X-ray light curve on 2014 Mar. 10.

\paragraph{The synchrotron$-$synchrotron process (SYN-SYN)\\}
For synchrotron emission of NIR and X-ray photons by accelerated electrons in the flaring region, the electron acceleration has to be high enough to directly emit X-ray photons.
It is difficult, however, to explain how to reach the required Lorentz factor of $\gamma=10^6$ \citep{marrone08,yusef-zadeh12,eckart12}.
Moreover, the synchrotron cooling time scale $\tau_\mathrm{sync}=8 \times \left( B/30~\mathrm{G}\right)^{-3/2} \times \left( \nu/10^{14}~\mathrm{Hz}\right)^{-1/2}$~min \citep{dodds-eden09} is very short for X-ray photons ($\approx 1$~s for $B=100$~G and $\nu=4\times 10^{17}$~Hz).
Thus, we must have continuous injection of accelerated electrons to maintain the X-ray flare during the decay phase, which lasts $\sim30$~min.
If the NIR and X-ray flares are created by the same population of electrons, whose energy distribution is described by a powerlaw as $N(E)=K\,E^{-p}$, the difference between the NIR and X-ray flux can be explained if the synchrotron spectrum has a cooling break frequency between the NIR and X-rays \citep{dodds-eden09}.
In this scenario, the X-ray spectrum has a spectral index of $\alpha=p/2$, whereas the NIR spectral index is $\alpha=(p-1)/2$ (with $S_{\nu} \propto \nu^{-\alpha}$; \citealt{dodds-eden09}). 
Knowing that the X-ray photons are produced by the electrons from the tail of the power law distribution, during the first part of the flare there are many more electrons that create NIR photons than those creating X-ray photons.
Then, the acceleration mechanism has to become more efficient, accelerating more electrons to the tail (and thus increasing $p$) of the distribution and thus changing the ratio between the NIR and the X-ray flux.
Hence the production of X-ray photons increases, which explains the second part of the flare.

\paragraph{The synchrotron$-$synchrotron self-Compton process (SYN-SSC)\\}
During synchrotron self-Compton emission, X-ray photons are produced by the scattering of the synchrotron radiation from radio to NIR on their own electron population.
If we compare the fluxes produced by the synchrotron and SSC emissions, the variation of the X-ray/NIR ratio constrains the size evolution of the flaring source.
Let us consider a spherical source of radius $R$ with a power law energy distribution of relativistic electrons.
Following \citet{vanderlaan66}, the radiative transfer for the synchrotron radiation can be computed as
\begin{equation}
S_\mathrm{SYN}=\int_{0}^{R}{\frac{\epsilon_\nu}{\kappa_\nu}\,\left(1-e^{-\tau_\nu(r)}\right)\,2\pi r\,dr}\, ,
\end{equation}
with $\kappa_\nu \propto B^{(p+2)/2}\,\nu^{-(p+4)/2}$ the absorption coefficient, $\epsilon_\nu \propto B^{(p+1)/2}\nu^{-(p-1)/2}$ the emission coefficient, $B$ the magnetic field \citep{lang99} and $\tau_\nu(r)$ the optical depth, which can be computed at each distance $r$ from the sphere center as:
\begin{equation}
 \tau_\nu(r)=\int_{0}^{2\sqrt{R^2-r^2}}{\kappa_\nu dl}\, .
\end{equation}
Assuming that we are in the optically thin regime (i.e., $\tau_\nu(r) << 1$), we utilize formula 3 of \citet{marrone08}: $S_\mathrm{SYN} \propto B^{(p+1)/2}\,\nu^{-(p-1)/2}\,R^3$.
For synchrotron radiation, we have $B \propto R^4\,\nu_\mathrm{m}^{5}\,S_\mathrm{m}^{-2}$ with $S_\mathrm{m}$ the maximum flux density of the spectral energy distribution occurring at frequency $\nu_\mathrm{m}$ \citep{marscher83}.
Finally, the synchrotron radiation can be expressed using $p=2\alpha+1$ as
\begin{equation}
 S_\mathrm{SYN} \propto R^{4\alpha+7}\,\nu_\mathrm{m}^{5(\alpha+1)}\,S_\mathrm{m}^{-2(\alpha+1)}\,\nu^{-\alpha}\, .
\end{equation}
The SSC radiation of X-ray photons is (formula 4 of \citealt{marscher83}):
\begin{equation}
 S_\mathrm{SSC} \propto R^{-2(2\alpha+3)}\,\nu_\mathrm{m}^{-(3\alpha+5)}\,S_\mathrm{m}^{2(\alpha+2)}\,\ln \left(\frac{\nu_2}{\nu_\mathrm{m}}\right)\,\nu^{-\alpha}\, .
\end{equation}
The natural logarithm in this equation could be approximated by $c_1\,(\nu_2/\nu_\mathrm{m})^{c_2}$ with $c_1=1.8$ and $c_2=0.201$ \citep{eckart12}.
The synchrotron-to-SSC flux ratio is
\begin{equation}
 \frac{S_\mathrm{SSC}}{S_\mathrm{SYN}} \propto R^{-(8\alpha+13)}\,\nu_\mathrm{m}^{-(8\alpha+10+c_2)}\,S_\mathrm{m}^{4\alpha+6}\, .
\end{equation}

We therefore have three parameters that may vary during the flare to explain the increased ratio of X-ray and NIR flux (Fig.~\ref{ir_x_ratio}).
Considering the plasmon model, for which a spherical source of relativistic electrons expands and cools adiabatically, we have \citep{vanderlaan66}: $\nu_\mathrm{m} \propto R^{-(8\alpha+10)/(2\alpha+5)}$ and $S_\mathrm{m} \propto R^{-(14\alpha+10)/(2\alpha+5)}$.
Thus, $S_\mathrm{SSC}/S_\mathrm{SYN} \propto R^{-\beta}$ with $\beta \equiv (8\alpha^2+(30-8\,c_2)\alpha+25-10\,c_2)/(2\alpha+5)$.
We first consider the adiabatic expansion.
For our observation, the ratio between the X-ray and the NIR flux increases during the 2014 Mar. 10 flare, implying that $R^{-\beta}$ must increase as the radius $R$ increases.
This condition is satisfied if the exponent $\beta$ is negative and thus if the $\alpha$ value is lower than $-2.5$ or is between $-2.3$ and $-1.25$, which is inconsistent because $\alpha$ must be positive.
The expansion case is thus likely to be rejected under the hypothesis of an optically thin plasmon that expands adiabatically.
We can also consider the case where the plasmon is compressed during its motion through a bottle-neck configuration of the magnetic field.
We can still use the equations of \citet{vanderlaan66}, since the conservation of the magnetic flux is explicitly taken into account.
The compression case is thus preferred, because it allows positive values of $\alpha$ for $\beta>5.4$.
Thus, for the SYN-SSC process, the plasmon must be adiabatically compressed with at least $S_\mathrm{SSC}/S_\mathrm{SYN} \propto R^{-5.4}$.
Therefore, the observed increase of the X-ray-to-NIR flux ratio by a factor of 10 in 1.2~h implies a decrease of the radius by a factor of about 0.6.
The average compression velocity is estimated as $V_\mathrm{comp}=\Delta R/\Delta t$, leading to $|V_\mathrm{comp}|/c<0.0034\,R/R\mathrm{_s}$ with $R\mathrm{_s}$ the Schwarzschild radius ($R\mathrm{_s}=1.2\times 10^{12}$~cm for \sgra, which corresponds to 0.08~au).
For comparison, the expansion velocities computed with this model in the literature range between 0.0028 and 0.15$c$ \citep{yusef-zadeh06c,eckart08,yusef-zadeh09}, which is of the same order as the compression velocity computed here.
Thus, the model of an adiabatic compression of a plasmon is the likely hypothesis to explain the variation of the ratio between X-ray and NIR flux, in the context of the SYN-SSC process.

\paragraph{The synchrotron$-$inverse Compton process (SYN-IC)\\}
\label{syn_ic}
In the case of inverse Compton emission, X-ray photons are produced by the scattering of either the NIR photons produced by synchrotron
emission from the thermal electron population associated with the accretion flow that produces the sub-millimeter photons, or the
sub-millimeter photons of the accretion flow on the electron population of the external source that produces the NIR photons by synchrotron radiation.
\begin{figure}
\centering
\includegraphics[width=3.6cm,angle=90]{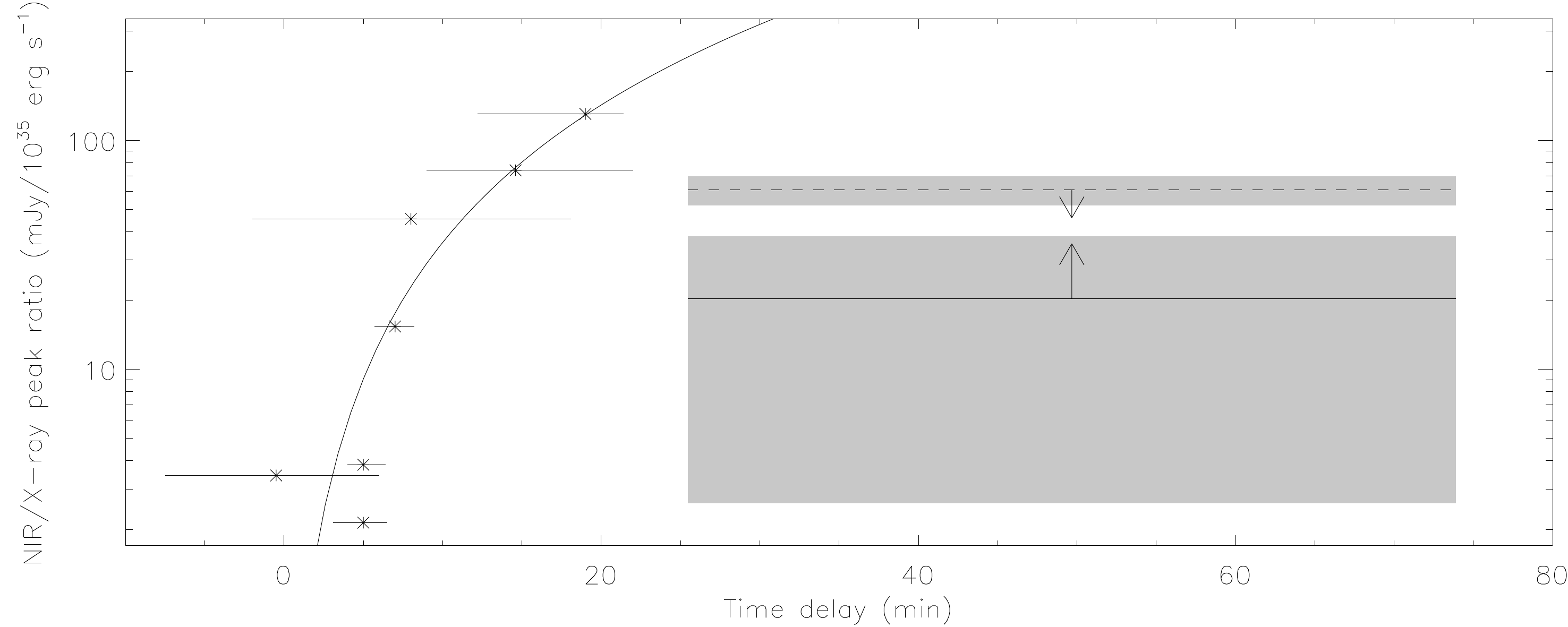}
\caption{NIR/X-ray peak ratio vs. time delay for the synchrotron-inverse Compton process.
The asterisks are the results reported by \citet{yusef-zadeh12}, the solid line is a parabolic fit.
The horizontal solid line and gray box are the lower limit and error bar on the NIR/X-ray peak ratio of the flare~\textrm{I}/1 on 2014 Mar.\ 10 and the corresponding time-delay range.
The dashed line is the ratio between the maximum NIR amplitude reported by \citet{witzel12} and the X-ray peak of flare~1, with the gray box being the corresponding error bar (see text for details).}
\label{time_delay}
\end{figure}

For the former process, the accretion flow is optically thin in the NIR, allowing all of the thermal electron population of the accretion flow to upscatter the NIR photons.
\citet{yusef-zadeh12} estimated the X-ray to NIR time delay for seven NIR/X-ray flares, which is due to upscattering of the NIR photons in the accretion flow.
They identified a trend between increasing time delay and the increase of the NIR/X-ray peak ratio that is consistent with the SYN-IC process.
The X-ray peak of flare~1 is defined as the maximum of the pn smoothed light curve (Table~\ref{table:2}).
We have only a lower limit on the NIR peak of flare~\textrm{I}, which results in an estimated time delay of 25.5$-$73.9~min, because we
have an observational gap in the HST data.
Figure~\ref{time_delay} shows a comparison of the peak ratio lower limit and time delay range of flare~\textrm{I}/1 (horizontal solid line) with those reported in Table~2 of \citet{yusef-zadeh12}.
This peak ratio lower limit is located below the observed trend.
Assuming that the actual NIR peak can not be larger than the maximum observed amplitude \citep[i.e., 32~mJy;][]{witzel12}, the actual peak ratio (dashed line) would be at least four times smaller than the value predicted by the SYN-IC process.
If the actual NIR peak corresponds to this predicted value, this NIR flare would be four times brighter than the brightest flare ever observed and its shape would be completely unusual.
We therefore consider this process to be very unlikely.

For the latter process, the accretion flow is optically thick in sub-millimeter, reducing the number of available sub-millimeter photons produced by the thermal electron population to be upscattered \citep{yusef-zadeh06,wardle11,yusef-zadeh12}.
If the sub-millimeter flux of the accretion flow is constant, the X-ray flare should have the same shape as the NIR flare.
But in flare~\textrm{I}/1 the X-ray flux increases while the NIR flux decreases.
Since the decay of the NIR flux can only be due to the decrease of the number of accelerated electrons, the rise of the X-ray flux would require a simultaneous large increase of the sub-millimeter flux, which appears rather fine tuned.
Therefore, we do not favor the SYN-IC process to explain the variation of the NIR/X-ray flux of flare~\textrm{I}/1.

\subsubsection{Two distinct flares with simultaneous NIR and X-ray peaks} 
\label{two_flares}
\begin{figure}
\centering
\includegraphics[width=5.7cm,angle=90]{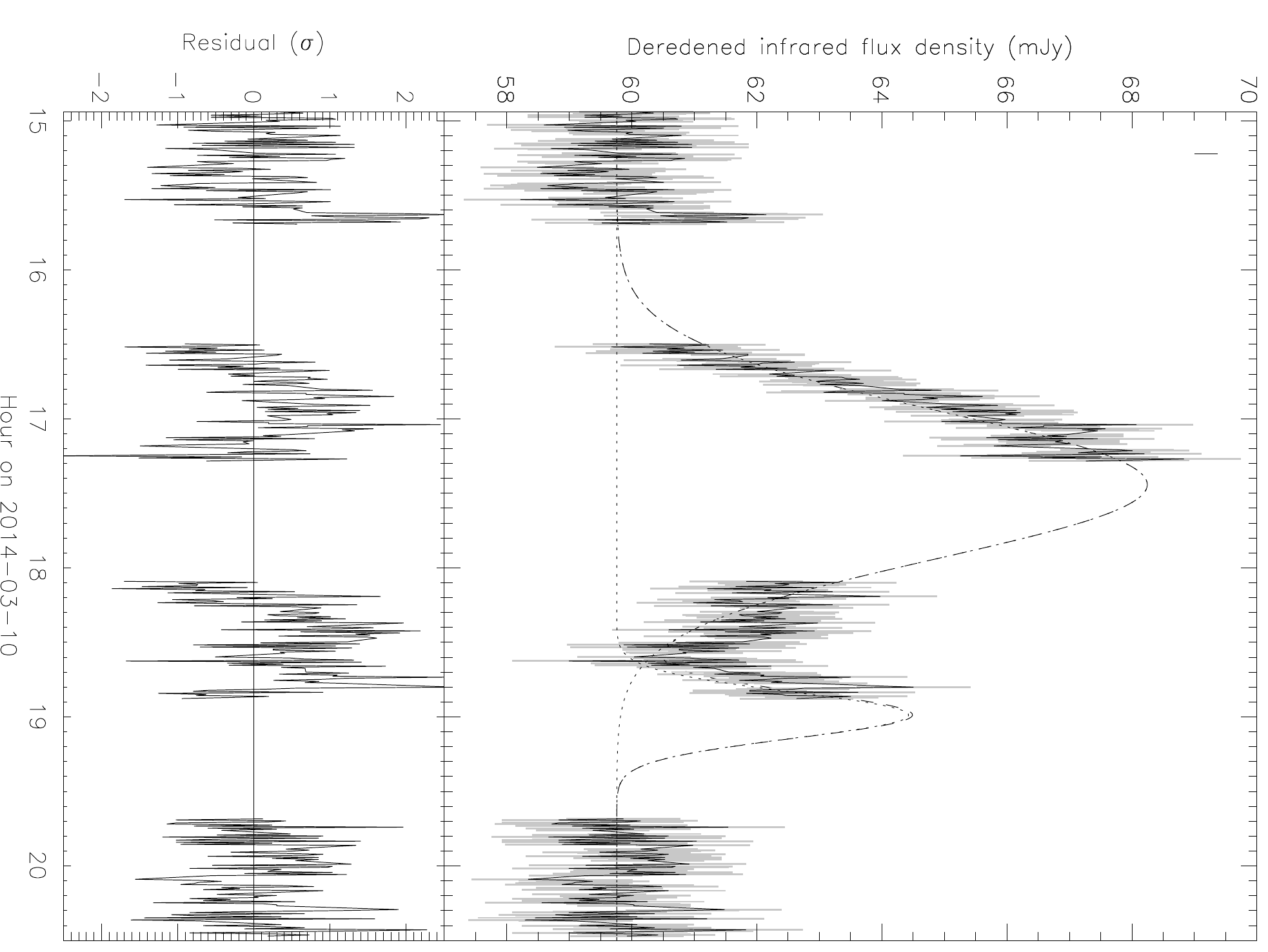}
\includegraphics[width=5.7cm,angle=90]{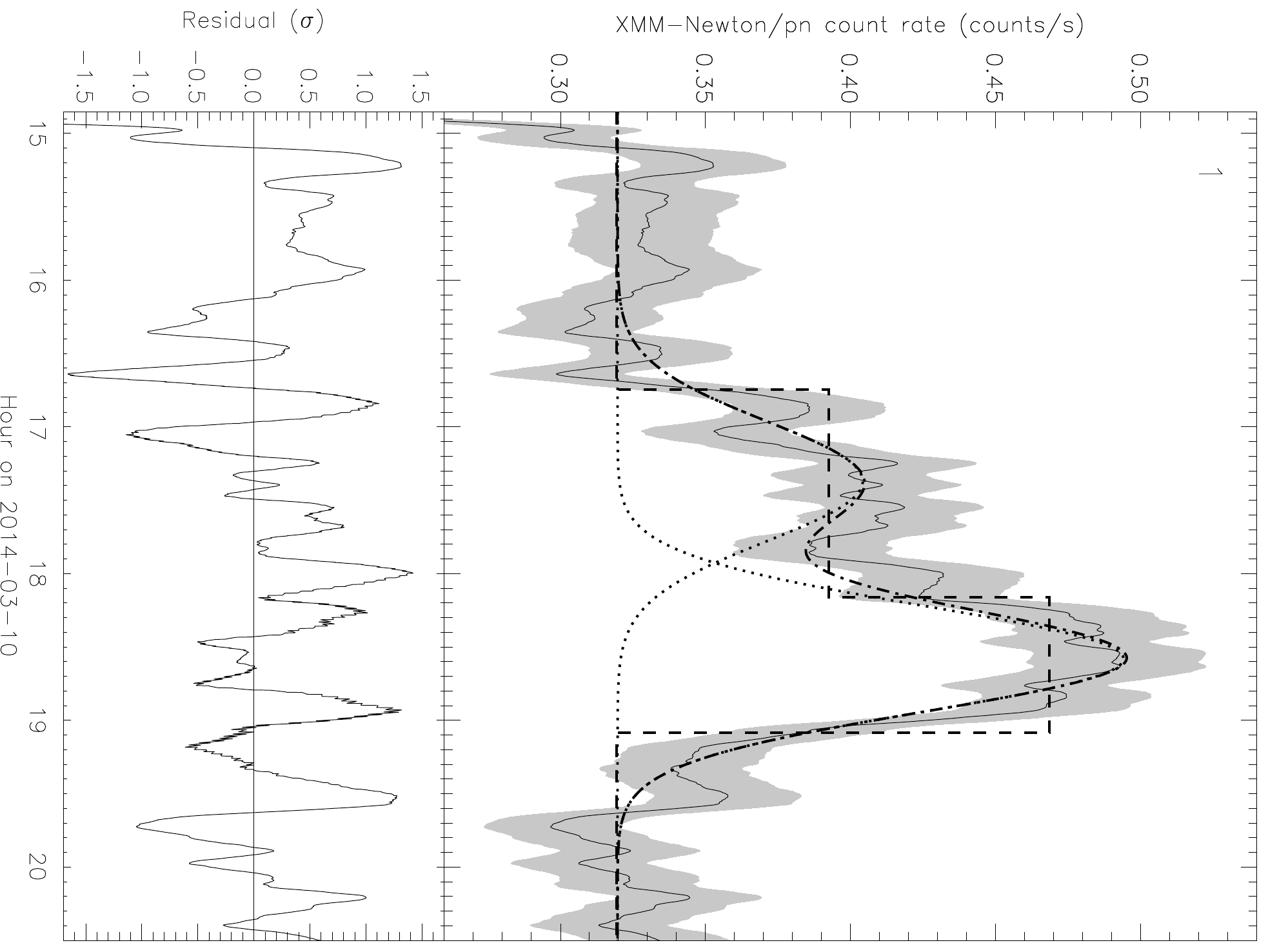}\\
\includegraphics[width=5.7cm,angle=90]{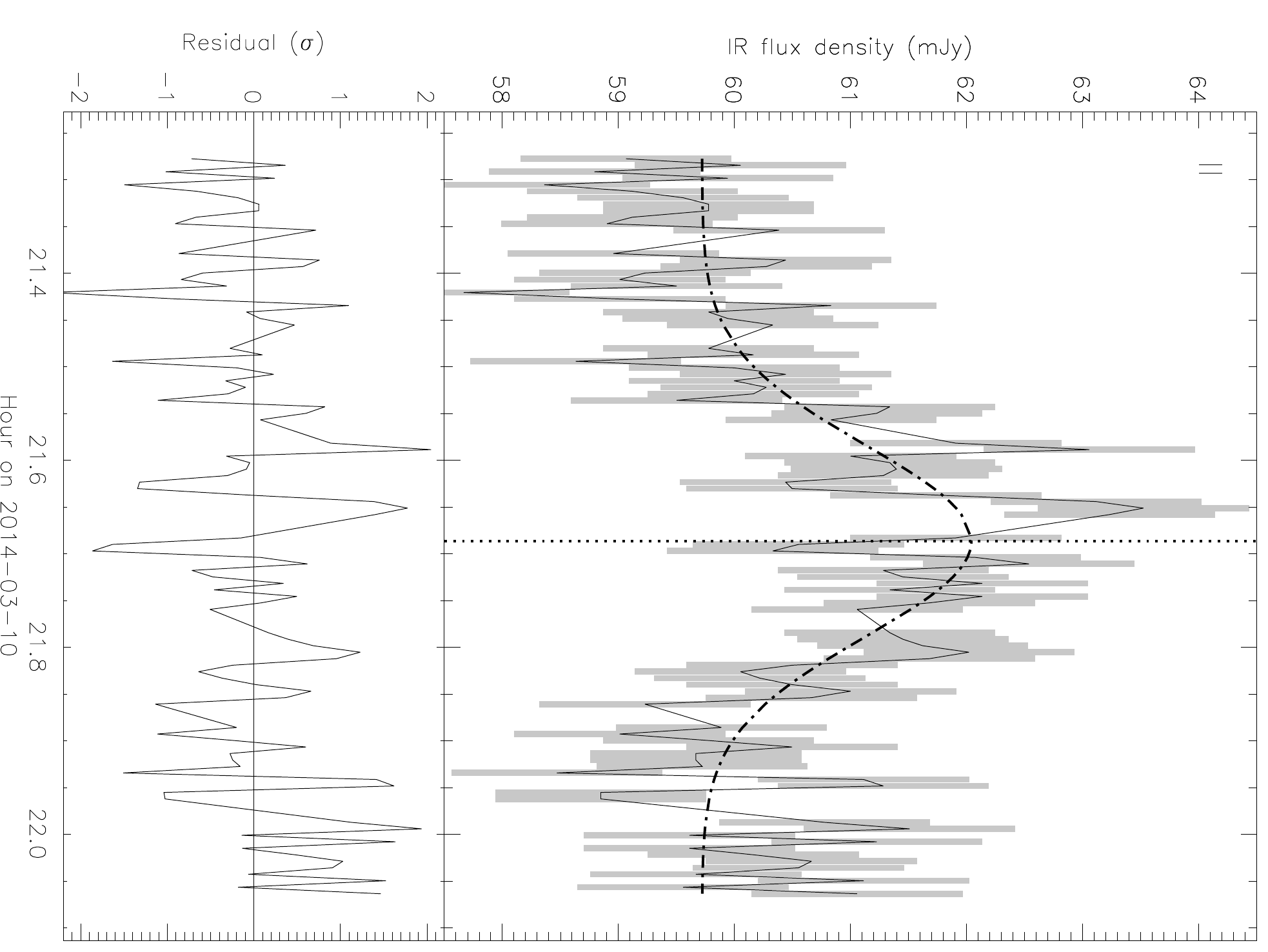}
\includegraphics[width=5.7cm,angle=90]{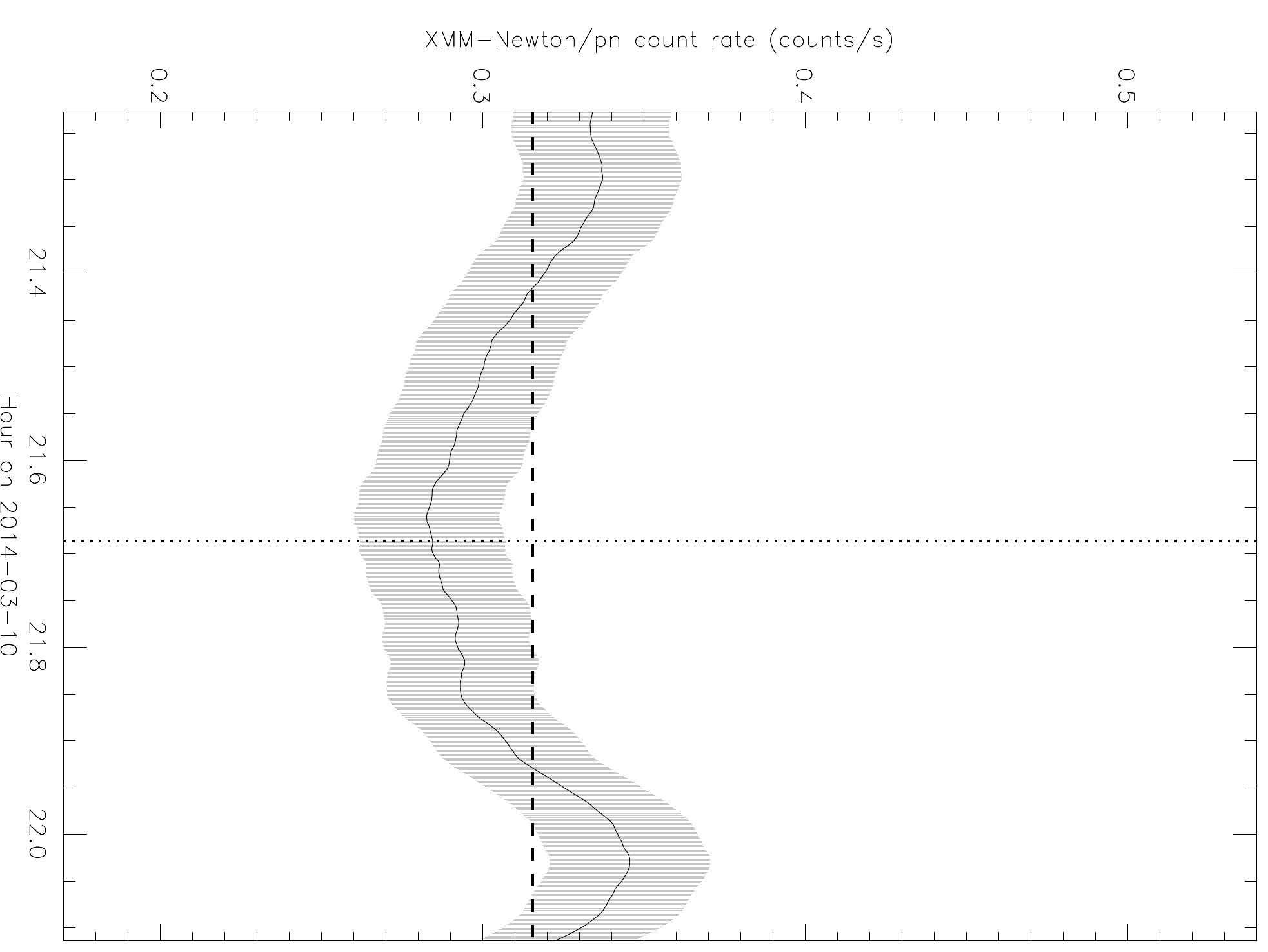}\\
\includegraphics[width=5.7cm,angle=90]{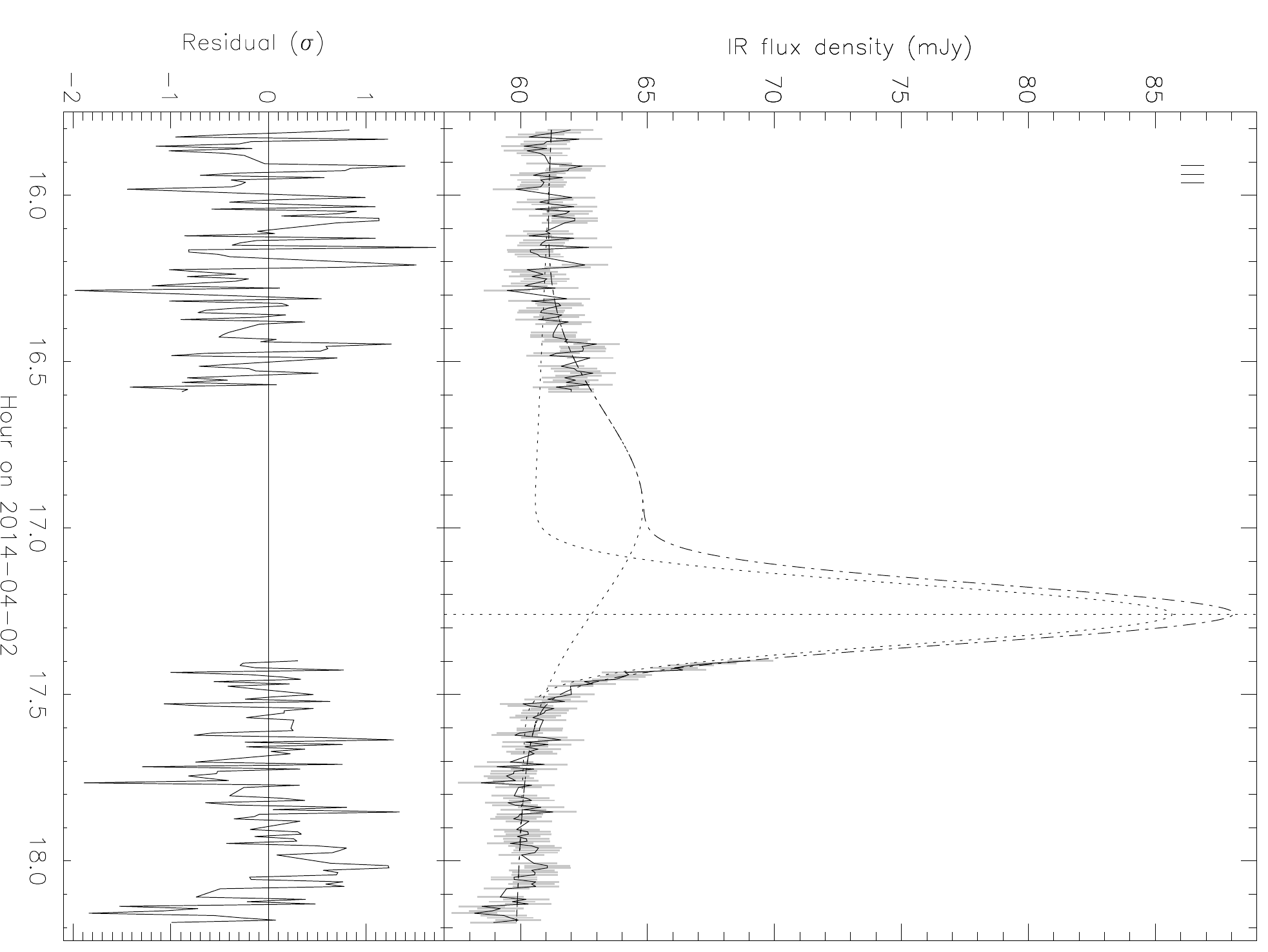}
\includegraphics[width=5.7cm,angle=90]{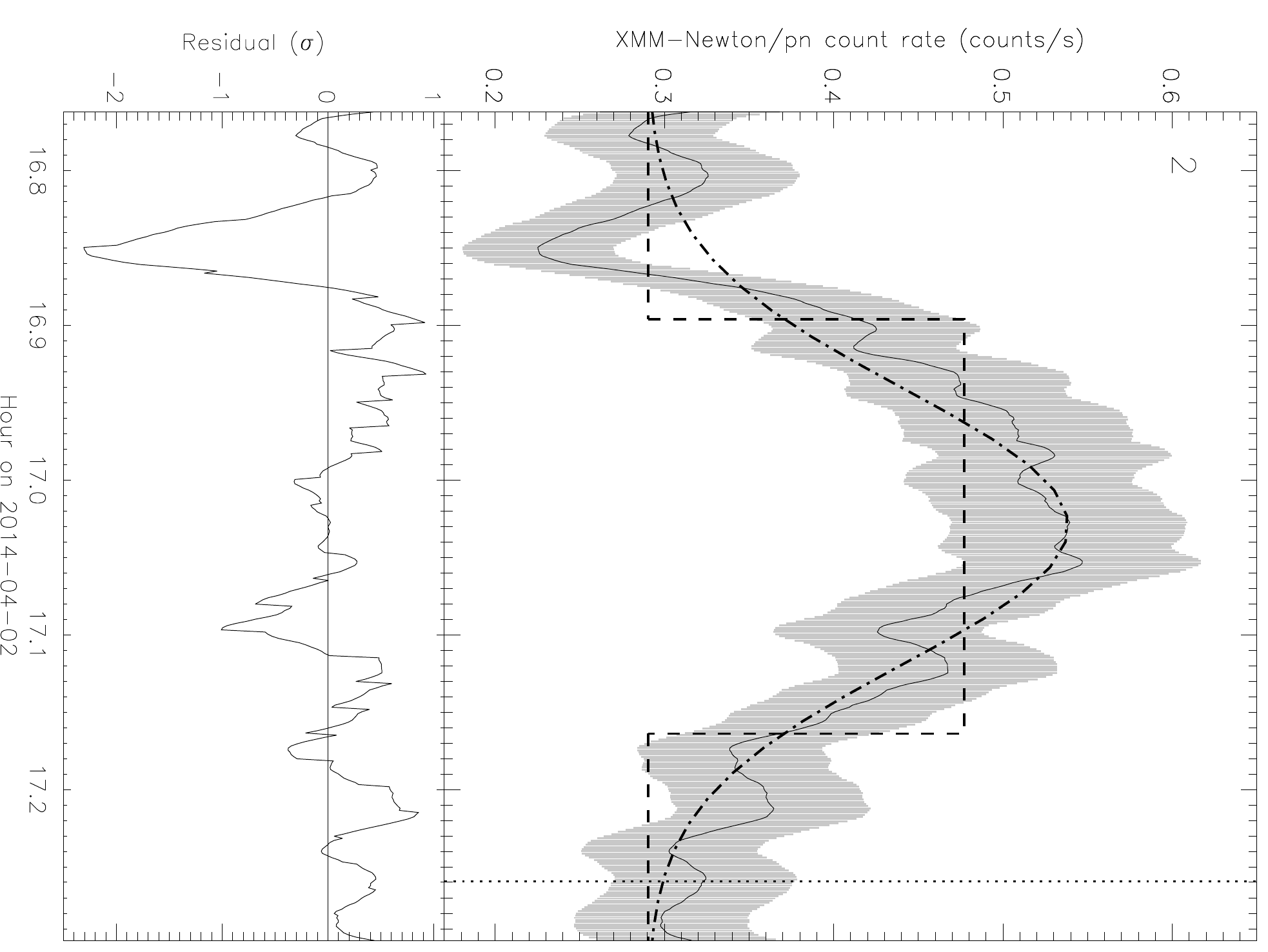}
\caption{Light curve fitting of the HST NIR flares (left panels) and the X-ray (right panels) counterparts.
The solid lines are the observed light curves with the error bars in gray.
The dashed lines in right panels are the Bayesian blocks.
The X-ray light curves are smoothed with a window width of 500~s and 100~s for 2014 Mar. 10 and Apr. 2, respectively.
The dotted lines are the individual Gaussians and the dot-dashed line is the sum of the Gaussians.
The vertical dotted lines are the time of the NIR flare peak when there is no detected X-ray counterpart.
The residuals are in units of $\sigma$. }
\label{fig:gauss_fit}
\end{figure}

The 2014 Mar. 10 flare could be decomposed into two flaring components (called \textrm{I}a/1a and \textrm{I}b/1b).
Each NIR/X-ray flaring component is produced by its own population of accelerated electrons.
We introduce here a general model that will be used in the next subsections to fit the NIR and X-ray light curves.
The model is composed of a linear part (if needed), representing the non-flaring level, plus one or two Gaussian flares:
\begin{equation}
 F(t)=F_0+F_1\,(t-t_0)+\sum_{i=1}^2 A_\mathrm{i}\,e^{-(t-t_\mathrm{i})^2/2\,\sigma_\mathrm{i}^2}\, ,
 \label{eq:gauss}
\end{equation}
with $A_\mathrm{i}$ the amplitude above the non-flaring level and $t_\mathrm{i}$ and $\sigma_\mathrm{i}$ the center and the standard deviation of each Gaussian.
For the X-rays, the non-flaring level is fixed to the Bayesian-block value.
The results of the fit are given in Table~\ref{table:gauss_fit} and the corresponding light curves and residuals are shown in Fig.~\ref{fig:gauss_fit} (top panels).

The time of the first and second peaks of the NIR and X-ray flares are consistent with each other within the $1\sigma$ errors.
Flare~1b appears broader in X-rays than in the NIR, but their widths are consistent with each other within $1.5\sigma$.
The delay time between the two X-ray maxima is about 5000~s, which is longer than the time between two X-ray flares observed during the 2012 \textit{Chandra XVP} campaign (about 4000~s; see Fig.~1 of \citealt{neilsen13}).
This argument, in addition to the change of flux ratio between the two flares, favors the interpretation of two distinct flares.

From the unabsorbed-flux-to-count-rate ratio derived in Section~\ref{spectrum_xmm}, we compute the unabsorbed total energy of these flares using the total number of counts in each Gaussian.
The start and stop times of the flares are defined as the $3\sigma$ distance from the time of the maximum, i.e., 16.0 and 17.6~h for flare~1a, and 17.4 and 19.8~h for flare~1b.
The unabsorbed total energy is $(12.7 \pm 6.7) \times 10^{37}$ and $(21.2 \pm 6.5) \times 10^{37}\ \mathrm{ergs}$ ($1\sigma$ error) for flares 1a and 1b, respectively.
The unabsorbed total energy of flare 1 is thus split nearly equally between its two components.
The peak amplitude of flare~1a is close to the smallest amplitudes of flares observed (Fig.~\ref{fig:neilsen13}).

\begin{table*}
\caption{Gaussian fitting of the NIR and X-ray flares observed during the 2014 campaign.}
\centering
\scalebox{.87}{
\label{table:gauss_fit}
\begin{tabular}{@{}rccccrrrrrr@{}}
\hline
\hline
\multicolumn{3}{c}{Flare} & \multicolumn{3}{c}{Non-flaring level} & \multicolumn{4}{c}{Gaussian flare} & \multicolumn{1}{c}{$\chi^2_\mathrm{red}$ (d.o.f)}\\
\cmidrule(l){1-3}  \cmidrule(l){4-6} \cmidrule(l){7-10}  \cmidrule(l){11-11}
\multicolumn{1}{c}{Date} & \multicolumn{1}{c}{Type} & \multicolumn{1}{c}{\#} & \multicolumn{1}{c}{$F_0$} & \multicolumn{1}{c}{$F_1$} & \multicolumn{1}{c}{$t_0$} & \multicolumn{2}{c}{$A_\mathrm{i}$} & \multicolumn{1}{c}{$t_\mathrm{i}$} & \multicolumn{1}{c}{$\sigma_\mathrm{i}$} & \\
\cmidrule(l){7-8}
\multicolumn{1}{c}{2014} & & & \multicolumn{1}{c}{(mJy)} & \multicolumn{1}{c}{(mJy$\mathrm{\ h^{-1}}$)} & \multicolumn{1}{c}{(h)} & \multicolumn{1}{c}{($\,$\tablefootmark{a})} & \multicolumn{1}{c}{(mJy)$\ $\tablefootmark{b}} & \multicolumn{1}{c}{(h)} & \multicolumn{1}{c}{(h)} & \\
\midrule
Mar. 10 & IR & \textrm{I}a & $59.8\pm0.5$ & \dots\dots\dots\dots\dots & \dots\dots & $8.64\pm0.03$ & $10.58\pm0.03$ & $17.4\pm0.1$ & $0.49\pm0.09$ & $1.52\ (648)$ \\
\rule[0.5ex]{3.1em}{0.55pt} & IR & \textrm{I}b & \rule[0.5ex]{4em}{0.55pt} & \dots\dots\dots\dots\dots & \dots\dots & $4.05\pm0.06$ & $4.97\pm0.06$ & $18.9\pm0.1$ & $0.2\pm0.1$ & \rule[0.5ex]{4em}{0.55pt} \\
\rule[0.5ex]{3.1em}{0.55pt} & X & 1a & [BB] & \dots\dots\dots\dots\dots & \dots\dots & $0.08\pm0.02$ & $(2.8\pm0.8) \times 10^{-4}$ & $17.37\pm0.09$ & $0.3\pm0.1$ & $0.39\ (10796)$ \\
\rule[0.5ex]{3.1em}{0.55pt} & X & 1b & [BB] & \dots\dots\dots\dots\dots & \dots\dots & $0.17\pm0.02$ & $(6.7\pm0.8) \times 10^{-4}$ & $18.58\pm0.07$ & $0.36\pm0.07$ & \rule[0.5ex]{5em}{0.55pt} \\
\midrule
Mar. 10 & IR & \textrm{II} & $59.7\pm0.1$ & \dots\dots\dots\dots\dots & \dots\dots & $2.3\pm0.2$ & $2.8\pm0.2$ & $21.68\pm0.01$ & $0.10\pm0.01$ & $0.67\ (96)$ \\
\rule[0.5ex]{3.1em}{0.55pt} & X & \dots & \dots\dots\dots\dots & \dots\dots\dots\dots\dots & \dots\dots & $<${\bf0.028} & $<1.1 \times 10^{-4}$ & [$21.67$] & \dots\dots\dots\dots & \dots\dots\dots\dots \\
\midrule
Apr. 2 & IR & \textrm{III}a & $61.21\pm0.05$ & $-0.577\pm0.003$ & [$15.8$] & $4.3\pm0.4$ & $4.6\pm0.4$ & $16.94\pm0.01$ & $0.29\pm0.02$ & $0.48\ (192)$ \\
\rule[0.5ex]{2.6em}{0.55pt} & IR & \textrm{III}b & \rule[0.5ex]{5em}{0.55pt} & \rule[0.5ex]{6em}{0.55pt} & \rule[0.5ex]{2.5em}{0.55pt} & $25.3\pm1.4$ & $27.5\pm1.4$ & $17.2\pm0.1$ & $0.09\pm0.03$ & \rule[0.5ex]{4em}{0.55pt} \\
\rule[0.5ex]{2.6em}{0.55pt} & X & 2 & [BB] & \dots\dots\dots\dots\dots & \dots\dots & $0.25\pm0.01$ & $(8.4\pm0.5) \times 10^{-4}$ & $17.03\pm0.04$ & $0.09\pm0.03$ & $1.11\ (1365)$ \\
\rule[0.5ex]{2.6em}{0.55pt} & X & \dots & \dots\dots\dots\dots & \dots\dots\dots\dots\dots & \dots\dots & $<0.030$ & $<1.2 \times 10^{-4}$  & [$17.2$] & \dots\dots\dots\dots & \dots\dots\dots\dots \\
\midrule
Apr. 3 & IR & \textrm{IV} & \dots\dots\dots\dots & \dots\dots\dots\dots\dots & \dots\dots & [$6.9\pm0.1$] & [$6.9\pm0.1$] & [$7.89$] & \dots\dots\dots\dots & \dots\dots\dots\dots \\
\rule[0.5ex]{2.6em}{0.55pt} & X & \dots & \dots\dots\dots\dots & \dots\dots\dots\dots\dots & \dots\dots & $<0.042$ & $<1.7 \times 10^{-4}$ & \rule[0.5ex]{2.5em}{0.55pt} & \dots\dots\dots\dots & \dots\dots\dots\dots \\
\midrule
Apr. 4 &  IR & \textrm{V} & \dots\dots\dots\dots & \dots\dots\dots\dots\dots & \dots\dots & [$5.30\pm0.09$] & [$5.30\pm0.09$] & [$8.82$] & \dots\dots\dots\dots & \dots\dots\dots\dots \\
\rule[0.5ex]{2.6em}{0.55pt} &  X & \dots & \dots\dots\dots\dots & \dots\dots\dots\dots\dots & \dots\dots & $<0.0093$ & $<3.7 \times 10^{-5}$ & \rule[0.5ex]{2.5em}{0.55pt} & \dots\dots\dots\dots & \dots\dots\dots\dots \\
\midrule
\end{tabular}
}
\tablefoot{
[BB] means that the value is fixed to the count rate level of the Bayesian block.
\tablefoottext{a}
{The units are $\mathrm{counts\ s^{-1}}$ for X-rays and mJy for NIR;}
\tablefoottext{b}
{In the $K_\mathrm{s}$-band.}
}
\normalsize
\end{table*}

\subsection{The NIR flare \textrm{II} on 2014 Mar. 10}
\label{march_10_2}
This flare is only detected in the NIR with HST.
We therefore fit the NIR light curve with a single Gaussian above a constant non-flaring level using Eq.~\ref{eq:gauss}.
The best fit parameters are reported in Table~\ref{table:gauss_fit}.

The upper limit on the amplitude of the undetected X-ray counterpart was computed using the Bayesian method for the determination of the confidence limits described by \citet[][see also \citealt{helene83}]{kraft91}.
We use the notations and equations of Sect.~\ref{flare_xmm}.
We first define a confidence limit $CL=0.95$ and the source $N$ as the number of counts during the time interval of the flare maximum (i.e., between $t_\mathrm{i}-3\sigma$ and $t_\mathrm{i}+3\sigma$ with $\sigma$ the error on $t_\mathrm{i}$ reported in Table~\ref{table:gauss_fit}).
The background $B$ is the number of counts in the non-flaring Bayesian-block at the time of the NIR flare peak.
We then determine $s_\mathrm{min}$ and $s_\mathrm{max}$ (see footnote 5 in Sect.~\ref{flare_xmm}) resolving the equation of $CL$.
For flare~\textrm{II}, $N=62$~counts and the non-flaring level is defined by the Bayesian blocks as $0.315\ \mathrm{counts\ s^{-1}}$ between 21.65 and 21.71~h, leading to $B=68$~counts and $S=-6$~counts.
Since $S$ is negative, $s_\mathrm{min}=0$, leading to $s_\mathrm{max}=6$.
The upper limit on the amplitude is thus $0.028\ \mathrm{counts\ s^{-1}}$ at a confidence level of 95\%.
The value of this upper limit is also reported in Table~\ref{table:gauss_fit}.

\subsection{The NIR flare \textrm{III} on 2014 Apr. 2}
We consider that two NIR flares happened during the occultation of \sgra\ by the Earth.
We thus fit the NIR light curve with two Gaussians (labeled \textrm{III}a and \textrm{III}b) above a linear component, which is used here to take into account the change in the non-flaring level between the last two HST orbits (Eq.~\ref{eq:gauss}).
The F-test strongly supports two Gaussian components, since this significanlty increases the goodness-of-fit  (p-value of $3 \times 10^{-4}$).
The best-fit parameters for the X-ray and NIR flares are given in Table~\ref{table:gauss_fit} and the resulting graphs are shown in the bottom panels of Fig.~\ref{fig:gauss_fit}.

We then fit the X-ray flare~{\bf2} with a Gaussian above a quiescent level equal to the Bayesian-block value.
The maximum of the X-ray flare has no time delay relative to the NIR flare~\textrm{III}a above the $3\sigma$ error bars, as usually observed for X-ray flares with NIR counterparts \citep{eckart06,yusef-zadeh06c,dodds-eden09}.
Moreover, the FHWM of the NIR flare~\textrm{III}a (2458~s) is about 3 times larger than that of the X-ray flare (762~s), which is reminiscent of the FWHM ratio of $\sim$2 observed by \citet{dodds-eden09} for the 2007 Apr. 4 NIR/X-ray flare.
The NIR flare~\textrm{III}a is thus probably the counterpart of the X-ray flare~2.
This conclusion is based on our Gaussian fitting of flare~\textrm{III}, but a more complex shape cannot be excluded due to the NIR observational gap. 
However, since the X-ray flare~2 and the previously observed NIR/X-ray flares also have a Gaussian shape \citep{eckart06,yusef-zadeh06c,dodds-eden09}, we consider that this conclusion is the simplest and thus the most likely.

The amplitude of the flare~\textrm{III}b is one of the largest observed when compared with the sample obtained with NACO \citep{witzel12}.
No X-ray counterpart is detected for this flare.
We thus obtain an upper limit on the X-ray amplitude using the same method as flare~\textrm{II} with $N=763$~counts between 16.9 and 17.5~h.
The background is defined as the sum of the number of counts in non-flaring Bayesian-block values ($626.4$~counts) and the number of counts in the Gaussian fit of flare~2 during the maximum of the flare ($121.7$~counts). 
We thus have $B=748.1$~counts, leading to $S=14.9$~counts.
The resulting $s_\mathrm{min}$ is 0, with $s_\mathrm{max}=65$~counts.
The upper limit on the amplitude of the undetected X-ray counterpart is thus $0.030\ \mathrm{counts\ s^{-1}}$ at the confidence level of 95\%.
This value is reported in Table~\ref{table:gauss_fit}.

\subsection{The NIR flare \textrm{IV} on 2014 Apr. 3}
\label{april_3}
The VLT light curves consist of bins of 400~s exposures.
The number of bins is too small and the bin size too large to fit a Gaussian to the VLT light curves.
We thus consider only the bin with the largest flux density as the peak of the flare~\textrm{IV}.
This value and the time of the maximum are reported in Table~\ref{table:gauss_fit}.

No X-ray counterpart is detected with XMM-Newton on Apr. 3.
We thus deduce an upper limit to the putative simultaneous X-ray flare using the same method that was used for flare~\textrm{II}.
The time interval of the maximum of flare~\textrm{IV} is defined as the bin length of the light curve, i.e., 400~s centered on 7.89~h.
The number of counts in this interval is $N=127$~counts and the background is $B=119.1\ \mathrm{counts}$, leading to $S=7.9$~counts.
The resulting $s_\mathrm{min}$ is 0, with $s_\mathrm{max}=17$~counts,
leading to an upper limit on the amplitude of $0.042\ \mathrm{counts\ s^{-1}}$.

\subsection{The NIR flare \textrm{V} on 2014 Apr. 4}
\label{april_4}
For flare \textrm{IV}, we do not fit the light curve with a Gaussian and we consider the maximum of the light curve as the peak flux density of the NIR flare (Table~\ref{table:gauss_fit}).

We have no XMM-Newton observation on 2014 Apr. 4.
However, as shown in Fig.~\ref{fig:obs_log}, there is a simultaneous legacy Chandra observation (ObsID: 16212; PI: D. Haggard) on this date.
We used the Chandra Interactive Analysis of Observations (CIAO; version 4.6) to analyze these data.
We worked with the level=2 event list of the \textit{ACIS-S} camera \citep{garmire03}, available in the primary package of the Chandra archive.
We extracted the source+background events in the 2$-$8~keV energy range in a 1$\farcs$25-radius circle centered on the radio coordinates of \sgra\  using the \texttt{dmcopy} task.
We used the Bayesian-blocks analysis with a false detection probability of $e^{-3.5}$ to detect any flaring event.
No X-ray counterpart to the NIR flare was detected during this observation.
Based on $N=1$~counts between 8.71 and 8.93~h and a non-flaring level of $0.0065\ \mathrm{counts\ s^{-1}}$, we compute $B=3$~counts
and $S=-2$~counts.
The resulting $s_\mathrm{min}$ is 0 with $s_\mathrm{max}=4$~counts.
The upper limit to the putative simultaneous X-ray amplitude is thus $0.01\ \mathrm{counts\ s^{-1}}$ at the confidence level of 95\% (see flare~\textrm{II} for explanations).

\section{Constraining the physical parameters of the flaring region}
\label{phys_param}
In this section we constrain the physical parameters of the flaring region by considering three radiative models for the NIR and X-ray emission.
After computing the NIR-to-X-ray simultaneous peak ratio sample detected during the 2014 campaign, we investigated the synchrotron$-$synchrotron (SYN-SYN),  synchrotron Self-Compton$-$synchrotron Self-Compton (SSC-SSC), and the synchrotron$-$synchrotron Self-Compton (SYN-SSC) radiation mechanisms.
These processes are called ``local'', because these emissions are produced only by the electrons accelerated in the flaring region.
The last subsection is dedicated to the Inverse Compton mechanism, which involves external electrons.

\subsection{The sample of NIR flares and the corresponding X-ray emission}
We compute the flux densities of the NIR and X-ray flare peaks to constrain the physical parameters of the flaring region needed to produce such fluxes.
We extrapolate the amplitude of each NIR peak to the $K_\mathrm{s}$-band using the $H-L$ spectral index computed in \citet{witzel13}.
The flux density of the X-ray flare peaks is computed from the spectral fitting in \texttt{ISIS}, using the typical spectral parameters of the X-ray bright flares (see Sect.~\ref{spectrum_xmm}).
The resulting conversion factor is 1~$\mathrm{pn\,count\,s^{-1}}$=3.935~$\mu$Jy at 4~keV.
The NIR peak flux density and corresponding values of the X-ray peaks (or upper limits) are reported in Table~\ref{table:gauss_fit}.

\begin{figure}
\centering
\includegraphics[width=6.5cm,angle=90]{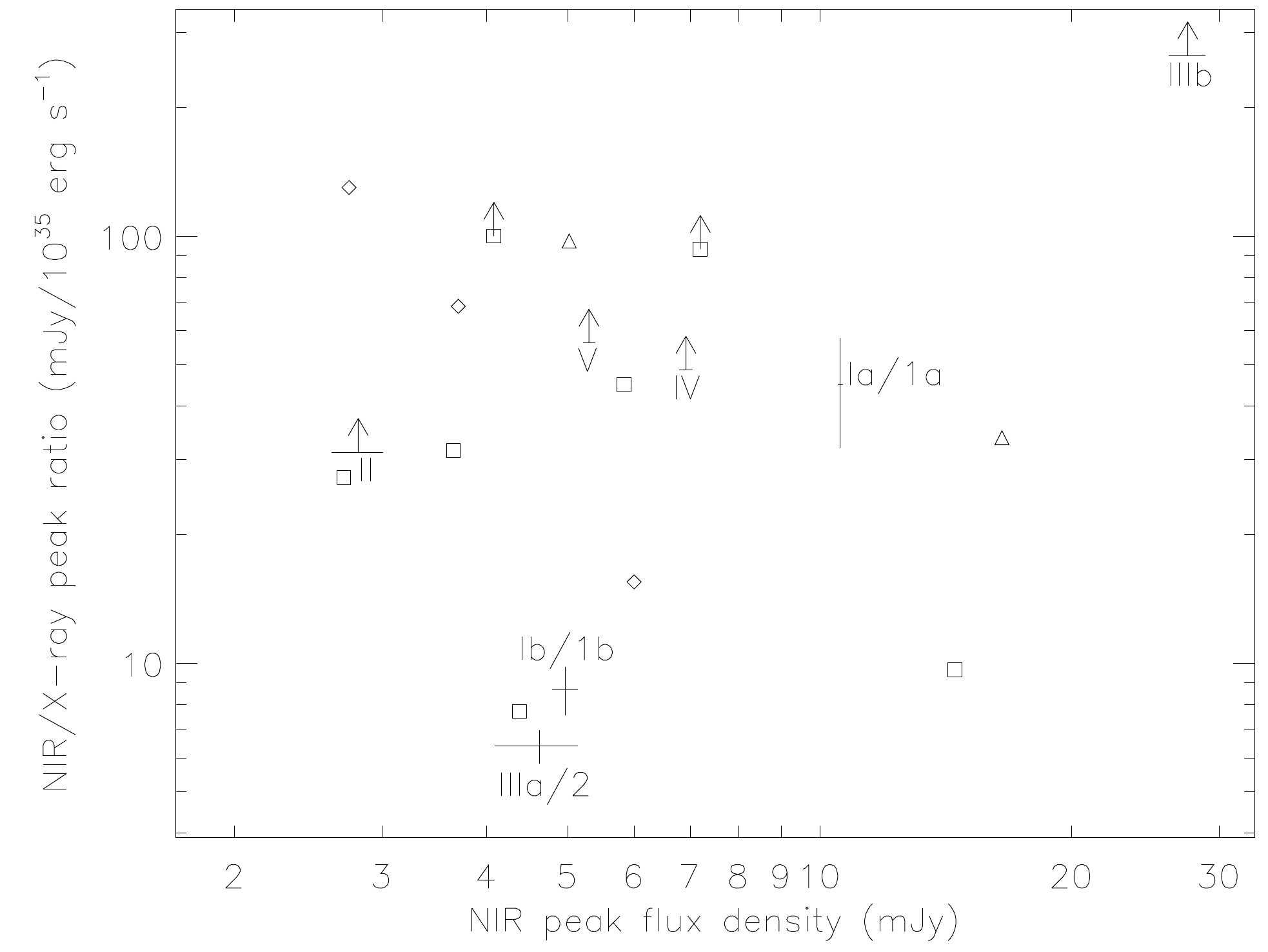}
\caption{NIR-to-X-ray peak ratio vs. amplitude of the NIR flares.
Squares  refer to the flares reported in Table 3 of \citet{eckart12}.
Triangles are the simultaneous NIR/X-ray flares detected on 2007 Apr. 4 and labeled $D$ and $E$ in Table 2 of \citet{trap11}.
Diamonds are the delayed flares of 2004 Jul. 7, 2008 Jul. 26+27 and 2008 May 5 reported in Table~2 of \citet{yusef-zadeh12}.
The labeled points are the NIR and X-ray flares observed during this campaign.}
\label{fig:ampl_flare}
\end{figure}

Figure~\ref{fig:ampl_flare} shows the NIR-to-X-ray peak ratio as a function of the amplitude of the NIR flares observed during the 2014 campaign.
The unabsorbed X-ray peak luminosities are computed using the conversion factor reported in Sect.~\ref{spectrum_xmm}.
The X-ray upper limit of NIR flare~\textrm{V} was obtained from Chandra data.
The corresponding unabsorbed-flux-to-count-rate ratio of $1.97 \times 10^{-10}\,\mathrm{erg\,s^{-1}\,cm^{-2}}/\mathrm{count\,s^{-1}}$ was computed with the same spectral parameters as for XMM-Newton.

\begin{figure*}
\centering
\begin{tabular}{@{}cc@{}}
\includegraphics[width=5.7cm,angle=90]{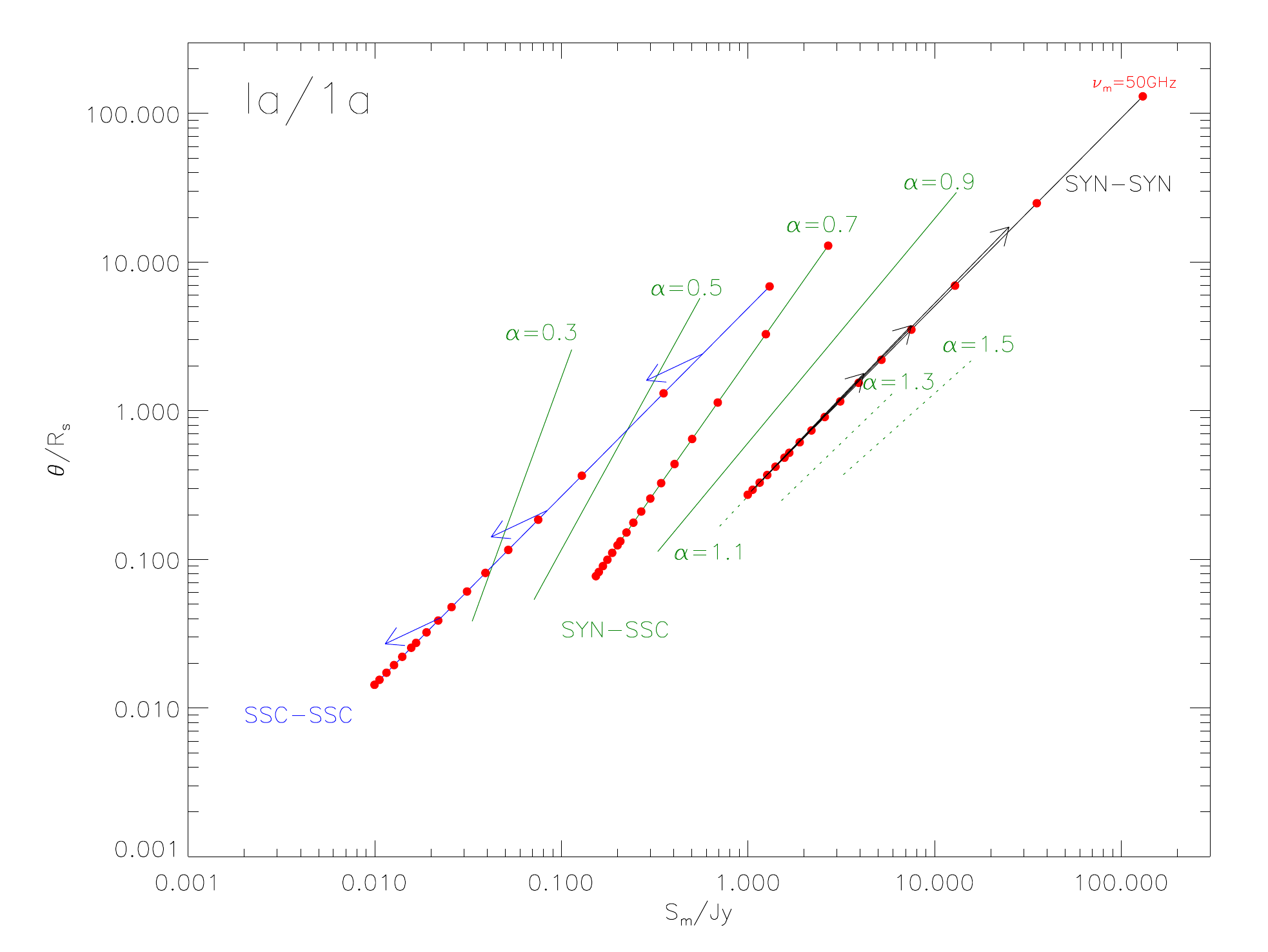}&
\includegraphics[width=5.7cm,angle=90]{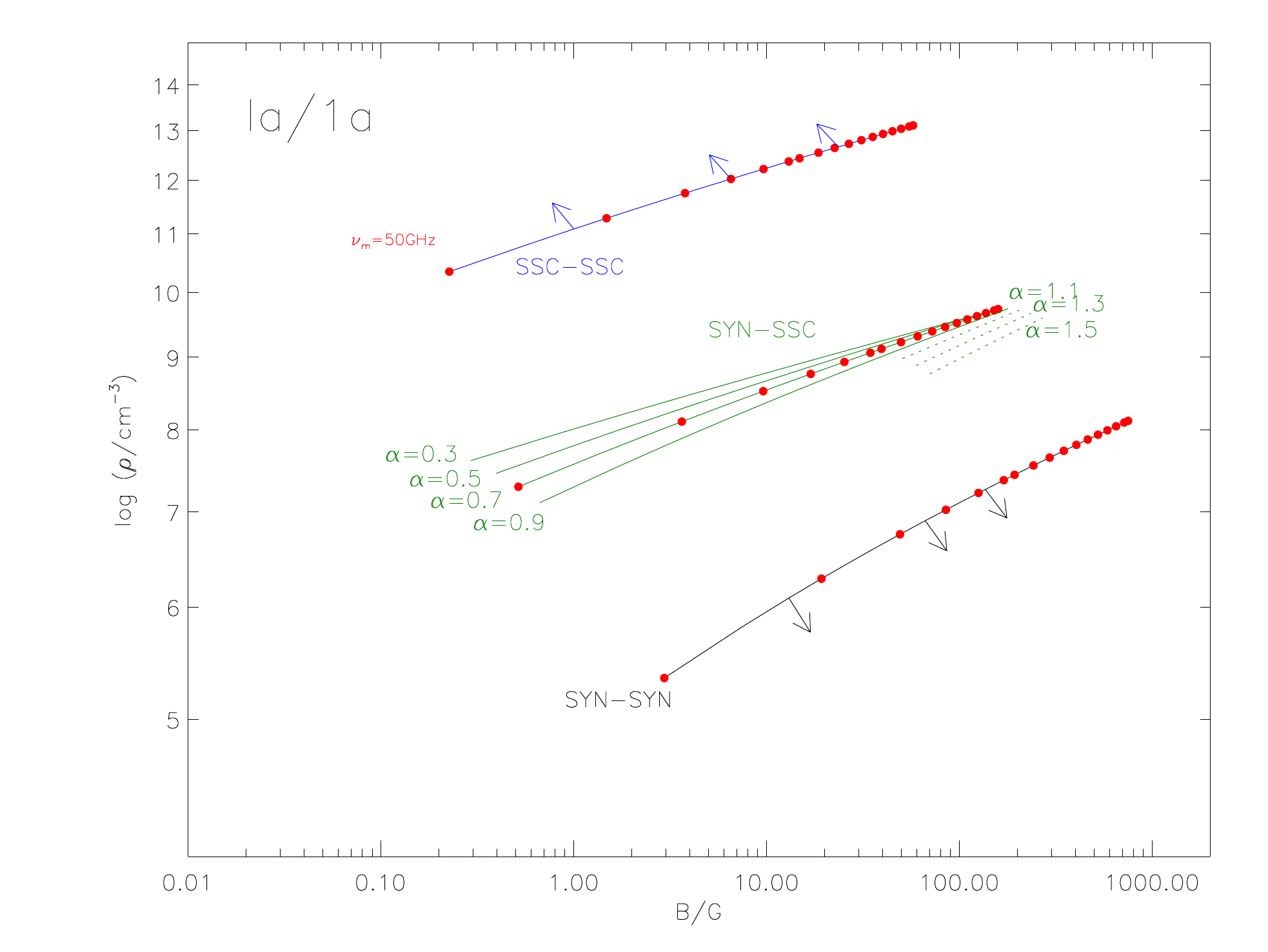}\\
\includegraphics[width=5.7cm,angle=90]{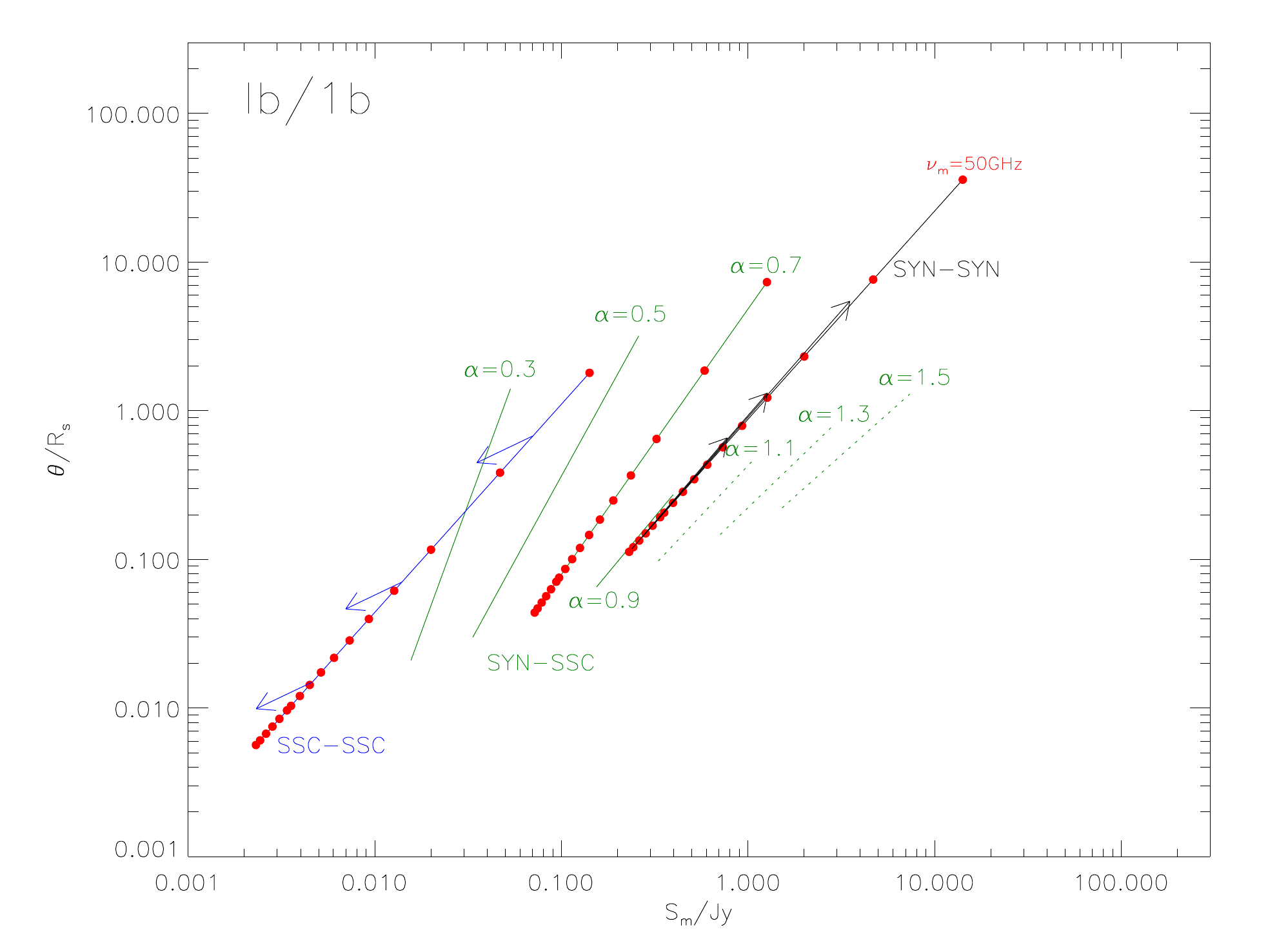}&
\includegraphics[width=5.7cm,angle=90]{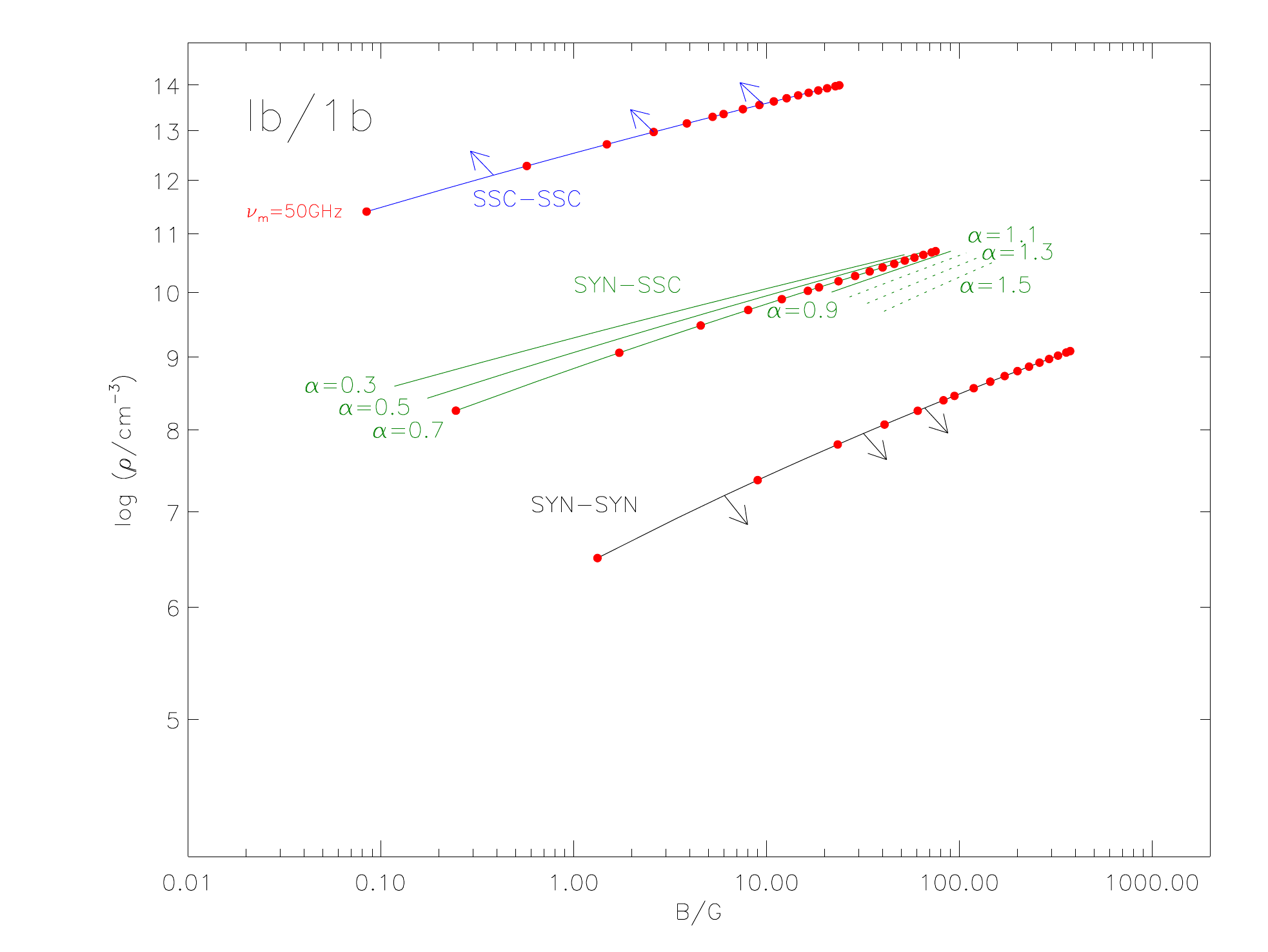}\\
\includegraphics[width=5.7cm,angle=90]{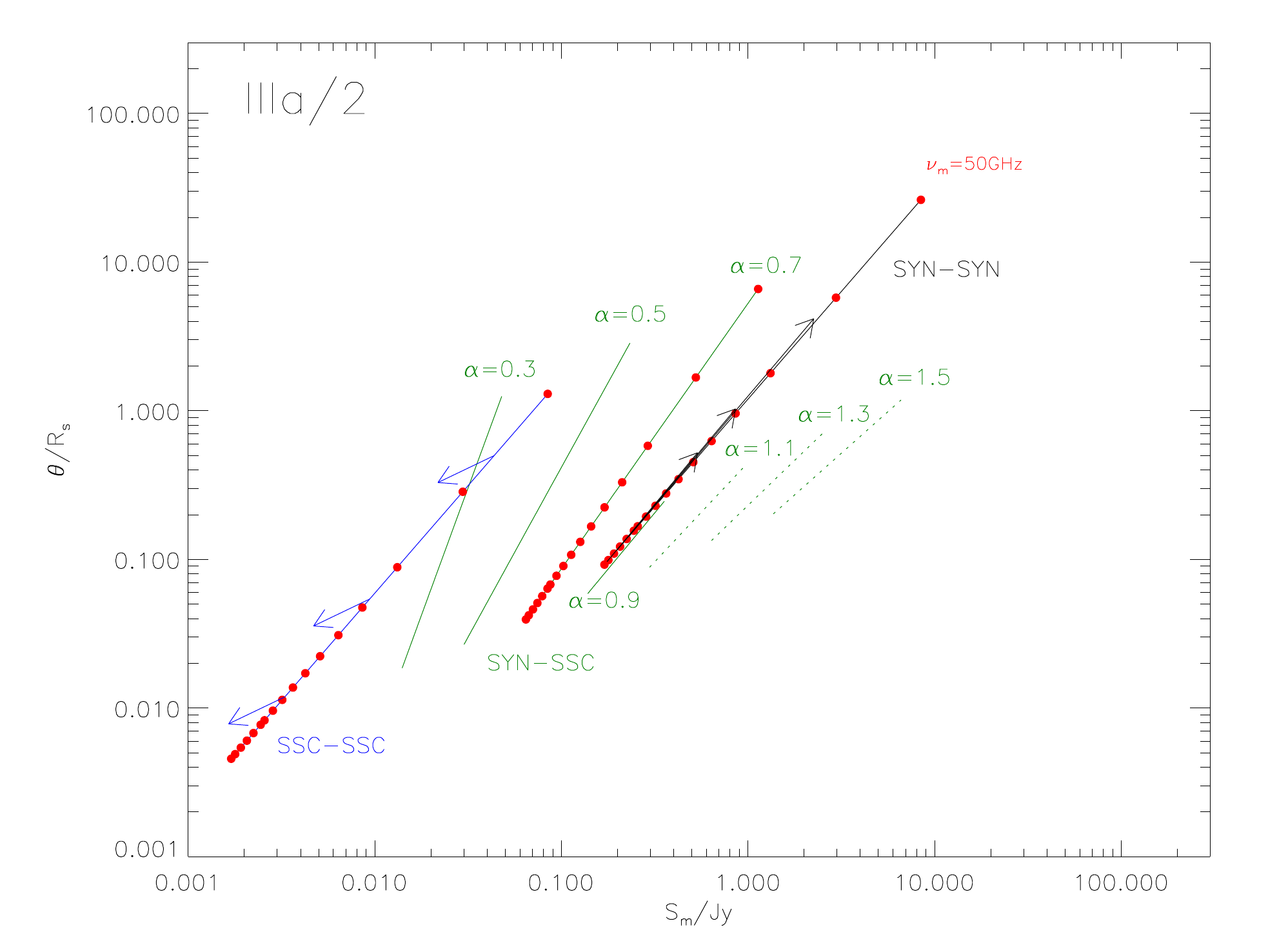}&
\includegraphics[width=5.7cm,angle=90]{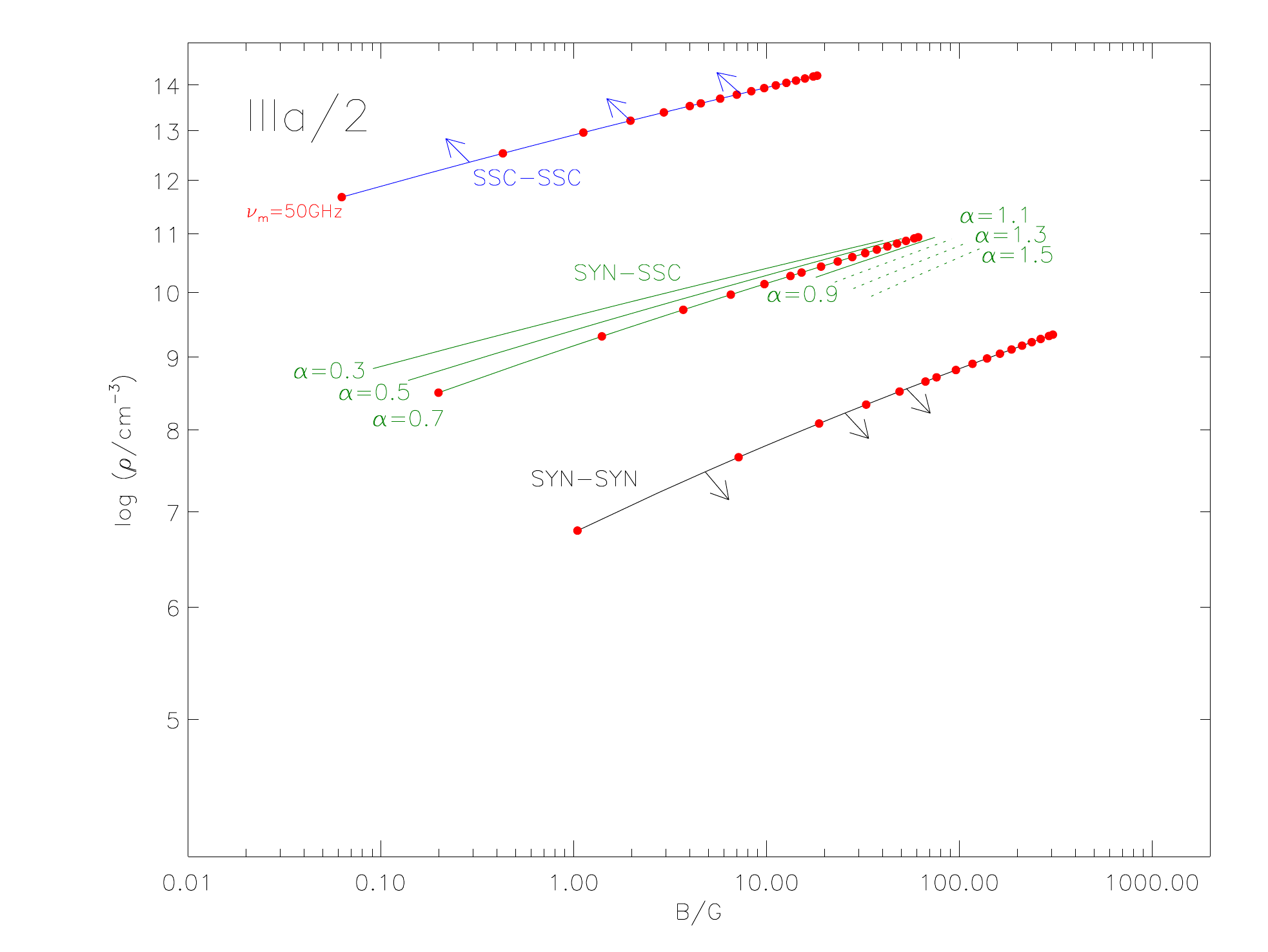}
\end{tabular}
\caption{Physical parameters of the flares observed simultaneously in X-rays and NIR for the three emission models.
The flare \textrm{I}a/1a, \textrm{I}b/1b and \textrm{III}a/2 are in the upper, middle and bottom panels, respectively.
Left panels are the size of the flaring-source region ($\theta$) vs. the peak of the spectrum ($S_\mathrm{m}$) at the frequency $\nu_\mathrm{m}$.
Right panels are the density of the relativistic electrons vs. the magnetic field.
The locii where the Synchrotron Self-Compton$-$Synchrotron Self-Compton (SSC-SSC), Synchrotron$-$Synchrotron Self-Compton (SYN-SSC) and Synchrotron-Synchrotron (SYN-SYN) are dominant are shown in black, blue and green, respectively.
The red dots represent the turnover frequencies from 50 to 3000~GHz by step of 200~GHz.
The arrows show the direction of the curves if the limit on the alternative emission processes is lowered.
Dotted lines are locii of SYN-SSC where the MIR emission is larger than the observed upper-limit values (see text for details). }
\label{fig:eckart12_detect}
\end{figure*}

We also show the flares reported by \citet{eckart12}, the two simultaneous flares on 2007 Apr. 4 \citep{porquet08,nowak12} labeled $D$ and $E$ in Table 2 of \citet{trap11}, and the delayed flares of 2004 Jul. 7, 2008 Jul. 26+27, and 2008 May 5 reported in Table~2 of \citet{yusef-zadeh12}.

The flare \textrm{I}a/1a lies within the bulk of NIR flare amplitudes and peak ratio, whereas the flare \textrm{III}a/2 has the lowest NIR-to-X-ray ratio ever observed.
The NIR flare~\textrm{III}b is amongst the largest NIR flares \citep[e.g.,][]{dodds-eden09,witzel12} and has the largest NIR-to-X-ray ratio ever observed.

\subsection{Investigation of the local radiative processes}
With the peak flux density of the flares in X-rays and NIR, we use the formalism developed by \citet{eckart12} to constrain the range of four physical parameters of the flaring emission: the size of the source region, the magnetic field, the density, and the maximum of the flux density spectrum.
\citet{eckart12} considered three cases, invoking the two radiative processes, implying the local electrons from the flaring source region: the SYN-SYN, SSC-SSC, and SYN-SSC emissions.
A radiative process is considered as dominant when the alternative emission processes are lower than 10\%. For example, the SYN-SYN case is dominant if both SSC contribution for NIR and X-rays are lower than 10\% of the synchrotron contribution.
Considering different values for the turnover frequency ($\nu_\mathrm{m}$), which defines the frequency at which the source becomes optically thin, we have four free physical parameters for each value of the spectral index ($\alpha$): the size of the emitting region ($\theta$), peak flux density at $\nu_\mathrm{m}$ ($S_\mathrm{m}$), number density of relativistic particles ($\rho$), and the magnetic field ($B$).
The spectral index is given by the ratio between the NIR and X-ray amplitudes for the SYN-SYN and SSC-SSC cases, and by seven different values of $\alpha$ from 0.3 to 1.5 for the SYN-SSC case.
Computing the SYN or SSC flux density with the equations given by \citet{eckart12}, we can constrain the values of the four physical parameters for each value of $\alpha$ and $\nu_\mathrm{m}$.

The resulting graphs for the flares detected in NIR and X-rays (labeled \textrm{I}a/1a, \textrm{I}b/1b and \textrm{III}a/2) are shown in Fig.~\ref{fig:eckart12_detect}.
Each line corresponds to one value of $\alpha$.
The red dots are the turnover frequencies from 50 to 3000~GHz in steps of 200~GHz.
The constraint on the MIR amplitude limit observed during the bright $L'$-band and X-ray flare on 2007 Apr. 4 of $57$~mJy at 11.88~$\mu$m \citep{dodds-eden09} is also used: the lines are dashed if this limit is exceeded.
This happens only for the SYN-SSC emission mechanism and for high values of $\alpha$.

The physical parameters are more constrained for flare \textrm{III}a/2, since the X-ray-to-NIR amplitude ratio is high.
For this flare, the SYN-SSC emission mechanism leads to a size of 0.03$-$7 times the Schwarzschild radius and an electron density of $10^{8.5}$--$10^{10.2}$~cm$^{-3}$ for a synchrotron spectral-index of 0.3$-$1.5.

From the magnetic field values deduced for these flares, one can infer the presence of sustained heating during the decay phase of the X-ray or NIR flares for the SYN-SYN and SYN-SSC case.
Indeed, if the synchrotron cooling timescale, defined as $\tau_\mathrm{sync}=8 \left(B/30~\mathrm{G}\right)^{-3/2} (\nu/10^{14}~\mathrm{Hz})^{-1/2}$~min \citep{dodds-eden09}, is shorter than the duration of the decay phase then sustained heating is needed.
We define the decay phase from the time of the maximum of the Gaussian fit (see Table~\ref{table:gauss_fit}) to the time leading to 10\% of the flare amplitude (corresponding to $2.1\sigma$ after the maximum) in order to still have a detectable emission of the flare.

For the SYN-SYN case, the synchrotron cooling timescale is shorter for the X-ray photons, leading to more constraints on the presence of sustained heating.
We thus consider the X-ray frequency ($\nu=10^{18}$~Hz) in the computation of the synchrotron cooling timescale.
The synchrotron cooling timescale is shorter than the decay time of flare~2 ({\bf695}~s) for $B$ larger than 1~G, implying that sustained heating must be present during the decay phase for these values of magnetic field.
A sustained heating is always needed for flares~\textrm{I}a/1a and \textrm{I}a/1b, since they have a minimum value of the magnetic field and a decay time larger than those of flare~2 (2318 and 2781~s, respectively).

For the SYN-SSC case, we consider the NIR frequency ($\nu=10^{14}$~Hz) in $\tau_\mathrm{sync}$ that we have to compare to the decay time of the NIR flares.
Sustained heating is now needed for flare~\textrm{III}a (whose decay time is 2240~s) with a magnetic field of greater than 11~G, corresponding to an electron density larger than $10^{10.1}\ \mathrm{cm^{-3}}$.
For flares~\textrm{I}a and \textrm{I}b (whose decay times are 1545 and 3785~s, respectively), sustained heating is needed for magnetic fields larger than 13 and 7~G, respectively.
The corresponding electron density is thus larger than $10^{8.4}$ and $10^{9.5}\ \mathrm{cm^{-3}}$.

\begin{figure*}
\centering
\begin{tabular}{@{}cc@{}}
\includegraphics[width=5.7cm,angle=90]{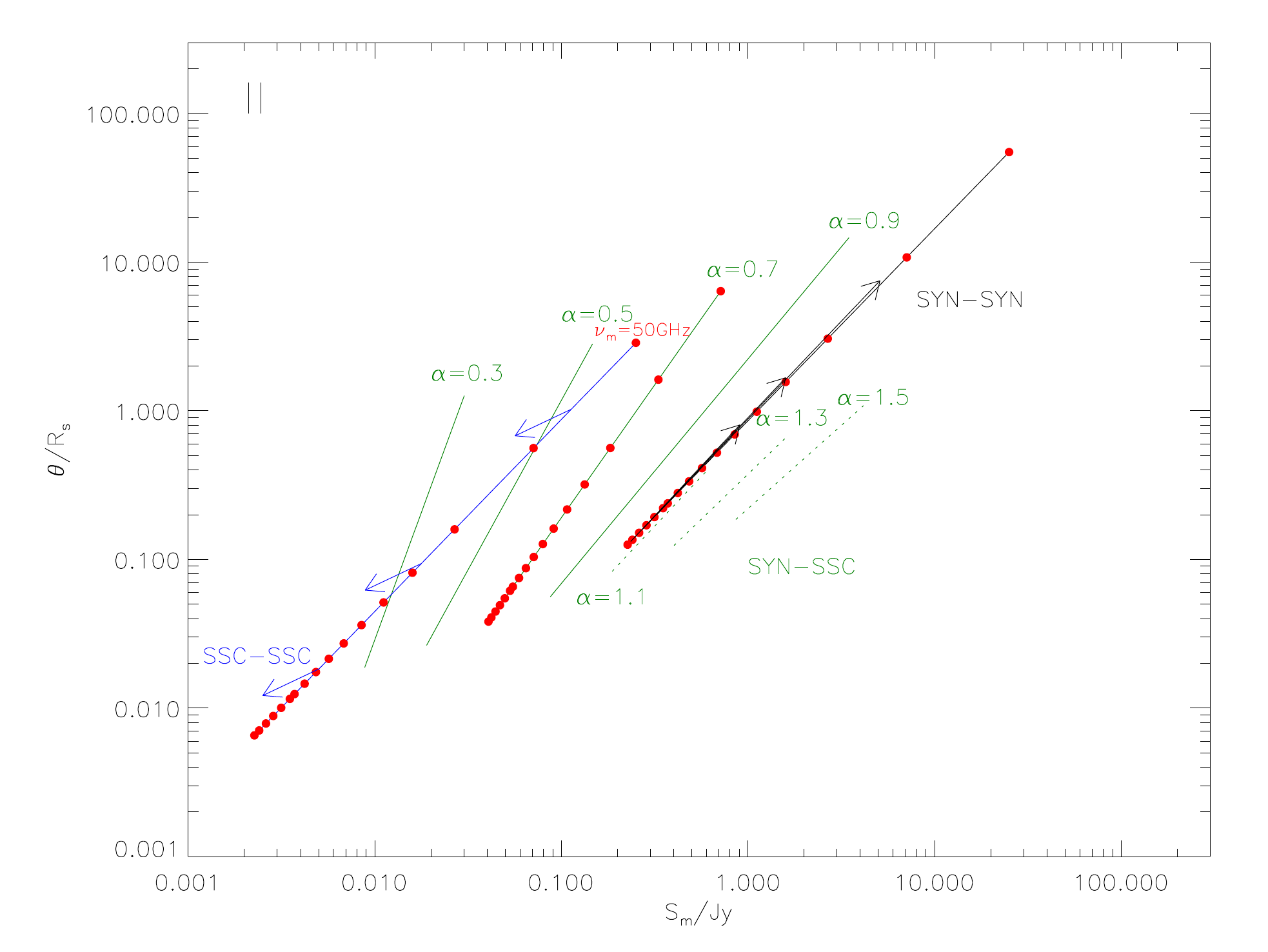}&
\includegraphics[width=5.7cm,angle=90]{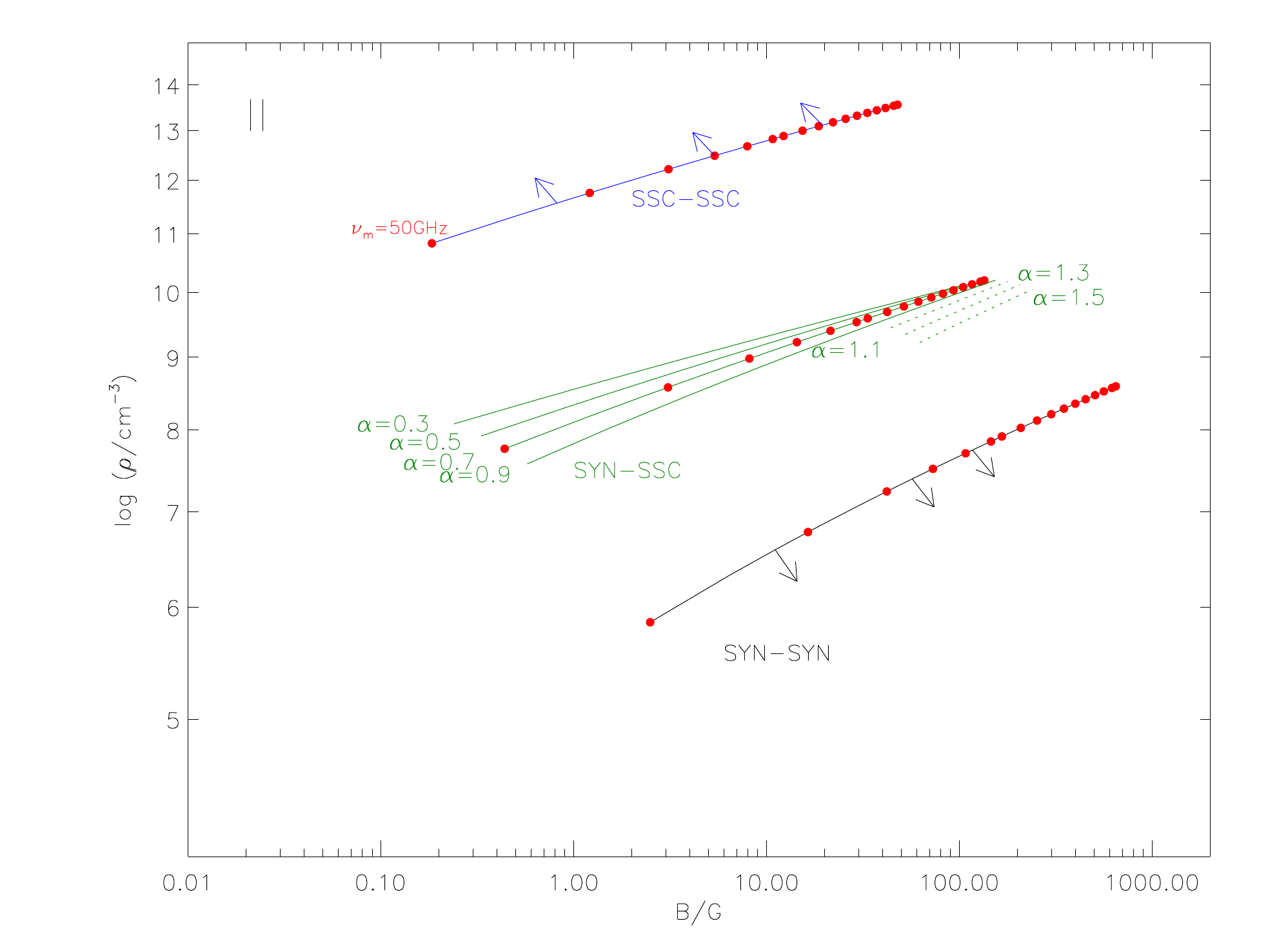}\\
\includegraphics[width=5.7cm,angle=90]{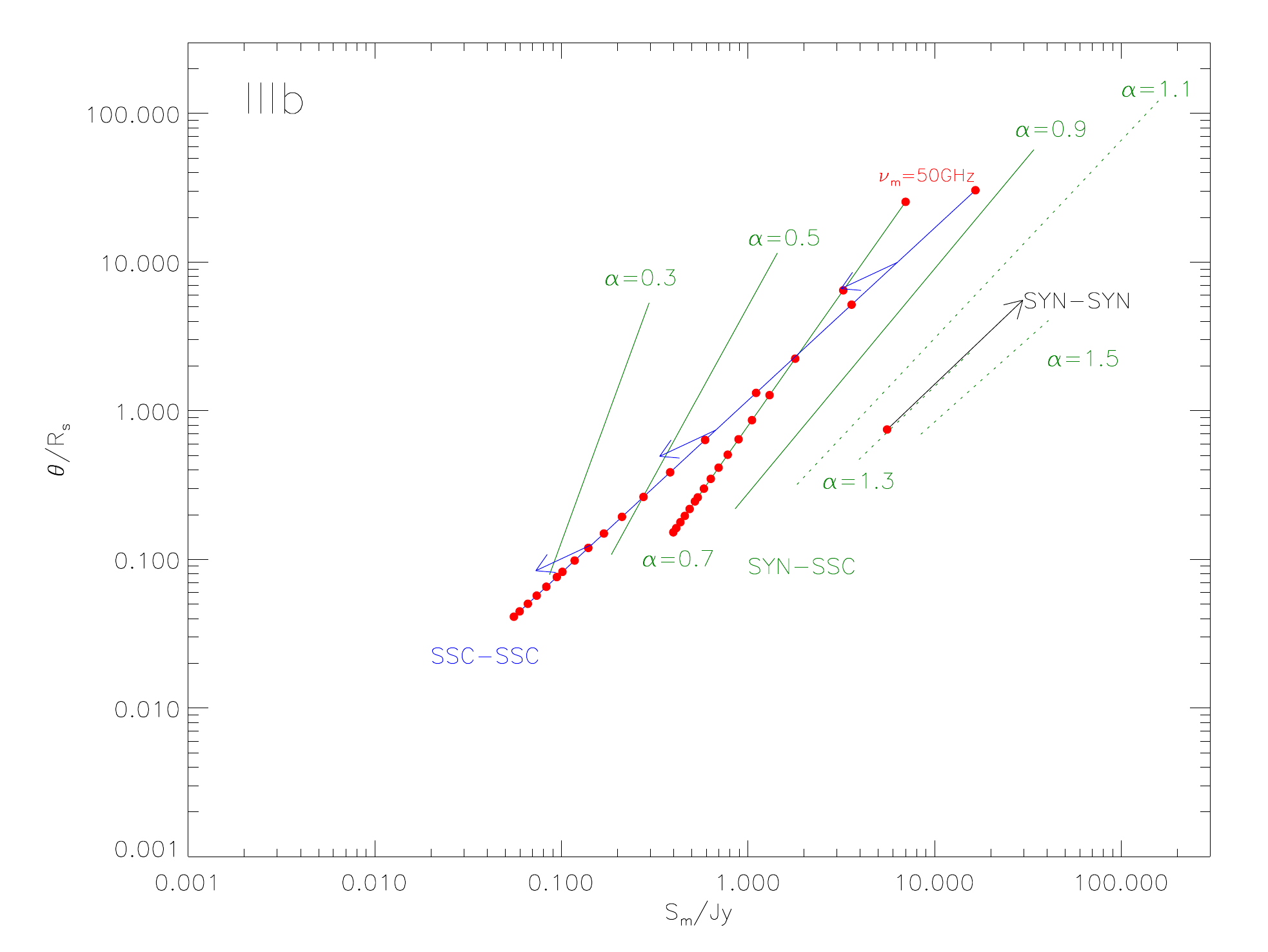}&
\includegraphics[width=5.7cm,angle=90]{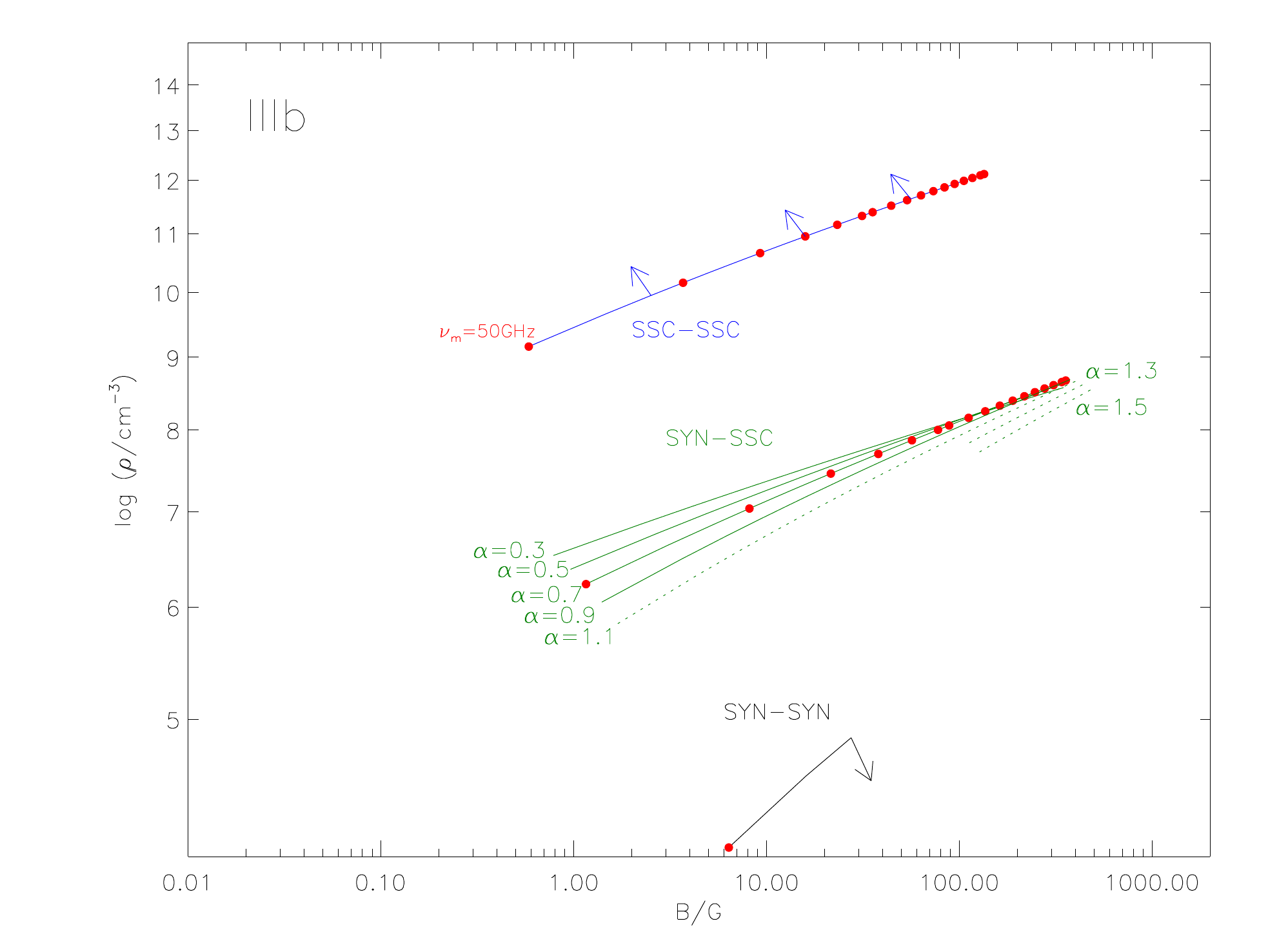}\\
\includegraphics[width=5.7cm,angle=90]{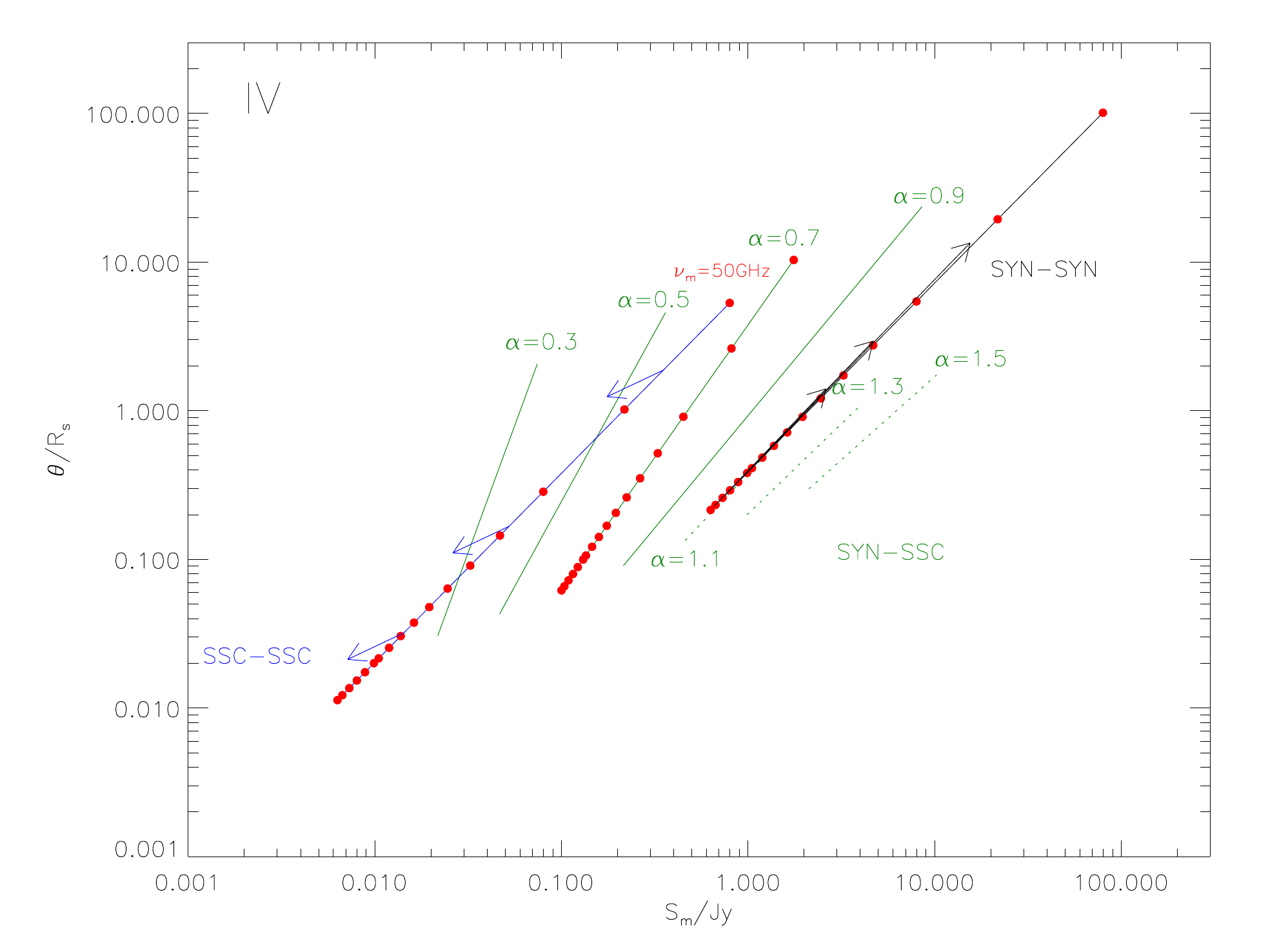}&
\includegraphics[width=5.7cm,angle=90]{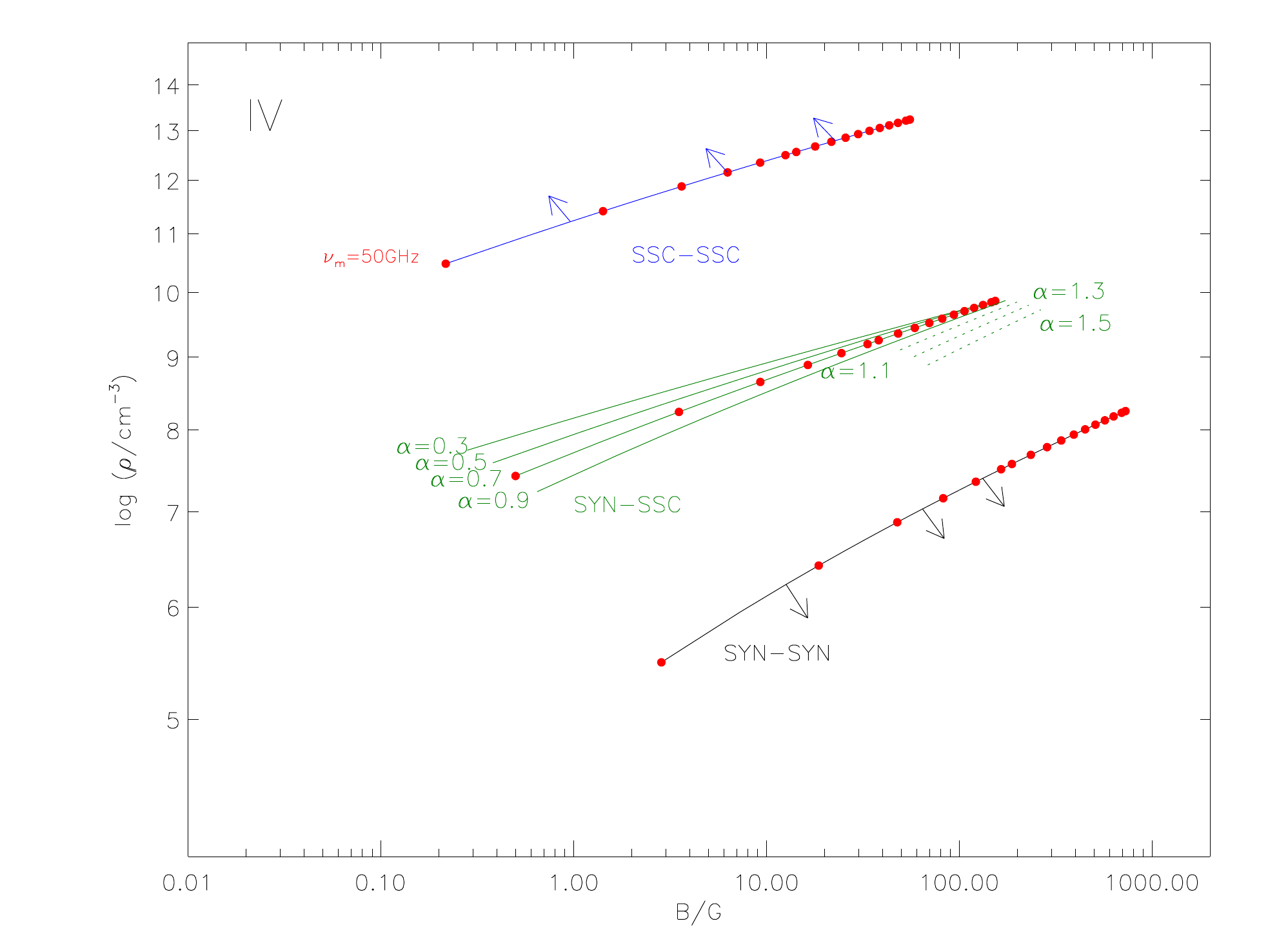}\\
\includegraphics[width=5.7cm,angle=90]{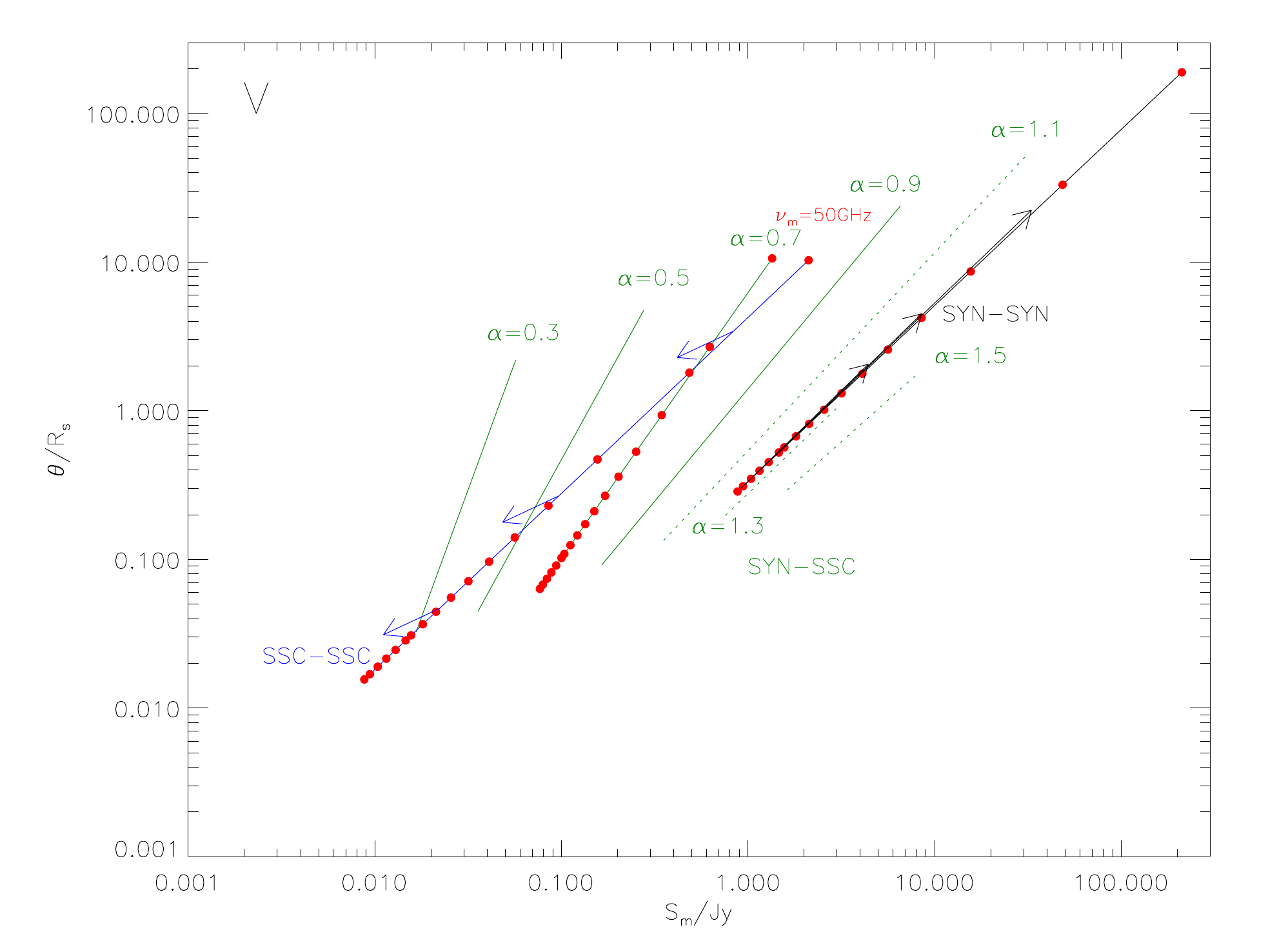}&
\includegraphics[width=5.7cm,angle=90]{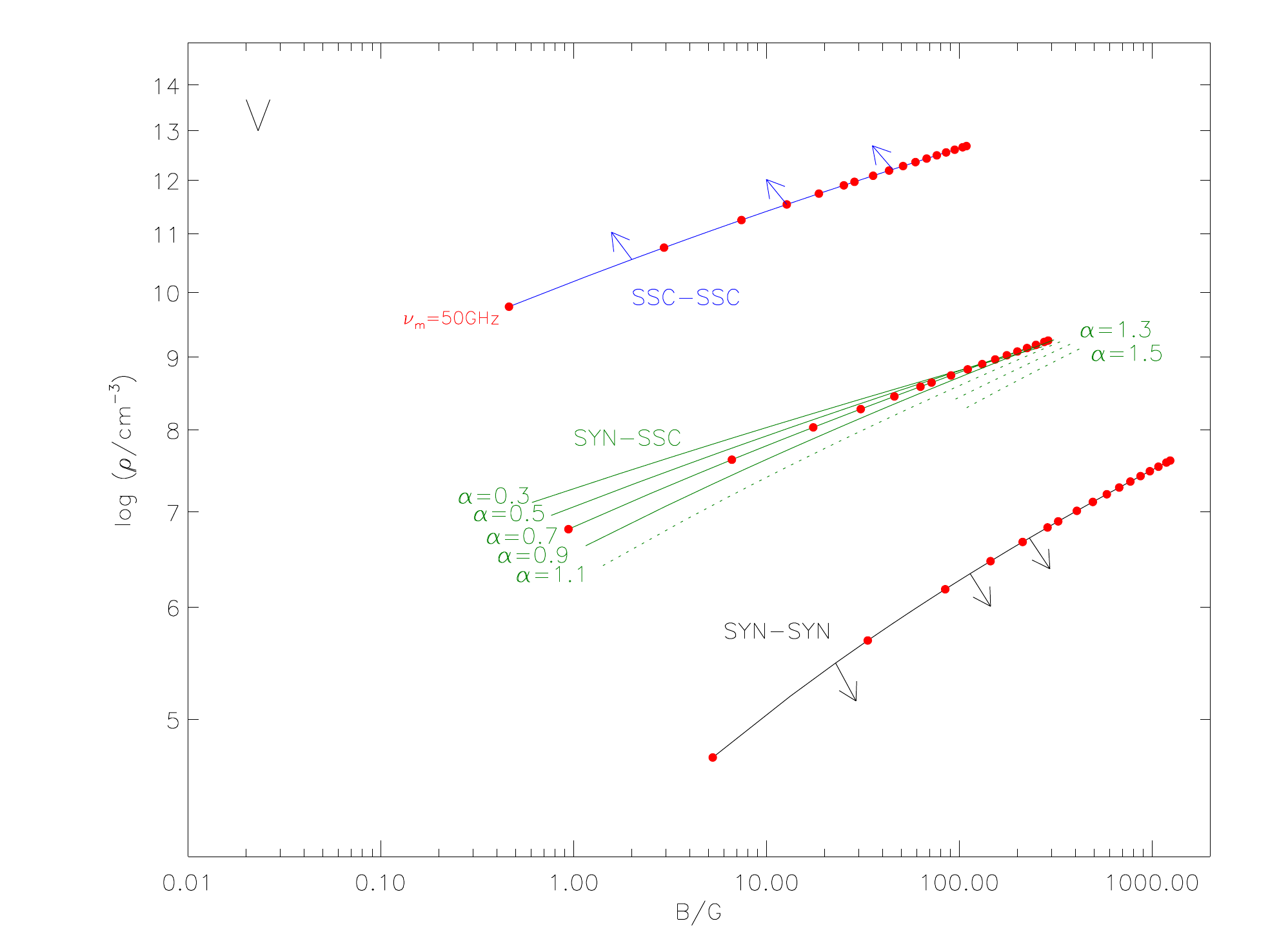}
\end{tabular}
\caption{Physical parameters of the flares only observed in NIR for the three emission models.
The NIR flare \textrm{II}, \textrm{III}b, \textrm{IV} and \textrm{V} are in the upper, second, third and bottom panels, respectively.
See Fig.~\ref{fig:eckart12_detect} caption for the panel description.}
\label{fig:eckart12_undetect}
\end{figure*}

We also apply the study of \citet{eckart12} to constrain the physical parameters of the flaring emission for the NIR flares that have no detected X-ray counterpart (flares \textrm{II}, \textrm{III}b, \textrm{IV} and \textrm{V}).
The resulting graphs are shown in Fig.~\ref{fig:eckart12_undetect}.
The necessary electron density and magnetic field ranges lie within lower values compared to those needed to produce detectable X-ray flares, since the efficiency of the production of X-ray photons is smaller.
Moreover, for flare \textrm{III}b, the SYN$-$SYN process is only dominant for small values of $\nu_\mathrm{m}$.
This is explained by the small X-ray-to-NIR amplitude ratio, since at large $\nu_\mathrm{m}$ the synchrotron process is too efficient for the production of a small number of X-ray photons.

We can also deduce the presence of sustained heating during the decay phase of NIR flares~\textrm{II} and \textrm{III}b for the SYN-SYN and SYN-SSC case.
The synchrotron cooling timescale is shorter than the decay time of flare~\textrm{II} (772~s) if $B$ is larger than 22~G, requiring sustained heating during the decay phase for these values of magnetic field.
For flare~\textrm{III}b (whose decay times are 695~s), sustained heating is required for $B$ larger than 1~G.

However, as argued by \citet{eckart12b}, alternative models such as different spectral indexes for the NIR and X-ray, due to inhomogeneities of the accretion disk, can also explain the data with larger numbers of free parameters.

\subsection{Investigation of the external radiative process}
As reported in Sect.~\ref{syn_ic}, \citet{yusef-zadeh12} investigated the upscattering of the NIR photons produced in the flaring region on electrons of the accretion flow.
The ratio between the Inverse Compton and the synchrotron emission is
\begin{equation}
  \frac{L_\mathrm{IC}}{L_\mathrm{SYN}}\propto \frac{U_\mathrm{ph}}{U_\mathrm{B}}\, ,
\end{equation}
with $U_\mathrm{ph}$ the photon energy density and $U_\mathrm{B}=B^2/8\pi$ the magnetic energy density.
Given the variation of $B$ with the distance from \sgra\  ($B=B_0\,(r/R_\mathrm{s})^{-1}$ with $B_0$ of several hundred of Gauss; \citealt{eatough13}), it is possible to create NIR and X-ray flares with a large range of NIR-to-X-ray ratio. 
Thus, we cannot identify the IC radiation by only considering the NIR-to-X-ray ratio.

However, using the estimation of the time delay between X-ray and NIR flare peaks as a function of the NIR-to-X-ray peak ratio reported in \citet{yusef-zadeh12} and shown in Fig.~\ref{time_delay}, we can estimate the time delays that we would observe during our 2014 campaign. 
For the detected X-ray flares, the NIR-to-X-ray peak ratio ranges between $6$ and $45\ \mathrm{mJy}/10^{35}\,\mathrm{erg\,s^{-1}}$ (see Fig.~\ref{fig:ampl_flare}) leading to a time delay less than 10~min, which is smaller than the error bars on the time of the maximum of the Gaussians.
The IC emission, therefore, is still a possible radiative process for the production of the X-ray flares observed during this campaign. 

For the undetected X-ray flares \textrm{II}, \textrm{III}b, \textrm{IV} and \textrm{V}, the NIR-to-X-ray ratio is larger than $32$, $269$, $48$, and $55\ \mathrm{mJy}/10^{35}\,\mathrm{erg\,s^{-1}}$, respectively.
The corresponding time delays are thus greater than  9, 26, 11, and 12~min, respectively.
These time delays are larger than the events with detected X-ray flares.
The efficiency of the flare detection with XMM-Newton and Chandra, however, does not allow us to detect such faint X-ray flares, which may have the largest delay in the inverse Compton framework.

Thus, the flares observed during the 2014 campaign leave the IC process as a possible emission mechanism for producing X-ray flares from the NIR photons.

\section{Discussing the X-ray flaring rate}
\label{discussion}
We can compare the X-ray flare frequency during our observations (three flares over $255.644$~ks) to the one derived from the \textit{Chandra XVP} campaign in 2012: 45 flares detected by Bayesian-block algorithm over $2983.93$~ks (1.5 flare per day).
Considering a sample of 45 flares having the same amplitude and duration distribution as those observed during the \textit{Chandra XVP} campaign superimposed on the non-flaring level observed with XMM-Newton during our campaign, the Bayesian-blocks algorithm detects 36 flares over $2983.93$~ks.
If we sum the number of flares that we can detect during the exposure time corresponding to each observation during the XMM-Newton 2014 campaign, we arrive at a prediction of 3.1 flares during this campaign.
We compare the flare rate observed during the \textit{Chandra XVP} campaign to those observed during this campaign (36 flares over $2983.93$~ks and 3 flares 
over $255.644$~ks), assuming a Poisson process \citep{gehrels86,fay10}.
The p-value for the null hypothesis that the flaring rate we have to observe and the rate we currently observe is the same, is 1, which implies that the flaring rate observed close to the pericenter passage of the DSO/G2 is consistent with that observed during the \textit{Chandra XVP} campaign.
The conclusion is the same if we consider only two X-ray flares instead of three (p-value=0.54).
To conclude that the measured flaring rate is statistically different from those observed during the \textit{Chandra XVP} campaign, we would have to detect at least 8 flares during our campaign (p-value=0.04), which corresponds to an increase of the flaring rate by a factor of 2.6 (95\% confidence interval of $1.0-5.7$).

Since the beginning of the observation of \sgra\  in X-rays, two temporary episodes of higher flaring rate were observed \citep{porquet08,neilsen13}.
\citet{porquet08} detected four flares on 2007 Apr. 04 with XMM-Newton.
Three of these flares happened during the last 39.6~ks of the observation, corresponding to a flaring rate of 8.8 flares per day.
We can compare this flaring rate to the 38 flares that should be detected by the Bayesian-block algorithm.
The ratio between the two rates is 5 and the 95\% confidence interval is 1.3$-$20 (p-value=0.03).
\citet{neilsen13} detected 4 flares during 23.6~ks with Chandra, which corresponds to a flaring rate of 14.6 flares per day.
We can directly compare this flaring rate to that computed during the 2012 \textit{Chandra XVP} campaign if we remove these 4 flares from the sample of 45 flares detected by the Bayesian-blocks algorithm.
Thus, we have to compare 41 flares over $2960.33$~ks and 4 flares over $23.6$~ks.
The ratio between the two rates is 13 and the 95\% confidence interval is 3.3$-$33.3 (p-value=$9\times 10^{-4}$).
This implies that some temporary increase of \sgra\  activity in X-ray may have been observed without an increase of the quiescent level due to an increase of the accretion rate.

The radio monitoring of \sgra\  with VLA between 2012 and 2014 May showed no change in the flux density or the spectrum \citep{bower15b,yusef-zadeh15d}.
Observations of \sgra\  after the DSO/G2 pericenter passage show that there is no increase of the flaring activity in radio/sub-mm \citep{tsuboi14,park15}.
The 2014 Feb.$-$June Chandra X-ray monitoring of \sgra\  shows no rise of the quiescent flux \citep{atel6242}.
The compactness of the object can explain the absence of any increase in the \sgra\ accretion rate during pericenter passage at $2014.39 \pm 0.14$ \citep{valencias15}, which corresponds to 2014 Apr. 20 (2014 Mar. 1$-$2014 Jun. 10).
Five flares with an absorbed fluence greater than $5\times 10^{-9}\ \mathrm{erg\ cm^{-2}}$ (corresponding to an unabsorbed fluence of $10.9\times 10^{-9}\ \mathrm{erg\ cm^{-2}}$ when using $\Gamma=2$ and $N_\mathrm{H}=14.3 \times 10^{22}\ \mathrm{cm^{-2}}$) were observed with XMM-Newton and Chandra between 2014 Aug. 30 and Oct. 20, implying an increase in the rate of energetic flares, but the overall flaring rate did not change \citep{ponti15}.

To assess the typical timescale for the accretion of fresh matter from the DSO/G2 object onto \sgra\  at pericenter, we compute the disk accretion timescale ($\tau_\mathrm{acc}$) for \sgra. 
It is governed by the viscous timescale, which is computed for an ADAF using the self-similar solution derived by \citet[][and references therein]{yuan14}.
At the distance $r$ of the SMBH, $\tau_\mathrm{acc}$ is defined as $r/V_\mathrm{rad}$ with $V_\mathrm{rad}$ the radial velocity for the self-similar solution, which gives us: $\tau_\mathrm{acc} \sim 3.0\,(r/2000R_\mathrm{s})^{1.5}\,(\alpha/0.1)^{-1}$~yrs with $\alpha\in[0,1]$ the efficiency of the mechanism of angular momentum transport introduced by \citet{shakura73}.
For a pericenter distance of about 2000~$R\mathrm{_s}$ \citep{pfuhl15,valencias15} and $\alpha=0.1$, we should not see any increase of the flux from \sgra\  before 2017.
Moreover, the large angular momentum of the gas and dust from DSO/G2 likely increases the true accretion timescale.

Some numerical simulations of the accretion of gas in a RIAF model were made, leading to a time range for the gas accretion of some months to several ten of years after the pericenter passage \citep{burkert12,schartmann12}.
However, these simulations modeled DSO/G2 as a gas cloud or a spherical shell of gas, but not as a young star with circumstellar material.
The accretion time when there is no central star may thus be lower than $\tau_\mathrm{acc}$, since the gas cloud is partially tidally disrupted before the pericenter passage.

\citet{zajacek14} modeled the DSO/G2 as an intermediate mass star of 2~$M_\odot$ moving in a RIAF.
They studied the tidal effects on a circumstellar dusty envelope and on a circumstellar accretion disk.
They showed that if the test particles are distributed in a disk-like structure, the number of particles that remain gravitationally bound to the star after the pericenter passage is larger than that for a spherical distribution of particles.
From their Fig.~13, we can also infer that the accretion onto \sgra\  begins earlier for a spherical distribution than for a disk-like model.
However, in these simulations, no circumstellar gas was taken into account.

\section{Conclusions}
\label{conclusion}
The pericenter passage of the DSO/G2 object at the beginning of 2014 was predicted to produce an increase of the flaring activity of \sgra\  in several wavelengths.
This 2014 Feb.$-$Apr. campaign was designed to follow an increase of its flaring activity simultaneously in X-rays, NIR, and radio/sub-mm.

Three NIR flares were detected with WFC3 on board HST: two on 2014 Mar. 10 (\textrm{I} and \textrm{II}) and one on Apr. 2 (\textrm{III}).
Two additional NIR flares were detected with SINFONI at ESO's VLT on 2014 Apr. 3 (\textrm{IV}) and 4 (\textrm{V}).
All of these NIR flares are within the top 8\% of the largest amplitude flares observed with NACO at ESO's VLT \citep{witzel12}.
Since the detection limit of WFC3 and SINFONI correspond to the 8 and 11\% amplitude levels of this sample, the fact that the observed NIR flares belong to the most luminous NIR flares is statistically expected and can not be taken as any indication for an increase of NIR activity.

Two X-ray flares were detected on 2014 Mar. 10 (1) and Apr. 2 (2) using the Bayesian-blocks method on the XMM-Newton observations.
The spectral parameters of these X-ray flares fitted with the MCMC method are consistent with those of the two brightest flares detected with XMM-Newton \citep{porquet03,porquet08}. 

The flare \textrm{I}/1 observed on 2014 Mar. 10 presents a change in the NIR to X-ray flux ratio, with an increase of the X-ray flux contribution during the second half of the flare.
We tested the three radiative processes that can explain the NIR/X-ray flares from \sgra\  as a single flare, considering energetic arguments.
The most likely interpretation is that the NIR and X-ray photons are produced in a plasmon in adiabatic compression by synchrotron and SSC emission mechanisms, respectively.
However, the flares \textrm{I} and 1 can also be decomposed into two Gaussian flares with a time separation of only 1.2~h.
We can thus associate the NIR flares \textrm{I}a and \textrm{I}b to the X-ray flares 1a and 1b, respectively.
They reproduce the characteristics observed in other simultaneous NIR/X-ray flares, i.e., no apparent delay between the maxima and a similar FWHM.
The flares \textrm{I}a/1a lie within the bulk of NIR flare amplitudes and peak ratio, but the flare \textrm{I}b/1b lies within the lowest peak ratio ever observed.

The NIR flare~\textrm{III} is actually composed of two close Gaussian flares (\textrm{III}a and \textrm{III}b).
The X-ray flare 2 is the counterpart of the NIR flare \textrm{III}a.
It has the lowest NIR-to-X-ray ratio ever observed.

The NIR flares \textrm{II}, \textrm{III}b, \textrm{IV}, and \textrm{V} have no detectable X-ray counterpart in our XMM-Newton observation or the legacy Chandra observation.
The upper limits on the X-ray amplitude were computed using the Bayesian method for the determination of the confidence limits described by \citet{helene83} and \citet{kraft91}.
The flare~\textrm{III}b lies within the largest NIR fluxes \citep[e.g.,][]{dodds-eden09,witzel12} and has the largest NIR-to-X-ray ratio ever observed.

In total, we detected seven NIR flares and three X-ray flares during the 2014 campaign.

On 2014 Mar.\ 10 we also identified an increase in the rising radio flux density at 13.37~GHz with the VLA, which could be the delayed radio emission from a NIR/X-ray flare that occurred before the start of our observation.

On 2014 Apr.\ 2 we identified a bump of the flux density on the rising 3.2-mm light curve observed with CARMA.
The time range of this observation does not allow us to associate this millimeter bump to a NIR/X-ray flare.
Moreover, we found no significant delay between the CARMA light curve and VLA $L-$ and $C-$band data.

On 2014 Apr.\ 3 two millimeter flares were identified above the decaying 3.2-mm light curve.
The former could be the delayed emission of the NIR flare~\textrm{IV}.

We derived physical parameters of the flaring emission for local radiative processes, as done previously by \citet{eckart12}, for each NIR/X-ray flare, and also for NIR flares with no detected X-ray counterpart.
Physical parameters for the flare \textrm{III}a/2 are better constrained when asssuming synchrotron and SSC emission mechanisms for the NIR and X-ray flares, respectively.
This flaring region has a size of 0.03$-$7 times the Schwarzschild radius and an electron density of $10^{8.5}$--$10^{10.2}$~cm$^{-3}$, for a synchrotron spectral-index of 0.3$-$1.5.
The derived physical parameters of the flaring emission associated with the undetected X-ray counterpart are poorly constrained, since the X-ray photon production efficiency is smaller.

We also tested the SYN-IC process using the NIR-to-X-ray peak amplitude ratio and the predicted time delay between the NIR and X-ray peaks.
This external radiative process is also a possible emission model for the emission of the flares observed during this campaign.

No significant increase in the X-ray flaring rate has been detected during this campaign, but continuous monitoring of \sgra\  is still important to detect any steady increase of its flaring activity that could be due to accreting material from the DSO/G2.
This may put some constrains on the physical properties of the G2 object and the ambient medium inside the Bondi radius of this SMBH.

\begin{acknowledgements}
This work has been financially supported by the Programme National Hautes Energies (PNHE).
The research leading to these results has received funding from the European Union Seventh Framework Program (FP7/2007-2013) under grant agreement n$^\circ$312789.
The XMM-Newton project is an ESA Science Mission with instruments and contributions directly funded by ESA Member States and the USA (NASA).
This work is based on observations made with the NASA/ESA Hubble Space Telescope obtained at the Space Telescope Science Institute, which is operated by the Association of Universities for Research in Astronomy, Inc., under NASA contract NAS 5-26555.
These HST observations are associated with programs 13403 and 13316.
This work is based on observations made with ESO Telescopes at the Paranal Observatory under programs 091.B-0183(H), 092.B-0920(A) and 093.B-0932(A).
Karl G. Jansky Very Large Array (VLA) of the National Radio Astronomy Observatory is a facility of the National Science Foundation, operated under a cooperative agreement by Associated Universities, Inc..
Support for CARMA construction was derived from the states of California, Illinois, and Maryland, the James S. McDonnell Foundation, the Gordon and Betty Moore Foundation, the Kenneth T. and Eileen L. Norris Foundation, the University of Chicago, the Associates of the California Institute of Technology, and the National Science Foundation. 
Ongoing CARMA development and operations are supported by the National Science Foundation under a cooperative agreement, and by the CARMA partner universities.
\end{acknowledgements}

\bibliographystyle{aa}
\bibliography{biblio_centre_galactique.bib}

\begin{appendix}
\section{The magnetar impact on the flare detection efficiency}
\label{appendix_b}
\begin{figure}
\centering
\includegraphics*[width=3.8cm, angle=90]{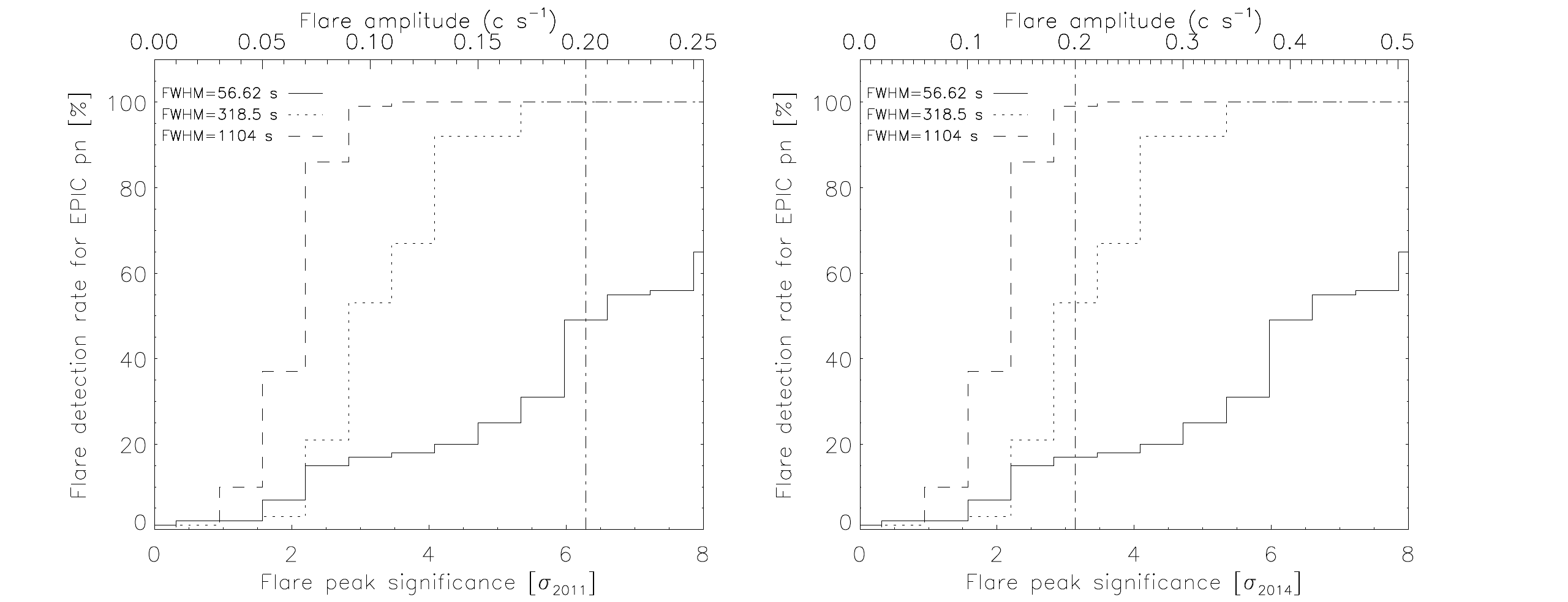}
\caption{Comparison of the flare detection level with the non-flaring level corresponding to those of the 2011 (\textit{left panel}) and February 2014 (\textit{right panel}) observations.
The vertical doted-dashed line represents an example flare with the same amplitude above the non-flaring level for a 2011 and 2014 Feb.$-$Apr. observations.}
\label{detection_level}
\centering
\includegraphics*[width=3.5cm,angle=90]{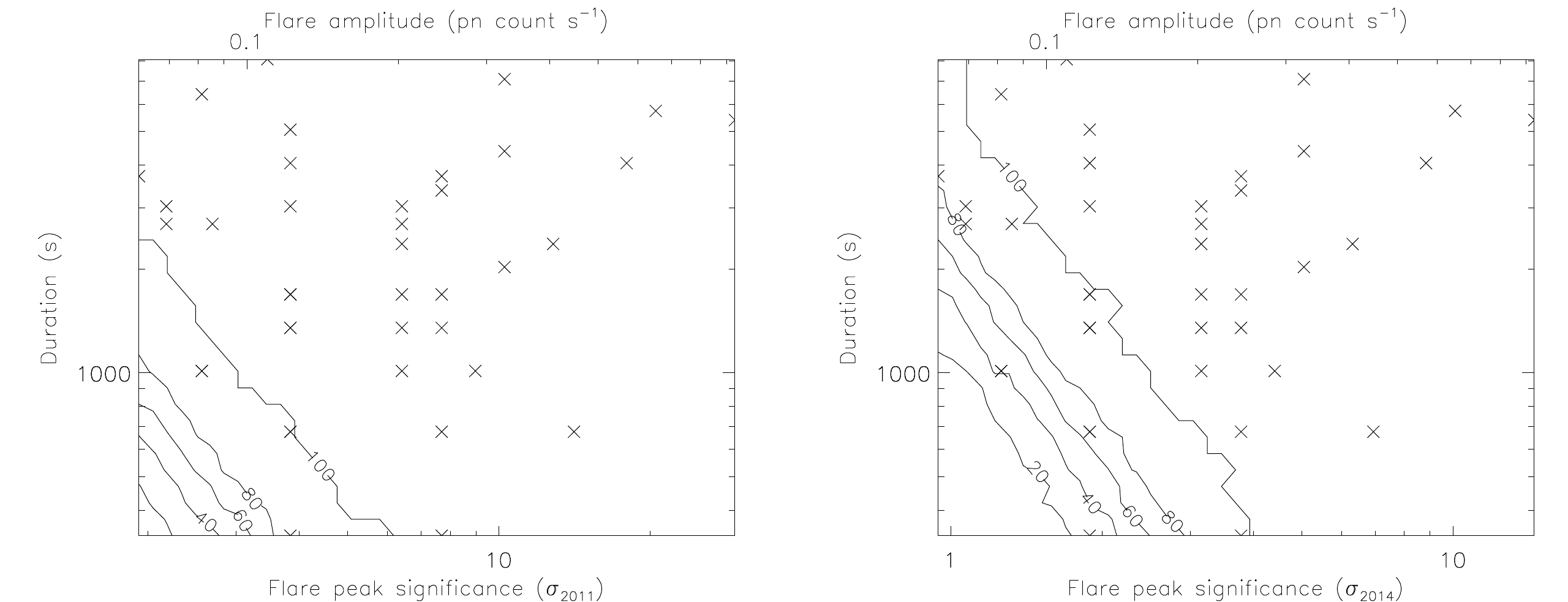}
\caption{Flare distribution seen by \textit{Chandra} and the detection probability of the Bayesian-blocks algorithm during an observation with XMM-Newton. 
The crosses are the X-ray flares detected  during the \textit{Chandra XVP} campaign of 2012.
\textit{Left panel:} The flare amplitude above the non-flaring level seen by EPIC/pn during the 2011 campaign.
\textit{Right panel:} The flare amplitude above the non-flaring level seen by EPIC/pn during the 2014 Feb. 28 observation.}
\label{distribution_flare}
\end{figure}
The contamination of the non-flaring level of \sgra\  by the Galactic center magnetar implies a decrease of the detection level of the faintest and shortest flares.
To assess the impact on our flare detection efficiency, we examine the flare detection rate (Fig. B.1. of \citealt{mossoux14}) versus the flare peak significance, i.e., the amplitude of the flare expressed in units of the standard deviation of the non-flaring level.
This scaling allows the comparison of observations with different non-flaring levels.
The flares used in these simulations have a Gaussian shape whose the Full Width at Half Maximum (FWHM) corresponds to the shortest, mean, and longest duration flares observed during the \textit{Chandra XVP} campaign of 2012 \citep{neilsen13}.
In Fig. \ref{detection_level}, we show the flare detection rate for the 2011 (left panel) and 2014 Feb. 28 (right panel) non-flaring levels for a false detection probability of $p_\mathrm{1}=\exp(-3.5)$.

We can see that because the non-flaring level in the 2014 Feb. 28 light curve has increased by a factor of about three by comparison with the 2011 campaign, the standard deviation is increased by a factor of about $\sqrt{3}$.
For example, if we consider a flare with an amplitude of 0.2$\ \mathrm{count}\ \mathrm{s^{-1}}$ above the non-flaring level, this corresponds to a peak significance of 6.3$\sigma$ for the 2011 light curves and this Gaussian shape flare is always detected if its duration is $\sim 320\ \mathrm{s}$ (FWHM).
A flare with the same amplitude in the 2014 Feb. 28 light curve corresponds to 3.2$\sigma$ and is only detected with a probability of 53\%.

In order to assess how many flares we cannot detect due to the magnetar contamination, we create a trial sample of flares following the duration and amplitude distribution determined during the \textit{Chandra XVP} campaign of 2012 \citep{neilsen13}.
We first compute a grid of 30 flare amplitudes and 30 flare durations in the range $[0.06-0.4]$ count s$^{-1}$ and $[337.5-8100]$~s, respectively, regularly distributed in the logarithmic scale.
For each point of the grid, we create 300 Gaussian flares characterized by the corresponding amplitude and duration (which is two times the standard deviation of the Gaussian).
We then apply the Bayesian blocks algorithm on all these flares superimposed above a non-flaring level corresponding to those of the 2011 XMM-Newton campaign seen with pn and each 2014 pn observation.
By computing how many flares are detected among the 300 simulated flares, we estimate the probability to detect a flare with a certain amplitude and duration.

Because \citet{neilsen13} detect 45 flares during a total time of $2983.93$~ks using the Bayesian-blocks method, we randomly select 100 sets of 45 flares following the amplitudes and durations distribution given by \citet{neilsen13}, i.e, $\mathrm{d}N/\mathrm{d}CR_\mathrm{Ch}=0.7\,CR_\mathrm{Ch}^{-1.9}e^{-CR_\mathrm{Ch}/0.3}$ and $\mathrm{d}N/\mathrm{d}T=0.05T^{-0.1}e^{-T/3000}$ with $CR_\mathrm{Ch}$ the peak count rate as observed by \textit{Chandra} and $T$ the flare duration\footnote{The cutoff value is given as a lower limit in \citet{neilsen13} but the specific value does not influence the result of our flare distribution because we are interested by flares characterized by small amplitude and short duration since these flares may suffer of the small detection rate.}.
In order to convert the \textit{Chandra} count rate to the XMM-Newton count rate ($CR_\mathrm{XMM}$), we can use the relation derived in \citet{mossoux14} between the Chandra HETG count rate (zero and first order) of the flare peak and the unabsorbed luminosity at the peak flare, i.e., $L^{\mathrm{unabs}}_{2-10\,\mathrm{keV}}/10^{34}\,\mathrm{erg\,s^{-1}}=-0.031+136.7CR_\mathrm{Ch}$.
This unabsorbed luminosity is obtained with the spectral index $\Gamma=2$ and the hydrogen column density $N_\mathrm{H}=14.3 \times 10^{22}\ \mathrm{cm^{-2}}$ \citep{neilsen13}.
We determine with the arf and rmf files of pn a count rate to unabsorbed luminosity ratio of $2.96 \times 10^{-36} \,\mathrm{pn\ count\,s^{-1}}/\mathrm{erg\,s^{-1}}$.
We can thus convert the Chandra count rate to the pn count rate assuming the same spectral parameters.
Since each flare can be associated to a detection probability between 0 and 1, the sum of the probability for the 45 flares give us the total number of flares that can be detected in average by the Bayesian-blocks method during a pn observation with an exposure time of $2.98393 \times 10^6$~s.
The distribution of the flare duration and amplitude seen during the \textit{Chandra XVP} campaign and the detection probability of the Bayesian-blocks algorithm is shown in Fig~\ref{distribution_flare}.
The left and right panels if this figure represents the detection probability corresponding to the mean non-flaring level seen by XMM-Newton during the 2011 campaign and to those observed during the 2014 Feb. 28 observation, respectively.

The mean of the number of detected flares for the 100 sets shows that considering the non-flaring level of the 2011 campaign, we can detect 85.4\% of the flares detected during the \textit{Chandra XVP} campaign.
The non-detected flares are the faintest and shortest ones.
For the 2014 Feb. 28, Mar. 10, Apr. 2 and Apr. 3, we detect 79.2\%, 79.4\%, 80.1\% and 79.8\% of the flares detected during the \textit{Chandra XVP} campaign, respectively.
Therefore, we estimate that we missed about 20.4\% of the flares from Sgr A*. Since we detected three flares this means that we lost no more than one flare.

\section{Filtering out of the magnetar pulsed emission}
\label{appendix_d}
To filter out the magnetar contamination, we first computed the period ($P$) and period ($\dot{P}$) derivative of \magn{} by folding the light curve of all XMM-Newton observations of this campaign in which gaps between observations, GTI and exposure correction were taken into account.
The relation between events arrival times $t$ in the barycentric referential (computed using the SAS task \texttt{barycen}) and the magnetar phase can be written as a Taylor series on the time:
\begin{equation}
 \phi(t)=\phi_0+\frac{t-t_0}{P}-0.5\, \frac{(t-t_0)^2}{P^2}\, \dot{P}\, ,
 \label{eq:phi}
\end{equation}
\begin{figure}
\centering
\includegraphics[width=5cm, angle=90]{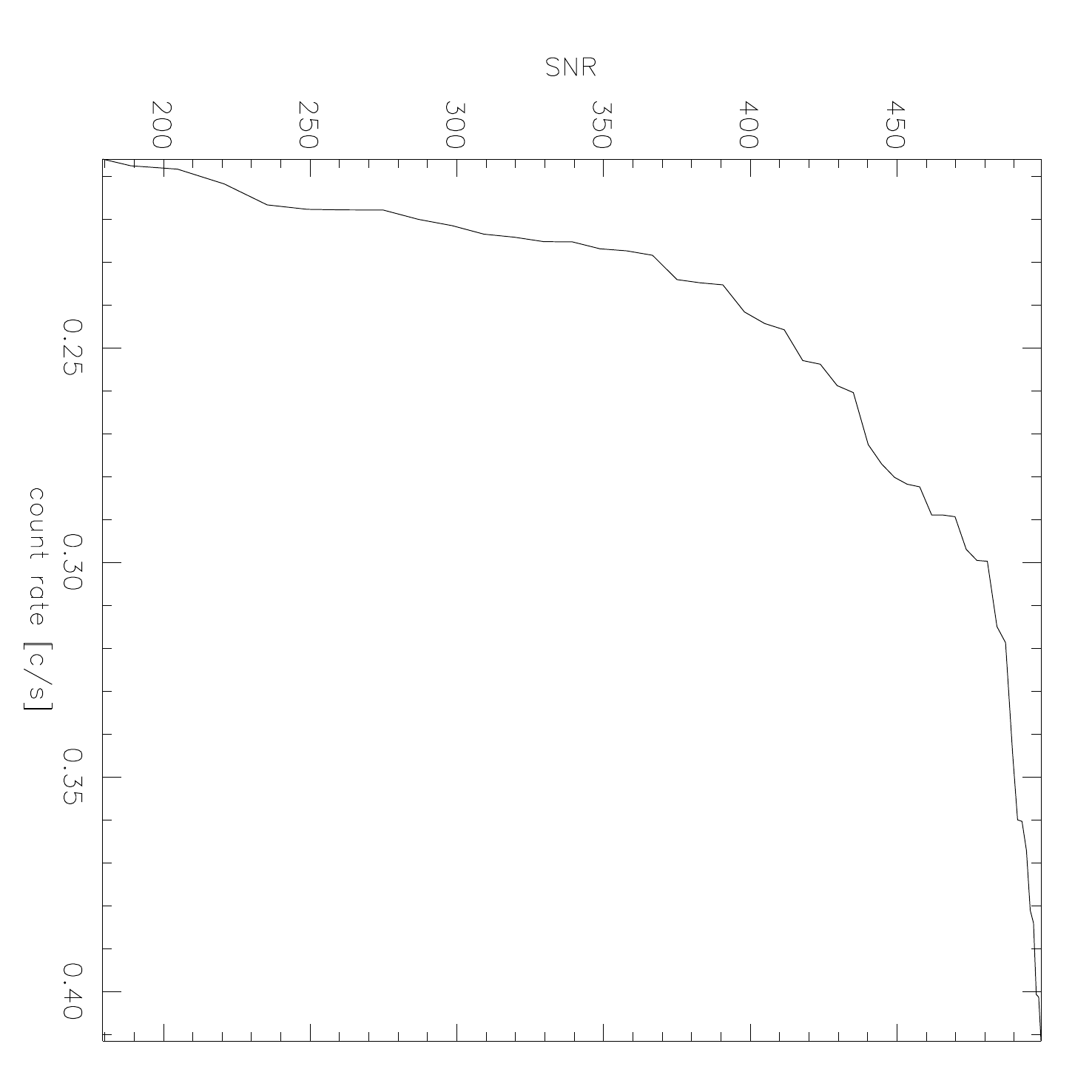}
\caption{Evolution of the $S/N$ as a function of $CR_{\mathrm{th}}$.}
\label{snr}
\end{figure}
\begin{figure}
\centering
\includegraphics[trim = 0.cm 0.cm 0.6cm 0.7cm, clip,height=4.2cm, angle=90]{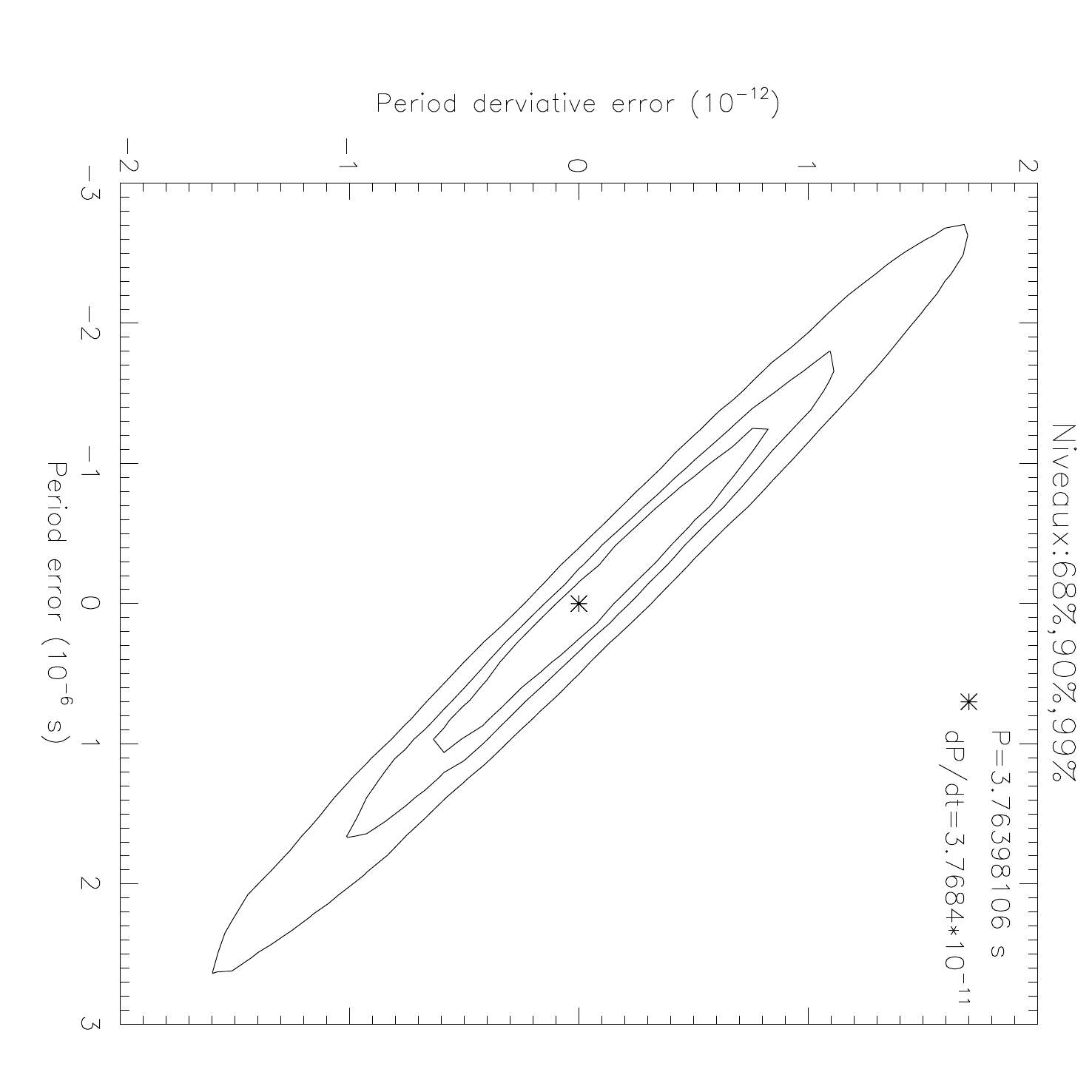}
\includegraphics[height=4.4cm, angle=90]{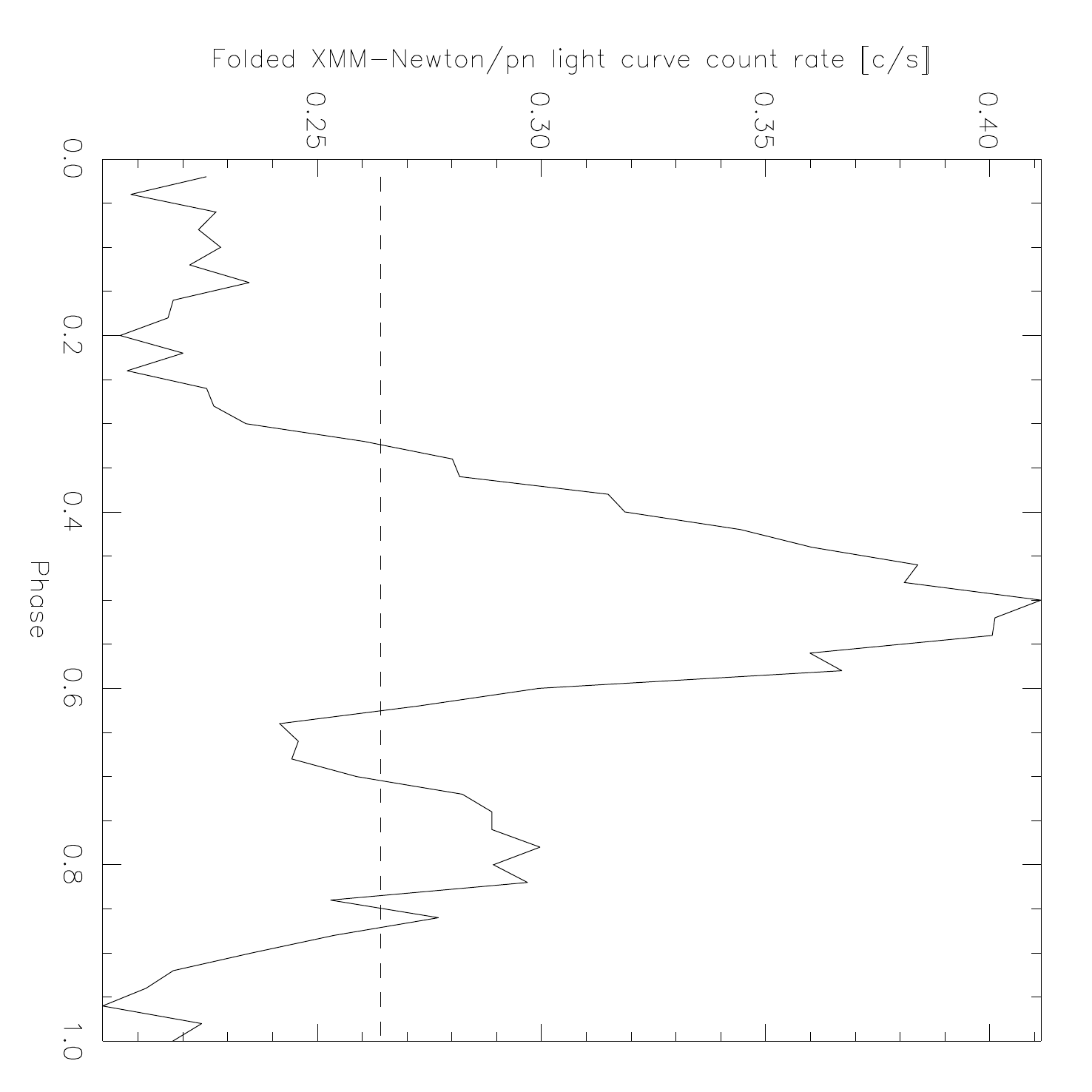}
\caption{\textit{Left panel:} $\chi^2$ distribution of the period and period derivative of the magnetar.
The contours are the 68\%, 90\% and 99\% of confidence level on the parameters.
\textit{Right panel:} Folded light curve on the four XMM-Newton observations with our best fit parameters (see Table~\ref{table:magnetar}).}
\label{chi_carre}
\end{figure}
with $t_0$ the start time of the first XMM-Newton observation and $\phi_0$ an arbitrary phase.
We choose $\phi_0$ in order to have the maximum of the pulse at $\phi=0.5$.
A $\chi^2$ fitting with a constant function was applied on the folded light curve.
The maximum $\chi^2$ give us the better period and period derivative and the corresponding 1-$\sigma$ errors which are reported on Table~\ref{table:magnetar}.
The confidence level of the $\chi^2$ distribution for these two parameters is given in Fig.~\ref{chi_carre} (left panel).
A comparison with the parameters derived from the literature is also shown.
For this comparison, we use the period and period derivative given in Table~\ref{table:magnetar}.
The folded light curve for these parameters is represented in Fig.~\ref{chi_carre} (right panel).
We consider only the EPIC/pn camera because it has a better time resolution (73.4 ms) than EPIC/MOS (2.6 s) \citep{XMM_UHB}.

\begin{table*}
\caption{Period and period derivative taken from the literature and from this work.}
\centering
\scalebox{.67}{
\label{table:magnetar}
\begin{tabular}{@{}ccccccc@{}}
\hline
\hline
References & \multicolumn{1}{c}{Period} & \multicolumn{1}{c}{Period derivative} & \multicolumn{1}{c}{Period second derivative} & \multicolumn{1}{c}{Epoch\tablefootmark{a}} & \multicolumn{1}{c}{Period on 56716 (MJD)\tablefootmark{b}} & \multicolumn{1}{c}{Period derivative on 56716 (MJD)\tablefootmark{c}} \\
 & (s) & \multicolumn{1}{c}{(s s$^{-1}$)} & \multicolumn{1}{c}{(s$^{-1}$)} & \multicolumn{1}{c}{(MJD)} & \multicolumn{1}{c}{(s)} & \multicolumn{1}{c}{(s s$^{-1}$)}\\
\hline  
 \citealt{mori13}$\ \ \ \ \ \ \ \ \ \ \ \ \ $  & $3.76354455 \pm 7.1 \times 10^{-7}$ & $\ \ \ \ 6.5 \times 10^{-12} \pm 1.4 \times 10^{-12}$ &  & $56409.2657$ & $\ \ \ \ 3.7637 \pm 6.18 \times 10^{-2}$ &  \\
 \citealt{rea13}$\ \ \ \ \ \ \ \ \ \ \ \ \ \ \ $ & $\ \ \ \ \ 3.7635537 \pm 2 \times 10^{-7}$ & $\ \ \ \ \ 6.61 \times 10^{-12} \pm 4 \times 10^{-14}$ &  & $\ \ \ \ 56424.55$ & $\ \ 3.76372 \pm 1.78 \times 10^{-3}$ &  \\
 \citealt{kaspi14}$\ \ \ \ \ \ \ \ \ \ \ \ $ & $3.76363824 \pm 1.3 \times 10^{-7}$ & $1.385 \times 10^{-11} \pm 1.5 \times 10^{-13}$ & $3.9 \times 10^{-19} \pm 6 \times 10^{-20}$ & $\ \ \ \ \ \ \ \ \ 56513$ & $3.7639871 \pm 6.2 \times 10^{-6}$ & $\ \ 2.05 \times 10^{-11} \pm 1.1 \times 10^{-12}$ \\
 \citealt{coti15} A & $\ \ \ 3.76363799 \pm 7 \times 10^{-8}$ & $\ \ \ 1.360 \times 10^{-11} \pm 6 \times 10^{-14}$ & $3.7 \times 10^{-19} \pm 2 \times 10^{-20}$ & $\ \ \ \ \ \ \ \ \ 56513$ & $3.7639903 \pm 1.1 \times 10^{-6}$ & $ 2.089 \times 10^{-11} \pm 3.5 \times 10^{-13}$ \\
 \citealt{coti15} B & $\ \ 3.7639772 \pm 1.2 \times 10^{-6}$ & $\ \ \ \ \ 3.27 \times 10^{-11} \pm 7 \times 10^{-13}$ &  & $\ \ \ \ \ \ \ \ \ 56710$ & $3.7639942 \pm 1.3 \times 10^{-6}$ &  \\
 This work\tablefootmark{d}$\ \ \ \ \ \ \ \ \ \ \ \ \ \ \ \ \ \ \ \ $ & $\ \ \ \ \ \ \ \ \ \period{}^{+\errDroiteP{}}_{-\errGaucheP{}}$ & $\ \ \ \ \ \ \ \dper{}^{+\errDroiteDp{}}_{-\errGaucheDp{}}$ &  & $\ \ \ \ \ \ \ \ \ 56716$ & $\ \ \ \ \ \ \ \period{}^{+\errDroiteP{}}_{-\errGaucheP{}}$ & $\ \ \ \ \ \ \dper{}^{+\errDroiteDp{}}_{-\errGaucheDp{}}$ \\
\hline
\end{tabular}
}
\tablefoot{
\tablefoottext{a}
{Reference epoch for computing the parameters. MJD=TJD+40000 days=JD-2400000.5 days;}
 \tablefoottext{b}
 {The period on $t=56716$ (MJD) is computed using $P=P_0+\dot{P}_0(t-t_0)+\ddot{P}_0(t-t_0)^2$ with $P_0$, $\dot{P}_0$, $\ddot{P}_0$ the period, period derivative and period second derivative given in the literature, $t_0$ the reference epoch in the literature.
Errors are propagated until $t=56716$ (MJD) thanks to $dP^2=\sum (\partial P/\partial p)^2\,dp^2$;}
 \tablefoottext{c}
 {The period derivative on $t=56716$ (MJD) is computed using $\dot{P}=\dot{P}_0+\ddot{P}_0(t-t_0)$ with the definitions given above.}
 \tablefoottext{d}
 {The errors are the 90\% confidence interval (see left panel of Fig.~\ref{chi_carre}).}
}
\normalsize
\end{table*}

We use this folded light curve to compute the count rate threshold which maximizes the signal-to-noise ratio.
As the magnetar flux is an additional noise on the \sgra\  light curve, magnetar flux contribution at each phase ($\tau$) of the folded
 light curve is $N_{\mathrm{magnetar}}(\tau)=\int^\tau_0 (CR_{\mathrm{fold}}(t)-CR_{\mathrm{Sgr\ A^*}})\,dt$ with $CR_{\mathrm{fold}}$ the
count rate of the folded light curve and $CR_{\mathrm{Sgr\ A^*}}=0.10\ \mathrm{count}\ \mathrm{s^{-1}}$ the non-flaring level of \sgra\  seen with pn \citep[e.g.,][]{mossoux14}.
The signal-to-noise ratio is
\begin{equation}
 S/N=\frac{CR_{\mathrm{Sgr\ A^*}}\, \tau}{\sqrt{N_{\mathrm{magnetar}}(\tau)}}\, .
\end{equation}
The phase $\tau$ which maximizes the \textit{S/N} allows us to compute the corresponding count rate threshold ($CR_{\mathrm{th}}$).
Figure \ref{snr} shows that there is no optimum value of the count rate threshold maximizing the \textit{S/N}.
Thus, we consider a count rate threshold which filters out 50\% of the magnetar flux.
This threshold is $0.27\ \mathrm{count}\ \mathrm{s^{-1}}$ and keeps 50\% of the observation time.
Then, from $P$ and $\dot{P}$, the time interval during which the count rate of the folded light curve is lower than $CR_{\mathrm{th}}$ can be computed for all observations from Eq.~\ref{eq:phi}.
Thus we can construct a new GTI file which is the combination of the GTI file from the event list of pn (which contains the time interval during which the cameras do not observe) and the GTI file created by removing the magnetar pulse using the SAS task \texttt{gtimerge}.

\section{The two X-ray flares seen in EPIC/pn, MOS1 and MOS2 cameras}
\label{appendix_a}
\begin{figure*}
\centering
\begin{tabular}{ccc}
\includegraphics[angle=90,width=6.2cm]{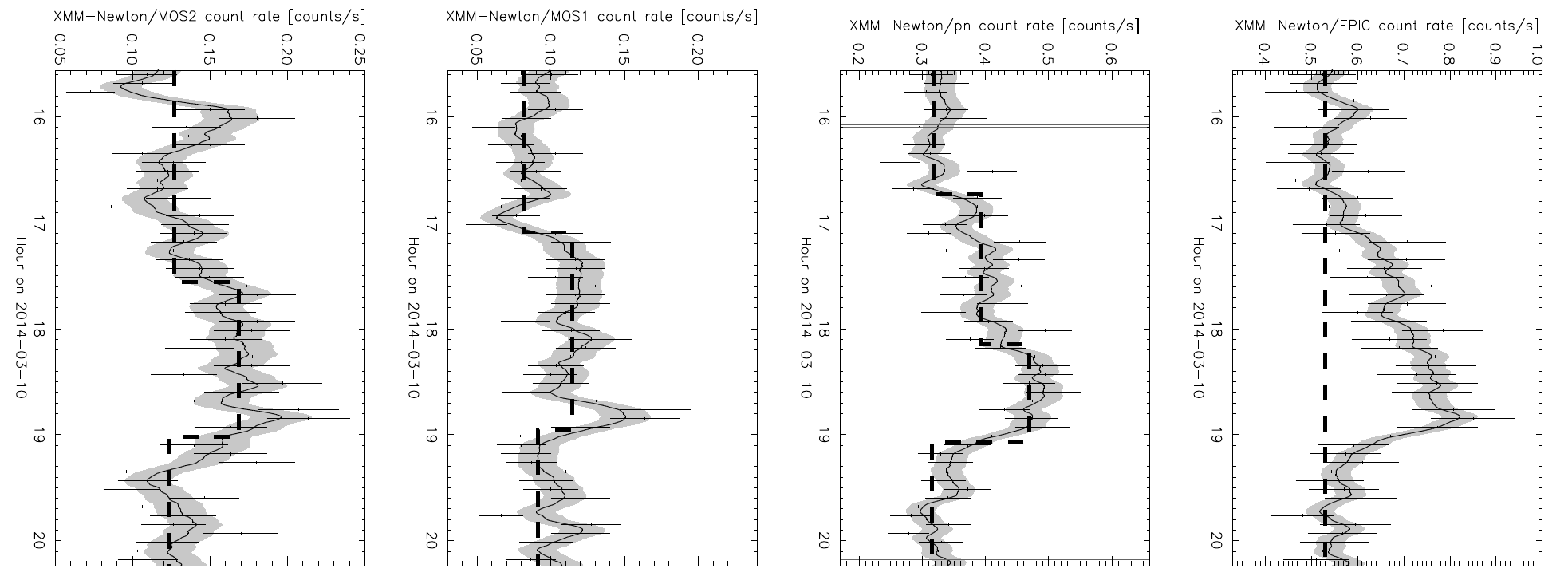} &
\hspace*{2cm}
\includegraphics[angle=90,width=6.2cm]{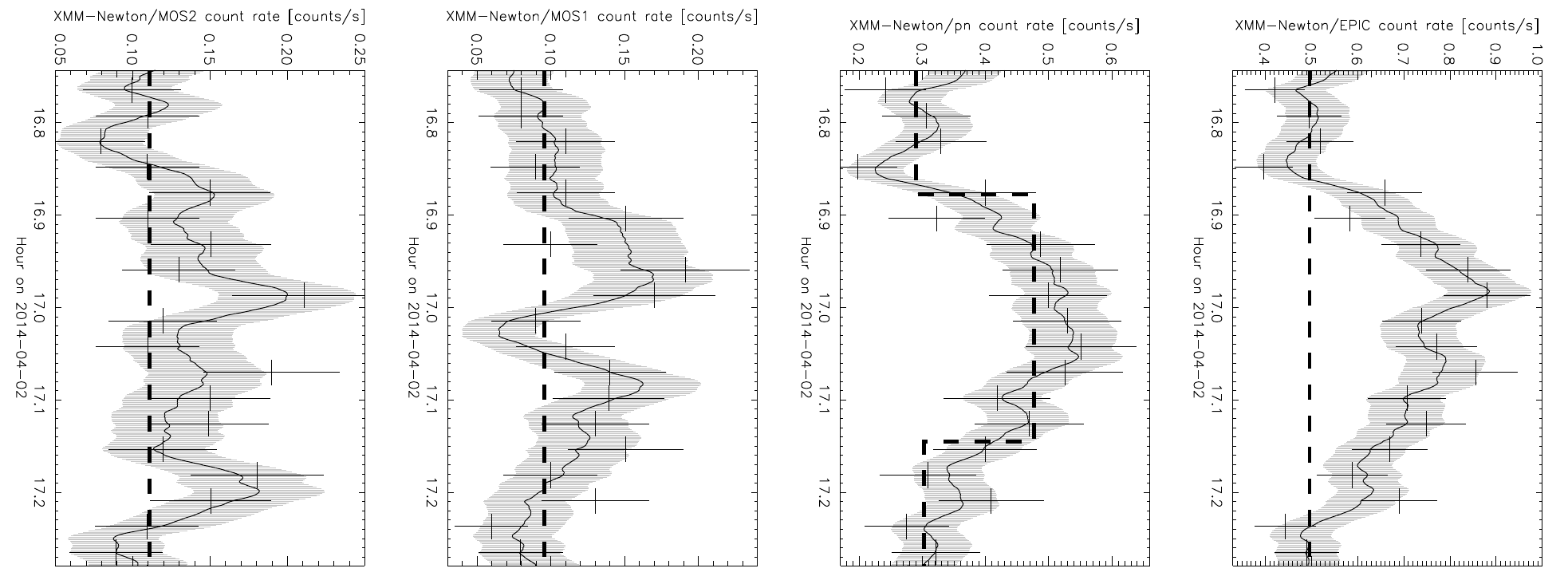}
\end{tabular}
\caption{XMM-Newton light curve of the X-ray from \sgra\  in the $2-10$~keV energy range. 
\textit{Left panels:} The light curve of flare 1 on 2014 Mar. 10 flare binned on 500s.
\textit{Right panels:} The light curve of flare 2 on 2014 Apr. 2 flare binned on 100s.
The total (pn+MOS1+MOS2) light curve is shown in the top panel.
The light curves of EPIC/pn, MOS1 and MOS2 are shown in the second, third and bottom panels.
The crosses are the data points of the total light curve.
The horizontal dashed line and the solid line are the sum of the non-flaring level and the smoothed light curve for each instrument.
The dashed lines represent the Bayesian blocks.
The solid line and the gray curve are the smoothed light curve and the associated errors ($h=500$ and 100s for flare 1 and 2, respectively).
The vertical gray stripe is the time during which the camera did not observe.}
\label{Fig6}
\end{figure*}

Figures \ref{Fig6} shows the flare light curves obtained with EPIC on board XMM-Newton on 2014 Mar. 10 (left panels) and Apr. 2 (right panels).
The Bayesian-blocks algorithm characterizes the 2014 Mar. 10 flare with two blocks in the pn light curve but only with one block in the MOS1 and MOS2 light curves.
Moreover, the duration of the flares seen in each camera is different (see Table~\ref{table:mos_mars}).
This can be explained by the lower number counts in MOS1 and MOS2 because of the RGS: the number of photons recorded by pn during the flare is larger and thus the accuracy on the determination of the beginning and end of the flare is better.
\begin{table}
\caption{Characteristics of the X-ray flare observed by EPIC/MOS on 2014 Mar. 10.}
\centering
\scalebox{.81}{
\label{table:mos_mars}
\begin{tabular}{@{}cccccc@{}}
\hline
\hline
Instrument  & Start Time\tablefootmark{a} & End Time\tablefootmark{a} & Duration & Total\tablefootmark{b} & Peak\tablefootmark{c} \\
 & (hh:mm:ss) & (hh:mm:ss)  & (s) & (counts) & ($\mathrm{count}\ \mathrm{s^{-1}}$) \\
\hline  
 MOS1 & 17:05:14 & 18:56:59 & 6705 & $780 \pm 28$ & $0.06 \pm 0.02$\\
 MOS2 & 17:33:32 & 19:01:11 & 5258 & $880 \pm 30$ & $0.07 \pm 0.02$ \\
\hline
\end{tabular}
}
\tablefoot{
\tablefoottext{a}
{Start and end times (UT) of the flare time interval defined by the Bayesian-blocks algorithm;}
\tablefoottext{b}
{Total counts in the 2$-$10~keV energy band obtained in the smoothed light curve during the flare interval after subtraction of the non-flaring level obtained with the Bayesian-blocks algorithm;}
\tablefoottext{c}
{Peak count rate in the 2$-$10~keV energy band at the flare peak (smoothed light curves) after subtraction of the non-flaring level.}
}
\normalsize
\end{table}

The flare on 2014 Apr. 2 is not detected by the Bayesian-blocks algorithm in MOS1 and MOS2 because the amplitude and the number of counts in this flare is rather small.

\end{appendix}

\end{document}